\documentclass[12pt,english,aps,preprint,prd,letterpaper,fleqn,nofootinbib,showpacs,showkeys,tightenlines,floatfix]{revtex4}
\usepackage[T1]{fontenc}
\usepackage[latin1]{inputenc}
\usepackage{float}
\usepackage{color}
\usepackage{graphicx}
\usepackage{amssymb}

\makeatletter

\usepackage{float}

\makeatletter

\usepackage{geometry}



\makeatother

\usepackage{babel}
\makeatother
\begin{document}
\newcommand{\barl}{\bar{\lambda}} \newcommand{\barp}{\bar{p}}

\preprint{ANL-HEP-PR-07-12, arXiv:0704.0001}

\title{Calculation of prompt diphoton production cross sections at Tevatron
and LHC energies}

\author{C. Bal\'{a}zs$^{1}$}

\thanks{balazs@hep.anl.gov; Current address: School of Physics, Monash University, Melbourne VIC 3800, Australia}

\author{E.~L.~Berger$^{1}$}

\thanks{berger@anl.gov}

\author{P. Nadolsky$^{1}$}

\thanks{nadolsky@hep.anl.gov}

\author{C.-P. Yuan$^{2}$}

\thanks{yuan@pa.msu.edu}

\affiliation{$^{1}$High Energy Physics Division, Argonne National Laboratory,
Argonne, IL 60439 \\
 $^{2}$Department of Physics and Astronomy, Michigan State University,
East Lansing, MI 48824}

\begin{abstract}
A fully differential calculation in perturbative quantum 
chromodynamics is presented for the production of massive photon pairs at hadron 
colliders. All next-to-leading order perturbative contributions from quark-antiquark, 
gluon-(anti)quark, and gluon-gluon subprocesses are included, as well as all-orders 
resummation of initial-state gluon radiation valid at next-to-next-to-leading 
logarithmic accuracy.  The region of phase space is specified in which 
the calculation is most reliable. Good agreement is demonstrated with data from 
the Fermilab Tevatron, and predictions are made for more detailed tests with CDF and
D\O~data.  Predictions are shown for distributions of diphoton pairs
produced at the energy of the Large Hadron Collider (LHC).  Distributions of the 
diphoton pairs from the decay of a Higgs boson are contrasted with those 
produced from QCD processes at the LHC, showing that enhanced sensitivity to 
the signal can be obtained with judicious selection of events.  
\end{abstract}

\date{May 3, 2007}



\maketitle

\section{Introduction}

The long-sought Higgs boson(s) $h$ of electroweak symmetry breaking
in particle physics may soon be observed at the CERN Large Hadron
Collider (LHC) through the diphoton decay mode ($h\rightarrow\gamma\gamma$).
Purely hadronic standard model processes are a copious source of diphotons,
and a narrow Higgs boson signal at relatively low masses will appear
as a small peak above this considerable background. A precise theoretical
understanding of the kinematic distributions for diphoton production
in the standard model could provide valuable guidance in the search
for the Higgs boson signal and assist in the important measurement 
of Higgs boson coupling strengths.

In this paper we address the theoretical calculation of the invariant
mass, transverse momentum, rapidity, and angular distributions of
continuum diphoton production in proton-antiproton and proton-proton
interactions at hadron collider energies. We compute all contributions
to diphoton production from parton-parton subprocesses through next-to-leading
order (NLO) in perturbative quantum chromodynamics (QCD). These higher-order
contributions are large at the LHC, and their inclusion is mandatory
for quantitatively trustworthy predictions. We resum initial-state
soft and collinear logarithmic terms associated with gluon radiation to all
orders in the strong coupling strength $\alpha_{s}$. This resummation
is essential for physically meaningful predictions of the transverse
momentum ($Q_{T}$) distribution of the diphotons at small and intermediate
values of $Q_{T}$, where the cross section is large. In addition,
we analyze the final-state collinearly-enhanced contributions, also
known as `fragmentation' contributions, in which one or both photons
are radiated from final-state partonic constituents. We compare the
results of our calculations with data on isolated diphoton production
from the Fermilab Tevatron~\cite{Acosta:2004sn}. The good agreement
we obtain with the Tevatron data adds confidence to our predictions
at the energy of the LHC. The present work expands on our recent abbreviated
report~\cite{Balazs:2006cc}, and it may be read in conjunction with
our detailed treatment of the contributions from the gluon-gluon subprocess~\cite{Nadolsky:2007ba}.
\begin{figure}
\includegraphics[width=8cm]{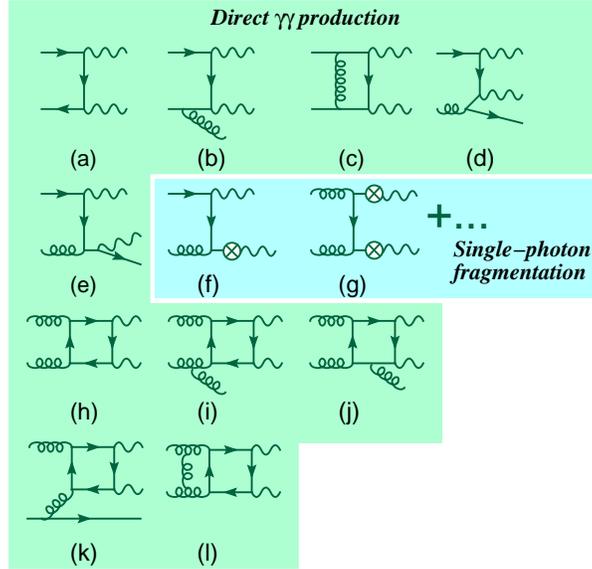}

\caption{Representative partonic subprocesses that contribute to continuum
diphoton production. All leading-order and next-to-leading order direct
production subprocesses, i.e., contributions (a)-(e) and (h)-(l),
are included in this study. Diagrams (f) and (g) are examples of single-photon
one- and two-fragmentation. \label{Fig:FeynDiag}}
\end{figure}

Our attention is focused on the production of isolated photons, \emph{i.e.},
high-energy photons observed at some distance from appreciable hadronic
remnants in the particle detector. The rare isolated photons tend
to originate directly in hard QCD scattering, in contrast to copiously
produced non-isolated photons that arise from nonperturbative processes
such as $\pi$ and $\eta$ decays, or from via quasi-collinear radiation
off final-state quarks and gluons.

We evaluate contributions to continuum diphoton production from the
basic short-distance channels for $\gamma\gamma$ production initiated
by quark-antiquark and (anti)quark-gluon scattering, as well as by
gluon-gluon and gluon-(anti)quark scattering proceeding through a
fermion-loop diagram. At lowest order in QCD, a photon pair is produced
from $q\bar{q}$ annihilation {[}Fig.~\ref{Fig:FeynDiag} (a)]. Representative
next-to-leading order (NLO) contributions to $q\bar{q}+qg$ scattering
are shown in Fig.~\ref{Fig:FeynDiag} (b)-(e). They are of ${\mathcal{O}}(\alpha_{s})$
in the strong coupling strength~\cite{Aurenche:1985yk,Bailey:1992br}.
Production of $\gamma\gamma$ pairs via a box diagram in $gg$ scattering
{[}Fig.~\ref{Fig:FeynDiag} (h)] is suppressed by two powers of $\alpha_{s}$
compared to the lowest-order $q\bar{q}$ contribution, but it is enhanced
by a product of two large gluon parton distribution functions (PDFs)
if typical momentum fractions $x$ are small~\cite{Berger:1983yi}.
The ${\mathcal{O}}(\alpha_{s}^{3})$ or NLO corrections to $gg$ scattering
include one-loop $gg\rightarrow\gamma\gamma g$ diagrams (i) and (j)
derived in Ref.~\cite{Balazs:1999yf,deFlorian:1999tp}, as well as
4-leg two-loop diagrams (l) computed in Ref.~\cite{Bern:2001df,Bern:2002jx}.
In this study we also include subleading contributions from the process
(k), $gq_{S}\rightarrow\gamma\gamma q_{S}$ via the quark loop, where
$q_{S}=\sum_{i=u,d,s,...}(q_{i}+\bar{q}_{i})$ denotes the flavor-singlet
combination of quark scattering channels.

Factorization is a central principle of hadronic calculations in perturbative
QCD, in which a high-energy scattering cross section is expressed as
a convolution of a perturbative partonic cross section with nonperturbative
parton distribution functions (PDFs), thus separating short-distance
from long-distance physics. The common factorization is a longitudinal
notion, in the sense that the convolution is an integral over longitudinal
momentum fractions, even if some partons in the hard-scattering process 
have transverse momenta that border the nonperturbative regime.
Unphysical features may then arise in the transverse momentum
($Q_{T}$) distribution of a color-neutral object with high invariant
mass ($Q$), such as a pair of photons produced in hadron-hadron collisions.
When calculated in the common factorization approach at any finite
order in perturbation theory, this distribution diverges as $Q_{T}\rightarrow0$,
signaling that infrared singularities associated with $Q_{T}\rightarrow0$
have not been properly isolated and regulated. These singularities
are associated with soft and collinear radiation from initial-state
partons shown by the diagrams in Figs.~\ref{Fig:FeynDiag} (b), (d),
and (i).

A generalized factorization approach that correctly describes the
small-$Q_{T}$ region was developed by Collins, Soper, and Sterman
(CSS)~\cite{Collins:1984kg} and applied to photon pair production~\cite{Balazs:1997hv,Balazs:1999yf,Nadolsky:2002gj}.
In this approach the hadronic cross section is expressed as an integral
over the transverse coordinate (impact parameter). The integrable singular functions present
in the finite-order differential distribution as $Q_{T}\rightarrow0$
are resummed, to all orders in the strong coupling $\alpha_{s}$,
into a Sudakov exponent, and a well-behaved cross section is obtained
for all $Q_{T}$ values. As explained in Sec.~II, our resummed calculation
is accurate to next-to-next-to-leading-logarithmic (NNLL) order. It
is applicable for values of diphoton transverse momentum that are less 
than the diphoton mass, i.e., for $Q_{T}<Q$.
When $Q_{T}\sim Q$, terms of the form $\ln^{n}(Q_{T}/Q)$ become
 small. A perturbative expansion with a single hard scale is then
applicable, and the cross section can be obtained from finite-order
perturbation theory.

In addition to the initial-state logarithmic singularities,
there is a set of important final-state singularities which arise
in the matrix elements when at least one photon's momentum is collinear
to the momentum of a final-state parton. They are sometimes referred
to as `fragmentation' singularities. At lowest order in $\alpha_{s}$,
the final-state singularity appears only in the $qg\rightarrow\gamma\gamma q$
diagrams, as in Fig.~\ref{Fig:FeynDiag} (e). There are various methods
used in the literature to deal with the final-state singularity, including
the introduction of explicit fragmentation functions $D_{\gamma}(z)$
for hard photon production, where $z$ is the light-cone fraction
of the intermediate parton's momentum carried by the photon. These
single-photon ``one-fragmentation'' and ``two-fragmentation'' 
contributions, corresponding to one or both photons produced
in independent fragmentation processes, are illustrated
by the diagrams in Figs.~\ref{Fig:FeynDiag} (f) and (g). In addition,
a fragmentation contribution of entirely different nature arises when
the $\gamma\gamma$ pair is relatively light and produced from
fragmentation of {\it one} parton, as discussed in
Secs.~\ref{sub:Low-Q-diphoton-fragmentation} and \ref{subsection:QT_gt_Q}.
A full and consistent treatment of the final-state logarithms beyond
lowest order would require a joint resummation of the initial- and
final-state logarithmic singularities. 

In the work reported here, we are guided by our interest in describing
the cross section for \emph{isolated} photons, in which the fragmentation
contributions are largely suppressed. A typical isolation condition
requires the hadronic activity to be minimal (e.g., comparable to
the underlying event) in the immediate neighborhood of each candidate
photon. Candidate photons can be rejected by energy deposit nearby 
in the hadronic calorimeter or the presence of hadronic
tracks near the photons. A theory calculation may approximate the
experimental isolation by requiring the full energy of the hadronic
remnants to be less than a threshold {}``isolation energy'' $E_{T}^{iso}$
in a cone of size $\Delta R$ around each photon. The two photons
must be also separated in the plane of the rapidity $\eta$ and azimuthal angle
$\varphi$ by an amount exceeding the resolution $\Delta R_{\gamma\gamma}$
of the detector. The values of $E_{T}^{iso},$ $\Delta R$, and $\Delta R_{\gamma\gamma}$
serve as crude characteristics of the actual measurement. The magnitude
of the final-state fragmentation contribution depends on the assumed 
values of $E_{T}^{iso},$ $\Delta R$, and $\Delta R_{\gamma\gamma}$.

An additional complication arises when the fragmentation radiation
is assumed to be exactly collinear to the photon's momentum, as implied
by the photon fragmentation functions $D_{\gamma}(z)$. 
The collinear approximation
constrains from below the values of $z$ accessible to $D_{\gamma}(z)$:
$z>z_{min}$. The size of the fragmentation contribution may depend
strongly on the values of $E_{T}^{iso}$ and $z_{min}$ as a consequence
of rapid variation of $D_{\gamma}(z)$ with $z$.

In our work we treat the final-state singularity using a prescription
that reproduces desirable features of the isolated cross sections
while bypassing some of the technical difficulties alluded to above.
For $Q_{T}>E_{T}^{iso}$, we avoid the final-state collinear singularity
in the $qg$ scattering channel by applying quasi-experimental isolation.
When $Q_{T}<E_{T}^{iso}$, we apply an auxiliary regulator which approximates
on average the full NLO rate from direct $qg$ and fragmentation cross
sections in this $Q_{T}$ range. Two prescriptions for the auxiliary
regulator (subtraction and smooth-cone isolation inside the photon's
isolation cone) are considered and lead to similar predictions at
the Tevatron and the LHC.

We begin with our notation in Sec.~\ref{sub:Notations}, followed by an overview
of the procedure for resummation of initial-state multiple parton
radiation in Sec.~\ref{subsection:ISRResummation}. The issue of
the final-state fragmentation singularity is discussed in
Sec.~\ref{subsection:Fragmentation-model}. Our approach is compared 
with that of the DIPHOX calculation~\cite{Binoth:1999qq},
in which explicit fragmentation function contributions are included
at NLO, but all-orders resummation is not performed.
Our theoretical framework is summarized in Sec.~\ref{sub:TheorySummary}.

In Sec.~\ref{Sec:Phenomenology} we compare the predictions of our
resummation calculation with Tevatron data.  Resummation is shown to 
be important for the successful description of physical $Q_{T}$ 
distributions, as well as for stable estimates of the effects of experimental 
acceptance on distributions in the diphoton invariant mass.
We compare our results with the DIPHOX calculation~\cite{Binoth:1999qq}
and demonstrate that
the requirement $Q_{T}<Q$ further suppresses the effects of the final-state
fragmentation contribution, beyond the reduction associated with isolation.
Next, we present our
predictions for distributions of diphoton pairs produced at the energy
of the LHC. Various distributions of the diphoton pairs
produced from the decay of a Higgs boson are contrasted 
with those produced from QCD continuum 
processes at the LHC, showing that enhanced sensitivity
to the signal can be obtained with judicious event selection. Our
conclusions are presented in Sec.~\ref{Sec:conclusions}.

\section{Theory overview \label{Sec:Theory}}

\subsection{Notation \label{sub:Notations}}

We consider the scattering process $h_{1}(P_{1})+h_{2}(P_{2})\rightarrow\gamma(P_{3})+\gamma(P_{4})+X$,
where $h_{1}$ and $h_{2}$ are the initial-state hadrons. In terms of the center-of-mass 
collision energy $\sqrt{S}$, the invariant mass $Q,$ transverse momentum $Q_{T}$, and 
rapidity $y$ of the $\gamma\gamma$ pair, the laboratory frame momenta $P_{1}^{\mu}$ and $P_{2}^{\mu}$ 
of the initial hadrons and $q^{\mu}\equiv P_{3}^{\mu}+P_{4}^{\mu}$ of
the $\gamma\gamma$ pair are \begin{eqnarray}
P_{1}^{\mu} & = & \frac{\sqrt{S}}{2}\left\{ 1,0,0,1\right\} ;\\
P_{2}^{\mu} & = & \frac{\sqrt{S}}{2}\left\{ 1,0,0,-1\right\} ;\\
q^{\mu} & = & \left\{ \sqrt{Q^{2}+Q_{T}^{2}}\cosh y,Q_{T},0,\sqrt{Q^{2}+Q_{T}^{2}}\sinh y\right\}. \end{eqnarray}
The light-cone momentum fractions for the boosted $2\rightarrow2$
scattering system are\begin{equation}
x_{1,2}\equiv\frac{2(P_{2,1}\cdot q)}{S}=\frac{\sqrt{Q^{2}+Q_{T}^{2}}e^{\pm y}}{\sqrt{S}}.\end{equation}
 Decay of the $\gamma\gamma$ pairs is described in the hadronic Collins-Soper
frame \cite{Collins:1977iv}. The Collins-Soper frame is a rest frame
of the $\gamma\gamma$ pair (with $q^{\mu}=\left\{ Q,0,0,0\right\} $
in this frame), chosen so that (a) the momenta $\vec{P}_{1}$ and
$\vec{P}_{2}$ of the initial hadrons lie in the $Oxz$ plane (with
zero azimuthal angle), and (b) the $z$ axis bisects the angle between
$\vec{P}_{1}$ and $-\vec{P}_{2}$. The photon momenta are antiparallel
in the Collins-Soper frame: \begin{eqnarray}
P_{3}^{\mu} & = & \frac{Q}{2}\left\{ 0,\sin\theta_{*}\cos\varphi_{*},\sin\theta_{*}\sin\varphi_{*},\cos\theta_{*}\right\} ,\label{p3CS}\\
P_{4}^{\mu} & = & \frac{Q}{2}\left\{ 0,-\sin\theta_{*}\cos\varphi_{*},-\sin\theta_{*}\sin\varphi_{*},-\cos\theta_{*}\right\} ,\label{p4CS}\end{eqnarray}
 where $\theta_{*}$ and $\varphi_{*}$ are the photon's polar
and azimuthal angles. In this section, we derive resummed predictions
for the fully differential $\gamma\gamma$ cross section $d\sigma/(dQ^{2}dydQ_{T}^{2}d\Omega_{*}),$
where $d\Omega_{*}=d\cos\theta_{*}d\varphi_{*}$ is a
solid angle element around the direction of $\vec{P}_{3}$ in the
Collins-Soper frame defined in Eq.~(\ref{p3CS}). The angles in the
Collins-Soper frame are denoted by a {}``$*$'' subscript, in
contrast to angles in the lab frame, which do not have such a subscript.
The parton momenta and helicities are denoted by lowercase $p_{i}$
and $\lambda_{i}$, respectively.

\subsection{Resummation of the initial-state QCD radiation \label{subsection:ISRResummation}}

For completeness, we present an overview of the finite-order and resummed
contributions associated with the direct production of diphotons.
At the lowest order in the strong coupling strength $\alpha_{s}$,
photon pairs are produced with zero transverse momentum $Q_{T}$.
The Born $q\bar{q}\rightarrow\gamma\gamma$ cross section corresponding
to Fig.~\ref{Fig:FeynDiag} (a) is \begin{equation}
\left.\frac{d\sigma_{q\bar{q}}}{dQ^{2}dy\, dQ_{T}^{2}d\Omega_{*}}\right|_{Born}=\delta(\vec{Q}_{T})\sum_{i=u,\bar{u},d,\bar{d},...}\frac{\Sigma_{i}(\theta_{*})}{S}f_{q_{i}/h_{1}}(x_{1},\mu_{F})f_{\bar{q}_{i}/h_{2}}(x_{2},\mu_{F}),\label{Bornqqbar}\end{equation}
 where $f_{q_{i}/h}(x,\mu_{F})$ denotes the parton distribution function
(PDF) for a quark of a flavor $i$, evaluated at a factorization scale
$\mu_{F}$ of order $Q$. The prefactor\begin{equation}
\Sigma_{i}(\theta_{*})\equiv\sigma_{i}^{(0)}\frac{1+\cos^{2}\theta_{*}}{1-\cos^{2}\theta_{*}},\label{Sigmaqqbar}\end{equation}
 with\begin{equation}
\sigma_{i}^{(0)}\equiv\frac{\alpha^{2}(Q)e_{i}^{4}\pi}{2N_{c}Q^{2}},\end{equation}
 is composed of the running electromagnetic coupling strength $\alpha\equiv e^{2}/4\pi$
evaluated at the scale $Q$, fractional quark charge $e_{i}=2/3$
or $-1/3$, and number of QCD colors $N_{c}=3$.

The lowest-order $gg\rightarrow\gamma\gamma$ scattering proceeds through
an amplitude with a virtual quark loop (a box diagram) shown in Fig.~\ref{Fig:FeynDiag} (h). 
Its cross section takes the form\begin{equation}
\left.\frac{d\sigma_{gg}}{dQ^{2}dy\, dQ_{T}^{2}d\Omega_{*}}\right|_{Born}=\delta(\vec{Q}_{T})\frac{\Sigma_{g}(\theta_{*})}{S}f_{g/h_{1}}(x_{1},\mu_{F})f_{g/h_{2}}(x_{2},\mu_{F}),\label{Borngg}\end{equation}
 where the prefactor\begin{equation}
\Sigma_{g}(\theta_{*})\equiv\sigma_{g}^{(0)}L_{g}(\theta_{*})\label{Sigmag}\end{equation}
 depends on the polar angle $\theta_{*}$ $ $ through a function
$L_{g}(\theta_{*})$ presented explicitly in Ref.~\cite{Nadolsky:2007ba}.
The overall normalization coefficient\begin{equation}
\sigma_{g}^{(0)}=\frac{\alpha^{2}(Q)\alpha_{s}^{2}(Q)}{32\pi Q^{2}(N_{c}^{2}-1)}\left(\sum_{i}e_{i}^{2}\right)^{2}\end{equation}
 involves the sum of the squared charges $e_{i}^{2}$ of the quarks
circulating in the loop.

The NLO direct contributions, represented by Figs.~\ref{Fig:FeynDiag} (b)-(e),
(i)-(l) and denoted as $P(Q,Q_{T},y,\Omega_{*})$, are computed
in Refs.~\cite{Aurenche:1985yk,Bailey:1992br,Balazs:1999yf,deFlorian:1999tp,Bern:2001df,Nadolsky:2007ba,Bern:2002jx}.
The NLO $2\rightarrow3$ differential cross section grows logarithmically
if the final-state parton is soft or collinear to the initial-state
quark or gluon, i.e., when $Q_{T}$ of the $\gamma\gamma$ pair is
much smaller than $Q$. These {}``initial-state'' logarithmic contributions
are summed to all orders later in this subsection. The NLO $qg$ cross section
also contains a large logarithm when one of the photons is produced
from a collinear $q\!\!\!\!^{^{(-)}}\rightarrow q\!\!\!\!^{^{(-)}}\gamma$
splitting in the final state. This {}``final-state'' collinear limit
is discussed in Section~\ref{subsection:Fragmentation-model}.

With contributions from the initial-state soft or collinear radiation
included, the NLO cross section is approximated in the small-$Q_{T}$
asymptotic limit by\begin{equation}
A_{q\bar{q}}(Q,Q_{T},y,\Omega_{*})=\sum_{i=u,\bar{u},d,\bar{d},...}\frac{\Sigma_{i}(\theta_{*})}{S}\left\{ \delta(\vec{Q}_{T})F_{i,\delta}(Q,y,\theta_{*})+F_{i,+}(Q,y,Q_{T})\right\} \label{ASYqqbar}\end{equation}
 in the $q\bar{q}+qg$ scattering channel, and by\begin{eqnarray}
A_{gg}(Q,Q_{T},y,\Omega_{*}) & = & \frac{1}{S}\Biggl\{\Sigma_{g}(\theta_{*})\left[\delta(\vec{Q}_{T})F_{g,\delta}(Q,y,\theta_{*})+F_{g,+}(Q,y,Q_{T})\right]\nonumber \\
 &  & \hspace{12pt}+\Sigma_{g}^{\prime}(\theta_{*},\varphi_{*})F_{g}^{\prime}(Q,y,Q_{T})\Biggr\}\label{ASYgg}\end{eqnarray}
 in the $gg+gq_{S}$ scattering channel. The functions $F_{a,\delta}(Q,y,\theta_{*})$
and $F_{a,+}^{(\prime)}(Q,y,Q_{T})$ for relevant parton flavors $a$
are listed in Appendix~\ref{Appendix:Asymptotic}. They include `plus
function' contributions of the type $\left[Q_{T}^{-2}\ln^{p}\left(Q^{2}/Q_{T}^{2}\right)\right]_{+}$ with
$p\geq0$, universal functions describing soft and collinear scattering,
and process-dependent corrections from NLO virtual diagrams.

The $q\bar{q}+qg$ asymptotic cross section $A_{q\bar{q}}(Q,Q_{T},y,\Omega_{*})$
is proportional to the angular function $\Sigma_{i}(\theta_{*}),$
the same as in the Born $q\bar{q}\rightarrow\gamma\gamma$ cross section,
cf. Eq.~(\ref{Bornqqbar}). Similarly, the $gg+gq_{S}$ asymptotic cross 
section $A_{gg}(Q,Q_{T},y,\Omega_{*})$
includes a term proportional to the Born angular function $\Sigma_{g}(\theta_{*}).$
In addition, $A_{gg}(Q,Q_{T},y,\Omega_{*})$ contains another
term proportional to $\Sigma_{g}^{\prime}(\theta_{*},\varphi_{*})\equiv L_{g}^{\prime}(\theta_{*})\cos2\varphi_{*}$,
where $L_{g}^{\prime}(\theta_{*})$ is derived in Ref.~\cite{Nadolsky:2007ba}.
This term arises due to the interference of Born amplitudes with incoming
gluons of opposite polarizations and affects the azimuthal angle ($\varphi_{*}$)
distribution of the photons in the Collins-Soper frame \cite{Nadolsky:2007ba}.

The small-$Q_{T}$ representations in Eqs.~(\ref{ASYqqbar}) and
(\ref{ASYgg}) can be used to compute fixed-order particle distributions
in the phase-space slicing method. In this method, we choose a small
$Q_{T}$ value $Q_{T}^{sep}$ in the range of validity of 
Eqs.~(\ref{ASYqqbar}) and (\ref{ASYgg}). If the
actual $Q_{T}$ in the computation exceeds $Q_{T}^{sep}$, we calculate
the differential cross section using the full $2\rightarrow3$ matrix
element. When $Q_{T}$ is smaller than $Q_{T}^{sep}$, we calculate
the event rate using the small-$Q_{T}$ asymptotic approximation $A(Q,Q_{T},y,\Omega_{*})$
and $2\rightarrow2$ phase space. Hence, the lowest bin of the $Q_{T}$
distribution is approximated in the NLO prediction by its \textit{average}
value in the interval $0\leq Q_{T}\leq Q_{T}^{sep}$, computed by
integration of the asymptotic approximations.

The phase-space slicing procedure is sufficient for predictions of
observables inclusive in $Q_{T}$, but not of the shape of $d\sigma/dQ_{T}$
distributions. The latter goal is met by all-orders summation of singular
asymptotic contributions with the help of the Collins-Soper-Sterman
(CSS) method \cite{Collins:1981va,Collins:1981uk,Collins:1984kg}.
The small-$Q_{T}$ resummed cross section is denoted as $W(Q,Q_{T},y,\Omega_{*})$
and given by a two-dimensional Fourier transform of a function $\widetilde{W}(Q,b,y,\Omega_{*})$
that depends on the impact parameter $\vec{b}$:\begin{eqnarray}
W(Q,Q_{T},y,\Omega_{*}) & = & \int\frac{d\vec{b}}{(2\pi)^{2}}e^{i\vec{Q}_{T}\cdot\vec{b}}\widetilde{W}(Q,b,y,\Omega_{*})\nonumber \\
 & \equiv & \int\frac{d\vec{b}}{(2\pi)^{2}}e^{i\vec{Q}_{T}\cdot\vec{b}}\widetilde{W}_{pert}(Q,b_{*},y,\Omega_{*})e^{-{\cal F}_{NP}(Q,b)}.\label{FourierIntegral}\end{eqnarray}
 In this equation, $\widetilde{W}(Q,b,y,\Omega_{*})$ is written
as a product of the perturbative part $\widetilde{W}_{pert}(Q,b_{*},y,\Omega_{*})$
and the nonperturbative exponent $\exp\left(-{\cal F}_{NP}(Q,b)\right),$
which describe the dynamics at small ($b\lesssim1\mbox{ GeV}^{-1}$)
and large ($b\gtrsim1\mbox{ GeV}^{-1}$) impact parameters, respectively. The purpose
of the variable $b_{*}$ is reviewed below.

If $Q$ is large, the perturbative form factor $\widetilde{W}_{pert}$
dominates the integral in Eq.~(\ref{FourierIntegral}). It is computed
at small $b$ as \begin{eqnarray}
\widetilde{W}_{pert}(Q,b,y,\theta_{*}) & = & \sum_{a}\frac{\Sigma_{a}(\theta_{*})}{S}h_{a}^{2}(Q,\theta_{*})e^{-\mathcal{S}_{a}(Q,b)}\nonumber \\
 & \times & \left[\mathcal{C}_{a/a_{1}}\otimes f_{a_{1}/h_{1}}\right](x_{1},b;\mu)\left[\mathcal{C}_{\bar{a}/a_{2}}\otimes f_{a_{2}/h_{2}}\right](x_{2},b;\mu).\label{UnpW2}\end{eqnarray}
 The {}``hard-vertex'' function $\Sigma_{a}(\theta_{*})h_{a}^{2}(Q,\theta_{*})$
is the normalized cross section for the Born scattering $a\bar{a}\rightarrow\gamma\gamma$,
with $a=u,\bar{u},d,\bar{d},...$ in $q\bar{q}\rightarrow\gamma\gamma$,
and $a=\bar{a}=g$ in $gg\rightarrow\gamma\gamma$. The Sudakov exponent\begin{eqnarray}
\mathcal{S}_{a}(Q,b)=\int_{C_{1}^{2}/b^{2}}^{C_{2}^{2}Q^{2}}\frac{d\bar{\mu}^{2}}{\bar{\mu}^{2}}\left[\mathcal{A}_{a}\left(C_{1},\bar{\mu}\right)\ln\left(\frac{C_{2}^{2}Q^{2}}{\bar{\mu}^{2}}\right)+\mathcal{B}_{a}\left(C_{1},C_{2},\bar{\mu}\right)\right]\label{Sudakov}\end{eqnarray}
 is an integral of two functions $\mathcal{A}_{a}\left(C_{1},\bar{\mu}\right)$
and $\mathcal{B}_{a}\left(C_{1},C_{2},\bar{\mu}\right)$ between momentum
scales $C_{1}/b$ and $C_{2}Q$, and $C_{1}$ and $C_{2}$ are constants
of order $c_{0}\equiv2e^{-\gamma_{E}}=1.123...$ and $1$, respectively.
The symbol $\left[\mathcal{C}_{a/a_{1}}\otimes f_{a_{1}/h}\right](x,b;\mu)$
stands for a convolution of the $k_{T}-$integrated PDF
$f_{a_{1}/h}(x,\mu)$ and Wilson coefficient function $\mathcal{C}_{a/a_{1}}(x,b;C_{1}/C_{2},\mu)$,
evaluated at a factorization scale $\mu$ and summed over intermediate
parton flavors $a_{1}$: \begin{eqnarray}
\left[\mathcal{C}_{a/a_{1}}\otimes f_{a_{1}/h}\right](x,b;\mu) & \equiv & \sum_{a_{1}}\left[\int_{x}^{1}{\frac{d\xi}{\xi}}\mathcal{C}_{a/a_{1}}\left(\frac{x}{\xi},b;\frac{C_{1}}{C_{2}},\mu\right)f_{a_{1}/h}(\xi,\mu)\right].\label{Eq:CalP}\end{eqnarray}
 We compute the functions $h_{a},$ $\mathcal{A}_{a}$, $\mathcal{B}_{a}$
and $\mathcal{C}_{a/a_{1}}$ up to orders $\alpha_{s},$ $\alpha_{s}^{3},$
$\alpha_{s}^{2},$ and $\alpha_{s},$ respectively, corresponding
to the NNLL accuracy of resummation. The perturbative coefficients
at these orders in $\alpha_{s}$ are listed in Appendix~\ref{Appendix:Summary}.

The subleading contribution from the nonperturbative region $b\gtrsim1\mbox{ GeV}^{-1}$
is included in our calculation using a revised {}``$b_{*}$'' model~\cite{Konychev:2005iy},
which provides excellent agreement with $p_{T}$-dependent data on Drell-Yan
pair and $Z$ boson production. In this model, the perturbative form
factor $\widetilde{W}_{pert}(Q,b_{*},y,\Omega_{*})$ in Eq.~(\ref{FourierIntegral})
is evaluated as a function of $b_{*}\equiv b/(1+b^{2}/b_{max}^{2})^{1/2},$
with $b_{max}=1.5\mbox{ GeV}^{-1}$. The factorization scale $\mu$
in $\left[{\mathcal{C}}\otimes f\right]$ is set equal to $c_{0}\sqrt{b^{-2}+Q_{ini}^{2}}$
, where $Q_{ini}$ is the initial scale of order 1 GeV in the parameterization
employed for $f_{a/h}(x,\mu)$, for instance, 1.3 GeV for the CTEQ6
PDFs~\cite{Pumplin:2002vw}. We have $\widetilde{W}_{pert}(b_{*})=\widetilde{W}_{pert}(b)$
at $b^{2}\ll b_{max}^{2},$ and $\widetilde{W}_{pert}(b_{*})=\widetilde{W}_{pert}(b_{max})$
at $b^{2}\gg b_{max}^{2}$. Hence, this ansatz preserves the exact
form of the perturbative form factor $\widetilde{W}_{pert}(Q,b,y,\Omega_{*})$
in the perturbative region of small $b$, while also incorporating
the leading nonperturbative contributions (described by a phenomenological
function $\mathcal{F}_{NP}(Q,b)$) at large $b$.

The form of $\mathcal{F}_{NP}(Q,b)$ found in the global $p_{T}$
fit in Ref.~\cite{Konychev:2005iy} suggests approximate independence
of $\mathcal{F}_{NP}(Q,b)$ from the type of $q\bar{q}$ scattering
process. It is used here to describe the nonperturbative terms in
the leading $q\bar{q}\rightarrow\gamma\gamma$ channel. We neglect
possible corrections to the nonperturbative contributions arising
from the final-state soft radiation in the $qg$ channel and additional
$\sqrt{S}$ dependence affecting Drell-Yan-like processes at $x\lesssim10^{-2}$
\cite{Berge:2004nt}, as these exceed the accuracy of the present
measurements at the Tevatron. The experimentally unknown $\mathcal{F}_{NP}(Q,b)$
in the $gg$ channel is approximated by $\mathcal{F}_{NP}(Q,b)$ for
the $q\bar{q}$ channel, multiplied by the ratio $C_{A}/C_{F}=9/4$.  This 
choice is motivated by the fact that the leading Sudakov color factors 
${\cal A}_a^{(k)}$ in
the $gg$ and $q\bar{q}$ channels are proportional to 
$C_{A}=3$ and $C_{F}=4/3$, respectively. 
The uncertainties in the $\gamma\gamma$
cross sections associated with $\mathcal{F}_{NP}(Q,b)$ are investigated
numerically in Ref.~\cite{Nadolsky:2007ba}.

In the region $Q_{T}\sim Q$, collinear QCD
factorization at a finite fixed order in $\alpha_{s}$ is applicable.
In order to include non-singular contributions important in this region,
we add to $W(Q,Q_{T},y,\Omega_{*})$ the regular piece $Y(Q,Q_{T},y,\Omega_{*}),$
defined as the difference between the NLO cross section $P(Q,Q_{T},y,\Omega_{*})$
and its small-$Q_{T}$ asymptotic approximation $A(Q,Q_{T},y,\Omega_{*})$:\begin{eqnarray}
\frac{d\sigma(h_{1}h_{2}\rightarrow\gamma\gamma)}{dQ\, dQ_{T}^{2}\, dy\, d\Omega_{*}} & = & W(Q,Q_{T},y,\Omega_{*})+P(Q,Q_{T},y,\Omega_{*})-A(Q,Q_{T},y,\Omega_{*})\nonumber \\
 & \equiv & W(Q,Q_{T},y,\Omega_{*})+Y(Q,Q_{T},y,\Omega_{*}).\label{Eq:Matched}\end{eqnarray}

At small $Q_{T},$ subtraction of $A(Q,Q_{T},y,\Omega_{*})$ in
Eq.~(\ref{Eq:Matched}) cancels large initial-state radiative corrections
in $P(Q,Q_{T},y,\Omega_{*}),$ which are incorporated in their resummed
form within $W(Q,Q_{T},y,\Omega_{*})$. At $Q_{T}$ comparable
to $Q$, $A(Q,Q_{T},y,\Omega_{*})$ cancels the leading terms
in $W(Q,Q_{T},y,\Omega_{*})$, but higher-order contributions remain 
from the infinite tower of logarithmic terms that are resummed in 
$W$. In this situation the $W+Y$ cross section drops below
the finite-order result $P(Q,Q_{T},y,\Omega_{*})$ at some value
of $Q_{T}$ (referred to as the {\em crossing point}) in both the
$q\bar{q}+qg$ and $gg+gq_{S}$ channels, for each $Q$ and $y$.
We use the $W+Y$ cross section as our final prediction at $Q_{T}$
values below the crossing point, and the NLO cross section $P$ at
$Q_{T}$ values above the crossing point.

A few comments are in order about our resummation calculation. The
hard-vertex contribution $\Sigma_{a}(\theta_{*})h_{a}^{2}(Q,\theta_{*})$
and the functions $\mathcal{B}_{a}\left(C_{1},C_{2},\bar{\mu}\right)$
and $\mathcal{C}_{a/a_{1}}(x,b;C/C_{2},\mu)$ can be varied in a mutually
compensating way while preserving the same value of the form factor
$W$ up to higher-order corrections in $\alpha_{s}$. This ambiguity,
or dependence on the chosen {}``resummation scheme'' \cite{Catani:2000vq}
within the CSS formalism, can be employed to explore the sensitivity
of theoretical predictions to further next-to-next-to-next-to-leading logarithmic 
(NNNLL) effects that are not accounted for explicitly.

The perturbative coefficients in Appendix~\ref{Appendix:Summary}
are presented in the CSS resummation scheme \cite{Collins:1984kg},
our default choice in numerical calculations, and in an alternative
scheme by Catani, de Florian and Grazzini (CFG) \cite{Catani:2000vq}.
In the original CSS resummation scheme, the ${{\cal B}}$ and ${{\cal C}}$
functions contain the finite virtual NLO corrections to the $2\rightarrow2$
scattering process, whereas in the CFG scheme the universal ${\mathcal{B}}$
and ${\mathcal{C}}$ depend only on the type of incident partons,
and the process-dependent virtual correction is included in the function
$h_{a}$. The difference between the CSS and CFG schemes is numerically
small in $\gamma\gamma$ production at both the Tevatron and the LHC \cite{Nadolsky:2007ba}.

In the $gg+gq_{S}$ scattering channel, the unpolarized resummed cross
section includes an additional contribution from elements of $k_{T}$-dependent
PDF spin matrices with opposite helicities of outgoing gluons \cite{Nadolsky:2007ba}.
The NLO expansion of this spin-flip resummed cross section generates
the term proportional to $\Sigma_{g}^{\prime}(\theta_{*},\varphi_{*})\propto\cos2\varphi_{*}$
in the small-$Q_{T}$ asymptotic cross section, cf.  Eq.~(\ref{ASYgg}).
Although the logarithmic spin-flip contribution must
be resummed in principle to all orders to predict the $\varphi_{*}$ dependence
in the $gg+gq_{S}$ channel, it is neglected in the present work in
view of its small effect on the full $\gamma\gamma$ cross section.

When integrated over $Q_T$ from 0 to scales of order $Q$, the resummed
cross section becomes approximately equal to the finite-order (NLO)
cross section, augmented typically by a few-percent correction from integrated
higher-order terms logarithmic in $Q_{T}$. Inclusive observables
that allow such integration (e.g., the large-$Q$ region of the $\gamma\gamma$
invariant mass distribution) are approximated well both by the resummed
and NLO calculations. However, the experimental acceptance constrains
the range of the integration over $Q_{T}$ in parts of phase space
and may break delicate cancellations between integrable singularities
present in the finite-order differential distribution. In this situation
(e.g., in the vicinity of the kinematic cutoff in $d\sigma/dQ$ discussed
in Sec.~\ref{Sec:Phenomenology}) the NLO cross section becomes unstable,
while the resummed cross section (free of discontinuities) continues
to depend smoothly on kinematic constraints. We see that the resummation
is essential not only for the prediction of physical $Q_{T}$ distributions
in $\gamma\gamma$ production, but also for credible estimates of
the effects of experimental acceptance on distributions in the diphoton
invariant mass and other variables.

\subsection{Final-state photon fragmentation \label{subsection:Fragmentation-model}}

\subsubsection{Single-photon fragmentation}

In addition to the QCD singularities associated with initial-state
radiation {[}described by the asymptotic terms in Eqs.~(\ref{ASYqqbar})
and (\ref{ASYgg})], other singularities arise in the ${\mathcal{O}}(\alpha_{s})$
process $q(p_{1})+g(p_{2})\rightarrow\gamma(p_{3})+\gamma(p_{4})+q(p_{5})$
{[}Fig.~\ref{Fig:FeynDiag} (e)] when a photon is collinear to the
final-state quark. In this limit, the $qg\rightarrow q\gamma\gamma$
squared matrix element grows as $1/s_{\gamma5}$, when $s_{\gamma5} \rightarrow 0$, 
where $s_{\gamma5}$ is the squared invariant mass of the collinear $\gamma q$
pair. In this limit, the squared matrix element factors as \begin{equation}
|{\mathcal{M}}(qg\rightarrow q\gamma\gamma)|^{2}\approx{\frac{2e^{2}e_{i}^{2}}{s_{\gamma5}}}P_{\gamma\leftarrow q}({z})|{\mathcal{M}}(qg\rightarrow q\gamma)|^{2}\label{CSsub}\end{equation}
 into the product of the squared matrix element $\left|{\mathcal{M}}(qg\rightarrow q\gamma)\right|^{2}$
for the production of a photon and an intermediate quark, and a splitting
function $P_{\gamma\leftarrow q}(z)=(1+(1-z)^{2})/z$ for fragmentation
of the intermediate quark into a collinear $\gamma q$ pair. In Eq.~(\ref{CSsub})
$z$ is the light-cone fraction of the intermediate quark's momentum
carried by the fragmentation photon, and $ee_{i}$ is the charge of
the intermediate quark. When the photon-quark separation $\Delta r=\sqrt{(\eta_{5}-\eta_{\gamma})^{2}+(\varphi_{5}-\varphi_{\gamma})^{2}}$
in the plane of pseudorapidity $\eta=-\log(\tan(\theta/2))$ and azimuthal
angle $\varphi$ in the lab frame is small, as in the collinear limit,
$s_{\gamma5}\approx E_{T\gamma}E_{T5}\Delta r^{2},$ where $E_{T\gamma}$
and $E_{T5}$ are the transverse energies of the photon and quark,
with $E_{T}\equiv E\sin\theta$. Note that $E_{T5}=Q_{T}$ at the
order in $\alpha_{s}$ at which we are working. Therefore, contributions
from the final-state collinear, or fragmentation, region
are most pronounced at small $\Delta r$ and relatively small $Q_{T}.$%
\footnote{In the soft, or $E_{5}\rightarrow0,$ limit, the final-state collinear
contribution is suppressed, reflecting the absence of the soft singularity
in the $qg\rightarrow q\gamma\gamma$ cross section. %
}

A fully consistent treatment of the initial- and final-state singularities
would require a joint initial- and final-state resummation. In the
approaches taken to date, the fragmentation singularity may be subtracted
from the direct cross section and replaced by a single-photon {}``one-fragmentation''
contribution $q+g\rightarrow(q\stackrel{frag}{\longrightarrow}\gamma)+\gamma$,
where {}``$(\stackrel{frag}{q\longrightarrow\gamma})$'' denotes
collinear production of one hard photon from a quark, described by
a function $D_{\gamma}(z,\mu)$ at a light-cone momentum fraction
$z$ and factorization scale $\mu$. Single-photon {}``two-fragmentation''
contributions arise in processes like $g+g\rightarrow(q\stackrel{frag}{\longrightarrow}\gamma)+(\bar{q}\stackrel{frag}{\longrightarrow}\gamma)$
and involve convolutions with two functions $D_{\gamma}(z,\mu)$ (one
per photon). The lowest-order Feynman diagrams for the one- and two-fragmentation
contributions are shown in Figs.~\ref{Fig:FeynDiag}(f) and \ref{Fig:FeynDiag}(g),
respectively. Parameterizations must be adopted for the nonperturbative
functions $D_{\gamma}(z,\mu)$ at an initial scale $\mu=\mu_{0}$.
This is the approach followed in the DIPHOX calculation~\cite{Binoth:1999qq},
in which the sum of real and virtual NLO corrections to direct and
single-$\gamma$ fragmentation cross sections is included. When explicit
fragmentation function contributions are included, the inclusive rate
is increased by higher-order contributions from photon production
within hadronic jets. However, much of the enhancement is suppressed
by isolation constraints imposed on the inclusive photon cross sections
before the comparison with data. Nevertheless, fragmentation contributions
surviving isolation may be moderately important in parts of phase
space.

An infrared-safe procedure can be formulated to apply isolation cuts
at each order of $\alpha_{s}$~\cite{Berger:1996vy,Catani:1998yh,Catani:2002ny}.
This procedure encounters difficulties in reproducing the effects
of isolation on fragmentation contributions, because theoretical models
reflect only basic features of the experimental isolation and may
introduce new logarithmic singularities near the edges of the isolation
cones.

As mentioned in the Introduction, the magnitude of the fragmentation
contribution depends on the values of isolation parameters $E_{T}^{iso},$
$\Delta R$, and $\Delta R_{\gamma\gamma}$, modeled only approximately
in a theoretical calculation. The collinear approximation constrains
from below the values of $z$ accessible to $D_{\gamma}(z,\mu)$:
$z>z_{min}\equiv(1+E_{T5}^{iso}/E_{T\gamma})^{-1}$. 
If $D_{\gamma}(z,\mu)$ varies rapidly with $z$, 
the fragmentation cross section is particularly sensitive to the
assumed values of $E_{T}^{iso}$ and $z_{min}$. 
For instance, if $D_{\gamma}(z,\mu)\sim1/z$,
the fragmentation cross section is roughly proportional to $E_{T}^{iso}$
under a typical condition $E_{T}^{iso}/E_{T\gamma}\ll1$. Such nearly
linear dependence on $E_{T}^{iso}$ of the fragmentation cross section 
$d\sigma/dQ_{T}$ is indeed observed in the DIPHOX calculation, as reviewed
in Sec.~\ref{Sec:Phenomenology}. In reality, some spread of the
parton radiation in the direction transverse to the photon's motion
is expected. The treatment of kinematics in parton showering programs
like PYTHIA results in somewhat different dependence on $z$~\cite{Balazs:1997hv}
compared to the collinear approximation, hence in a different magnitude
of the fragmentation cross section.

In this work we adopt a procedure that reproduces desirable features
of the isolated cross sections, while bypassing some of the difficulties
summarized above. To simulate experimental isolation, we reject an
event if (a) the separation $\Delta r$ between the final-state parton
and one of the photons is less than $\Delta R$, and (b) $E_{T5}$
of the parton is larger than $E_{T}^{iso}$. This condition is applied
to the NLO cross section $P(Q,Q_{T},y,\Omega_{*})$, but not to
$W(Q,Q_{T},y,\Omega_{*})$ and $A(Q,Q_{T},y,\theta_{*})$,
as these correspond to initial-state QCD radiation and are free of 
the final-state collinear singularity.

This quasi-experimental isolation excludes the singular final-state
direct contributions at $E_{T5}>E_{T}^{iso}$ and $\Delta r<\Delta R$
(or $s_{\gamma5}<E_{T\gamma}E_{T5}\Delta R^{2}$). It is effective
for $Q_{T}>E_{T}^{iso}$, but the collinear direct contributions survive
when $Q_{T}<E_{T}^{iso}$. The integrated (but not the differential)
fragmentation rate in the region $Q_{T}<E_{T}^{iso}$ may be estimated
from a calculation with explicit fragmentation functions. In our approach, 
we do not introduce
fragmentation functions, but we apply an auxiliary regulator to the direct
$qg$ cross section at $Q_{T}<E_{T}^{iso}$ and $\Delta r<\Delta R$.
In our numerical study we find that this prescription preserves a
continuous differential distribution except for a small finite discontinuity 
at $Q_{T} = E_{T}^{iso}$.  It approximately reproduces
the integrated $qg$ rate obtained in the DIPHOX calculation at small
$Q_{T}$, for the nominal $E_{T}^{iso}$.

Two forms of the auxiliary regulator are considered below, based on
subtraction of the leading collinear contribution and smooth-cone
isolation \cite{Frixione:1998jh}. In the first case, we subtract
the leading part Eq.~(\ref{CSsub}) of the direct $qg$ matrix element
when $E_{T5}<E_{T}^{iso}$ and $\Delta r<\Delta R.$ We take 
$z=1-p_{s}\cdot p_{5}/(p_{s}\cdot p_{f}+p_{s}\cdot p_{5}+p_{f}\cdot p_{5}),$
where $p_{f}^{\mu},$ $p_{5}^{\mu},$ and $p_{s}^{\mu}$ are the four-momenta
of the fragmentation photon, fragmentation quark, and spectator photon,
respectively~\cite{Catani:1996vz}. This prescription is used in
most of the numerical results in this paper.

In the second case, we suppress fragmentation contributions at $\Delta r<\Delta R$
and $E_{T5}<E_{T}^{iso}$ by rejecting events in the $\Delta R$ cone
that satisfy $E_{T5}<\chi(\Delta r)$, where $\chi(\Delta r)$ is
a smooth function satisfying $\chi(0)=0,$ $\chi(\Delta R)=E_{T}^{iso}$.
This {}``smooth-cone isolation''~\cite{Frixione:1998jh} transforms
the fragmentation singularity associated with $D_{\gamma}(z,\mu)$
into an integrable singularity, which depends on the assumed functional
form of $\chi(\Delta r)$. The cross section for direct contributions
is rendered finite by this prescription without explicit introduction
of fragmentation functions $D_{\gamma}(z,\mu)$. For our smooth function,
we choose $\chi(\Delta r)=E_{T}^{iso}(1-\cos\Delta r)^{2}/(1-\cos\Delta R)^{2}$,
which differs from the specific form considered in Ref.~\cite{Frixione:1998jh},
but still satisfies the condition $\chi(0)=0.$ Our earlier results
in Ref.~\cite{Balazs:2006cc} are computed with this prescription.
Here we employ it only in a few instances for comparison with the
subtraction method and obtain similar results.

Differences between the two prescriptions can be used to quantify
sensitivity of the predictions to the treatment of the $Q_{T}<E_{T}^{iso}$
and $\Delta r<\Delta R$ region. The two prescriptions yield identical
predictions outside of this restricted region, notably at $Q_{T}>E_{T}^{iso}$,
where our NLO perturbative expression $P(Q,Q_{T},y,\Omega_{*})$ in 
the $q\bar{q}+qg$ channel
is controlled only by quasi-experimental isolation and coincides with
the corresponding direct cross section in DIPHOX. The default subtraction
prescription predicts a vanishing $d\sigma/dQ_{T}$ in the extreme
$Q_{T}\rightarrow0$ limit, while the smooth-cone prescription has
an integrable singularity in this limit, avoided by an explicit small-$Q_{T}$
cutoff in the calculation of our $Y$-piece. Both prescriptions are
free of the logarithmic singularity at $Q_{T}=E_{T}^{iso}$ arising
in the fixed-order (DIPHOX) calculation.

\subsubsection{Low-$Q$ diphoton fragmentation \label{sub:Low-Q-diphoton-fragmentation}}

\begin{figure}
\begin{centering}\includegraphics{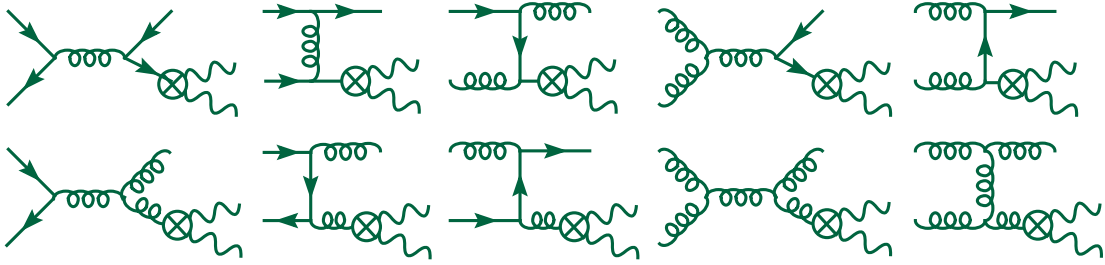}\par\end{centering}

\caption{Lowest-order Feynman diagrams describing fragmentation of the final-state
partons into photon pairs with relatively small mass $Q$. \label{fig:LowQfrag}}
\end{figure}

Another class of large radiative corrections arises when the $\gamma\gamma$
invariant mass $Q$ is smaller than the $\gamma\gamma$ transverse momentum
$Q_{T}$. In this case, one final-state quark or gluon fragments into
a low-mass $\gamma\gamma$ pair, e.g. as 
$q+g\rightarrow(q\stackrel{frag}{\longrightarrow}\gamma\gamma)+g$. The 
lowest-order contributions of this kind are shown in Fig.~\ref{fig:LowQfrag}.
The process is described by a $\gamma\gamma$-fragmentation function
$D_{\gamma\gamma}(z_{1},z_{2},\mu)$, different from the single-photon
fragmentation function $D_{\gamma}(z,\mu)$. This new {}``two-photons from one-fragmentation''
contribution is not included yet in existing calculations, even though
similar fragmentation mechanisms have been studied in large-$Q_{T}$
Drell-Yan pair production~\cite{Berger:1998ev,Berger:2001wr}. The
importance of low-$Q$ $\gamma\gamma$-fragmentation may be elevated
in some kinematic regions for typical experimental cuts. They can
be removed by adjustments in the experimental cuts, as discussed in
Sec.~\ref{Sec:Phenomenology}.

\subsection{Summary of the calculation \label{sub:TheorySummary}}

We conclude this section by summarizing the main features of our calculation.
Full direct NLO cross sections, represented by the graphs (a)-(e),
(h)-(l) in Fig.~\ref{Fig:FeynDiag}, are computed, and their initial-state 
soft/collinear logarithmic singularities are resummed at
small $Q_{T}$ in both the $q\bar{q}+qg$ and $gg+gq_{S}$ channels.
The perturbative Sudakov functions ${\mathcal{A}}$ and ${\mathcal{B}}$
and Wilson coefficient functions ${\mathcal{C}}$ in the resummed
cross section $W$ are computed up to orders $\alpha_{s}^{3},$ $\alpha_{s}^{2}$,
and $\alpha_{s}$, respectively, corresponding to resummation at NNLL
accuracy.

Our resummation calculation requires an integration over all values
of impact parameter $b$, including the nonperturbative region of
large $b$. In our default calculation of the resummed cross section,
we adopt the nonperturbative functions introduced in Ref.~\cite{Konychev:2005iy}.
We consider two resummation schemes, the traditional scheme introduced
in the CSS paper as well as an alternative scheme~\cite{Catani:2000vq}.
The comparison allows us to estimate the magnitude of yet higher-order
corrections that are not included. The size of these effects is different
in the $q\bar{q}+qg$ and $gg+gq_{S}$ channels but not particularly
significant in either \cite{Nadolsky:2007ba}.

The final-state collinear singularity in the $qg$ scattering channel
is avoided by applying quasi-experimental isolation when $Q_{T}>E_{T}^{iso}$
and an auxiliary regulator when $Q_{T}<E_{T}^{iso}$ to approximate
on average the full NLO rate from direct $qg$ and fragmentation cross
sections in this $Q_{T}$ range. Two prescriptions for the auxiliary
regulator (subtraction and smooth isolation inside the photon's
isolation cone) are considered and lead to similar predictions at
the Tevatron and LHC.

The singular logarithmic contributions associated with initial-state
radiation are subtracted from the NLO cross section $P$ to form a
regular piece $Y,$ which is added to the small-$Q_{T}$ resummed
cross section $W$ to predict the production rate for small and intermediate
values of $Q_{T}$. In the $gg+gq_{S}$ channel, we also subtract
from $P$ a new singular spin-flip contribution that affects azimuthal
angle ($\varphi_{*})$ dependence in the Collins-Soper reference frame.
We switch our prediction to the fixed-order perturbative result $P$
at the point in $Q_{T}$ where the cross section $W+Y$ drops below
$P$. This crossing point is located at $Q_{T}$ of order $Q$ in both 
$q\bar{q}+qg$ and $gg+gq_{S}$ channels.

\section{Comparisons with Data and Predictions \label{Sec:Phenomenology}}

Our calculation of the differential cross section $d\sigma/(dQdQ_{T}dyd\Omega_{*})$
is especially pertinent for the transverse momentum $Q_{T}$ distribution
in the region $Q_{T}\lesssim Q$, for fixed values of diphoton mass
$Q$ (cf. Section~\ref{subsection:KinematicsTev}). It would be best
to compare our \emph{multiple} differential distribution with experiment,
but published collider data tend to be presented in the form of singly
differential distributions in $Q$, $Q_{T}$, and $\Delta\varphi\equiv\varphi_{3}-\varphi_{4}$
in the lab frame, after integration over the other independent kinematic
variables. We follow suit in order to make comparisons with Tevatron
collider data, but we recommend that more differential studies be
made, and we comment on the features that can be explored. We show
results at the energy of the Tevatron collider and then make predictions
for the Large Hadron Collider.

The analytical results of Sec.~\ref{Sec:Theory} are implemented
in our computer code. As a first step, resummed and NLO $\gamma\gamma$
cross sections are computed on a grid of discrete values of $Q$,
$Q_{T}$, and $y$ by using the resummation program \textsc{Legacy}
described in Refs.~\cite{Ladinsky:1993zn,Landry:2002ix}. At the
second stage, matching of the resummed and NLO cross sections is performed,
and fully differential cross sections are evaluated by Monte-Carlo
integration of the matched grids in the latest version of the program
\textsc{ResBos}~\cite{Balazs:1997xd,Balazs:1999gh}. The calculation
is done for $N_{f}=5$ active quark flavors and the following values
of the electroweak and strong interaction parameters~\cite{Eidelman:2004wy}:
\begin{eqnarray}
 &  & G_{F}=1.16639\times10^{-5}~\textrm{GeV}^{-2},~~m_{Z}=91.1882~{\textrm{GeV}},\\
 &  & \alpha(m_{Z})=1/128.937,~~\alpha_{s}(m_{Z})=0.1187.\end{eqnarray}
 The following choices of the factorization constants are used: $C_{1}=C_{3}=2e^{-\gamma_{E}}\approx1.123...$,
and $C_{2}=C_{4}=1.$ The choice $C_{4}=1$ implies that we equate
the renormalization and factorization scales to the invariant mass
of the photon pair, $\mu_{R}=\mu_{F}=Q$, in the fixed-order and asymptotic
contributions $P(Q,Q_{T},y,\Omega_{*})$ and $A(Q,Q_{T},y,\Omega_{*})$.
We use two-loop expressions for the running electromagnetic and strong
couplings $\alpha(\mu)$ and $\alpha_{S}(\mu)$, as well as the NLO
parton distribution function set CTEQ6M~\cite{Pumplin:2002vw} with
$Q_{ini}=1.3$ GeV. For calculations with explicit final-state fragmentation
functions included, we use set 1 of the NLO photon fragmentation functions
from Ref.~\cite{Bourhis:1997yu}.

\subsection{Results for Run 2 at the Tevatron}

\subsubsection{Kinematic constraints \label{subsection:KinematicsTev}}

In this section, we present our results for the Tevatron $p\bar{p}$
collider operating at $\sqrt{S}=1.96$~TeV. In order to compare with
the data from the Collider Detector at Fermilab (CDF) collaboration~\cite{Acosta:2004sn},
we make the same restrictions on the final-state photons as those
used in the experimental measurement (unless stated otherwise): \begin{eqnarray}
 &  & {\textrm{transverse momentum}}~p_{T}^{\gamma}>p_{T\, min}^{\gamma}=14~(13)~{\textrm{GeV for the harder (softer) photon, }}\label{pTcutTev}\\
 &  & {\textrm{and rapidity}}~|y^{\gamma}|<0.9~{\textrm{for each photon}}.\end{eqnarray}
 We impose isolation conditions described in Section \ref{subsection:Fragmentation-model},
assuming the nominal isolation energy $E_{T}^{iso}=1$ GeV specified
in the CDF publication, along with $\Delta R=0.4,$ and $\Delta R_{\gamma\gamma}=0.3$.

We also show predictions for the constraints that approximate
event selection conditions used by the Fermilab 
D\O~Collaboration~\cite{Dyer:2006}:
$p_{T}^{\gamma}>p_{T\, min}^{\gamma}=21~(20)$ GeV for the harder
(softer) photon, $|y^{\gamma}|<1.1$, and $E_{T}^{iso}/E_{T}^{\gamma}=0.07$
for each photon, for the same $\Delta R$ and $\Delta R_{\gamma\gamma}$
values as in the CDF case.

\begin{figure*}
\includegraphics[width=0.7\columnwidth,keepaspectratio]{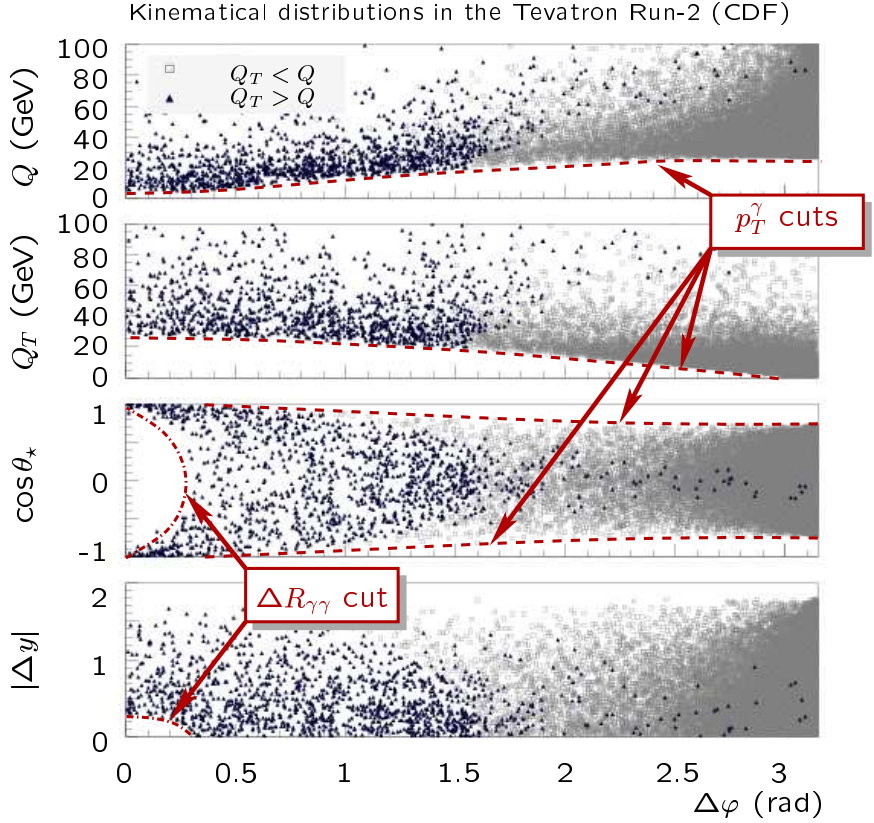}

\caption{The diphoton event distribution from the theoretical simulation for
$\sqrt{S}=1.96$\,GeV, with the selection criteria imposed in the
CDF measurement, as a function of the various kinematic variables described
in the text, shown for $Q_T < Q$ and $Q_T > Q$ separately. \label{Fig:KinCuts}}
\end{figure*}

A scatter plot of event distributions from our theoretical simulation
for CDF kinematic cuts and arbitrary luminosity is shown in Fig.~\ref{Fig:KinCuts}.
The events are plotted versus the invariant mass $Q$, transverse
momentum $Q_{T}$, rapidity separation 
$\left|\Delta y\right|\equiv\left|y_{hard}-y_{soft}\right|$,
and azimuthal separation $\Delta\varphi\equiv\left|\varphi_{hard}-\varphi_{soft}\right|$
(with $0\leq\Delta\varphi\leq\pi)$ between the harder and softer
photon in the lab frame, as well as the cosine of the polar angle
$\theta_{*}$ in the Collins-Soper frame. It can be seen from
the figure that $\Delta\varphi$ is correlated with the difference
$Q_{T}-Q$. Events with $Q_{T}<Q$ ($Q_{T}>Q$) tend to populate regions
with $\Delta\varphi>\pi/2$ ($\Delta\varphi<\pi/2$). The extreme
case $Q_{T}=0$ relevant to the Born approximation corresponds to
$\Delta\varphi=\pi$.

The $p_{T}^{\gamma}$ cuts suppress the mass region 
$Q\lesssim2\sqrt{p_{Tmin}^{\gamma_{3}}p_{Tmin}^{\gamma_{4}}}\approx27$
GeV at $\Delta\varphi\approx\pi$ and $Q_{T}\lesssim25$ GeV at $\Delta\varphi\approx0$,
leading to the appearance of a kinematic cutoff in the invariant mass
distribution and a {}``shoulder'' in the transverse momentum distribution,
as shown in later sections. Our theoretical framework is applicable
in the region $Q_{T}\lesssim Q$ (large $\Delta\varphi$), where the dominant
fraction of events occurs. The appearance of singularities in the
NLO calculation at $Q_{T}\rightarrow0$ and the fact that there are
two different hard scales, $Q_{T}$ and $Q$, relevant for the event
distributions in the low-$Q_{T}$ region require that we address and
resum large logarithmic terms of the form $\log(Q/Q_{T})$. 
Different and interesting physics becomes important in the complementary
region $Q_{T}>Q$ (small $\Delta\varphi$), a topic we address in  
Sec.~\ref{subsection:QT_gt_Q}.

\subsubsection{Tevatron cross sections \label{subsection:XSecTeV}}

\begin{figure*}
\includegraphics[height=9.5cm]{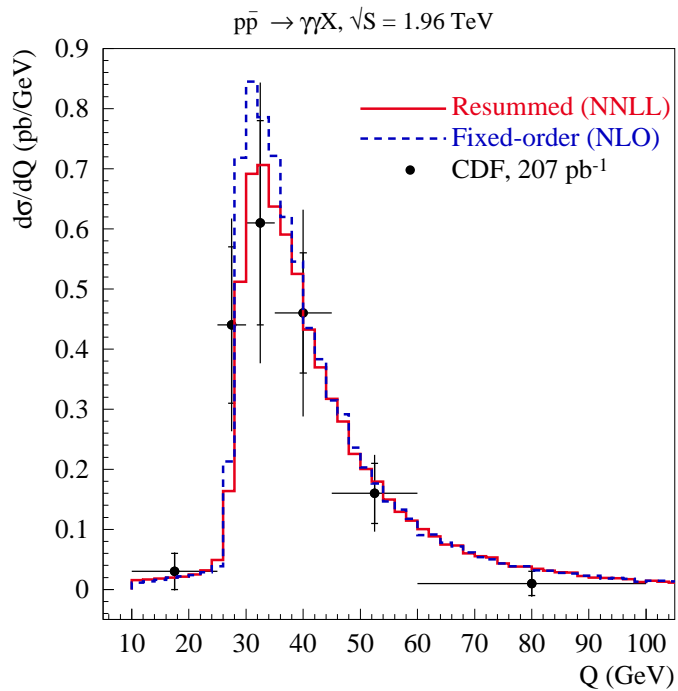}

\caption{Invariant mass distributions of photon pairs in $p\bar{p}\rightarrow\gamma\gamma X$
at $\sqrt{S}=1.96$~TeV with QCD contributions calculated in the
soft--gluon resummation formalism (red solid) and at NLO (blue dashed).
The calculations include the cuts used by the CDF collaboration whose
data are shown~\cite{Acosta:2004sn}. \label{Fig:QCDF}}
\end{figure*}

We compare our resummed and finite-order predictions for the invariant
mass ($Q$) distribution of photon pairs, shown in Fig.~\ref{Fig:QCDF}
as solid and dashed lines, respectively. The finite-order cross section
is evaluated at $O(\alpha_{s})$ accuracy in the $q\bar{q}+qg$ channel
and at $O(\alpha_{s}^{3})$ accuracy in the $gg+gq_{S}$ channel. These finite-order 
calculations are performed with the phase-space slicing method described
in Sec.~\ref{subsection:ISRResummation}. When integrated over all
$Q_{T}$, as in the $d\sigma/dQ$ distribution at large $Q$, the
resummed logarithmic terms from higher orders in $\alpha_{s}$ produce
a relatively small NNLO correction, such that the resummed and finite-order
mass distributions in Fig.~\ref{Fig:QCDF} are close to one another
in normalization and shape. Both distributions also agree with the
CDF data in this $Q$ range within experimental uncertainties.

The shape of $d\sigma/dQ$ at small $Q$ is affected by the cuts in
Eq.~(\ref{pTcutTev}) on the transverse momenta $p_{T}^{\gamma}$
of the two photons. In addition to being responsible for the characteristic
cutoff at $Q\approx27$ GeV explained in the previous subsection,
the cuts on the individual transverse momenta $p_{T}^{\gamma}$ also
introduce a dependence of the invariant mass distribution on the shape
of the $Q_{T}$ spectrum of the $\gamma\gamma$ pairs. Because of
this correlation between the $Q$ and $Q_{T}$ distributions, the
discontinuities in $d\sigma/dQ_{T}$ as $Q_{T}\rightarrow0$, when
computed at finite order, make finite-order predictions for $d\sigma/dQ$
somewhat unstable.

\begin{figure*}
\begin{centering}\includegraphics[width=0.43\columnwidth]{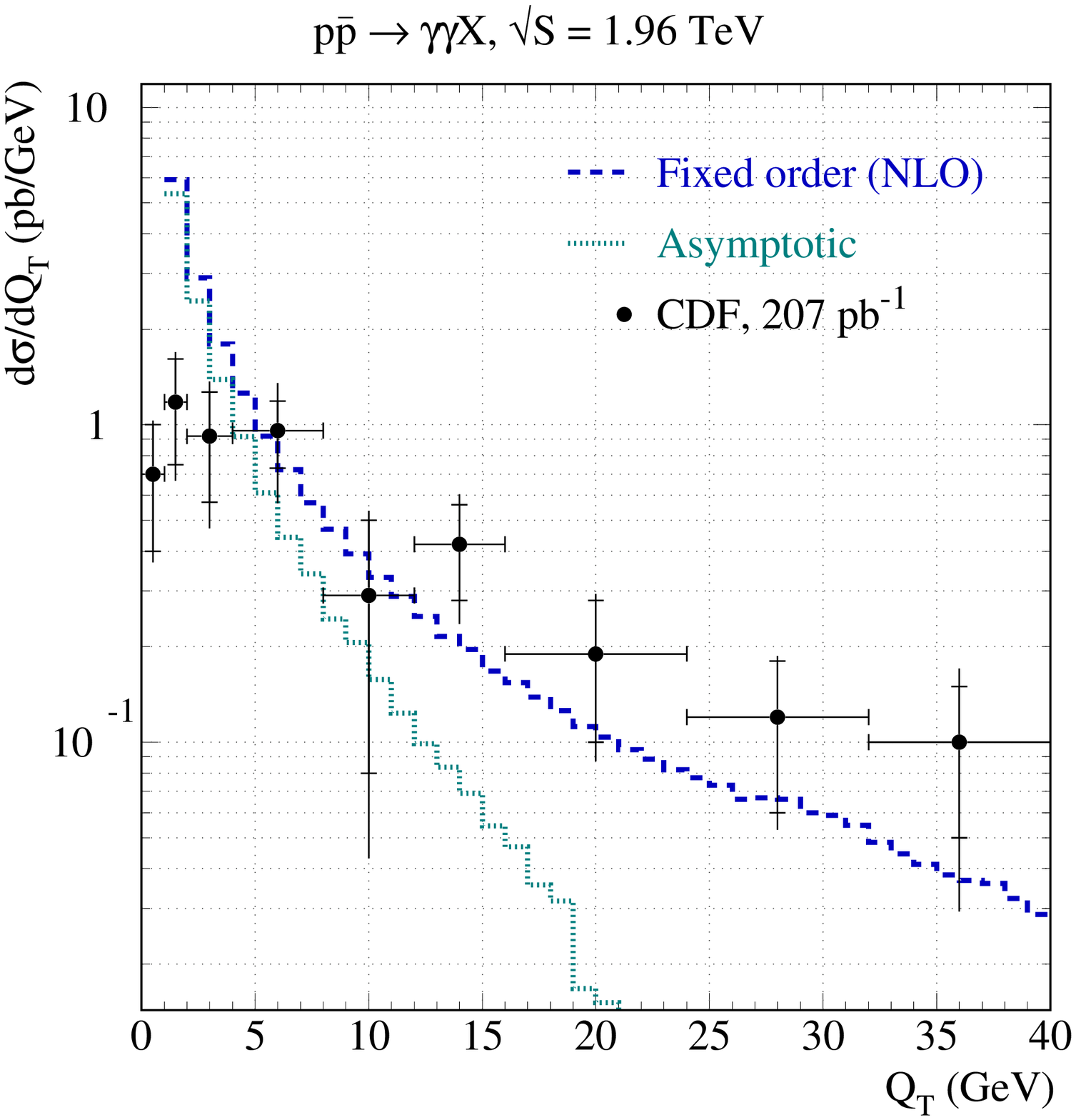}
\includegraphics[width=0.43\columnwidth]{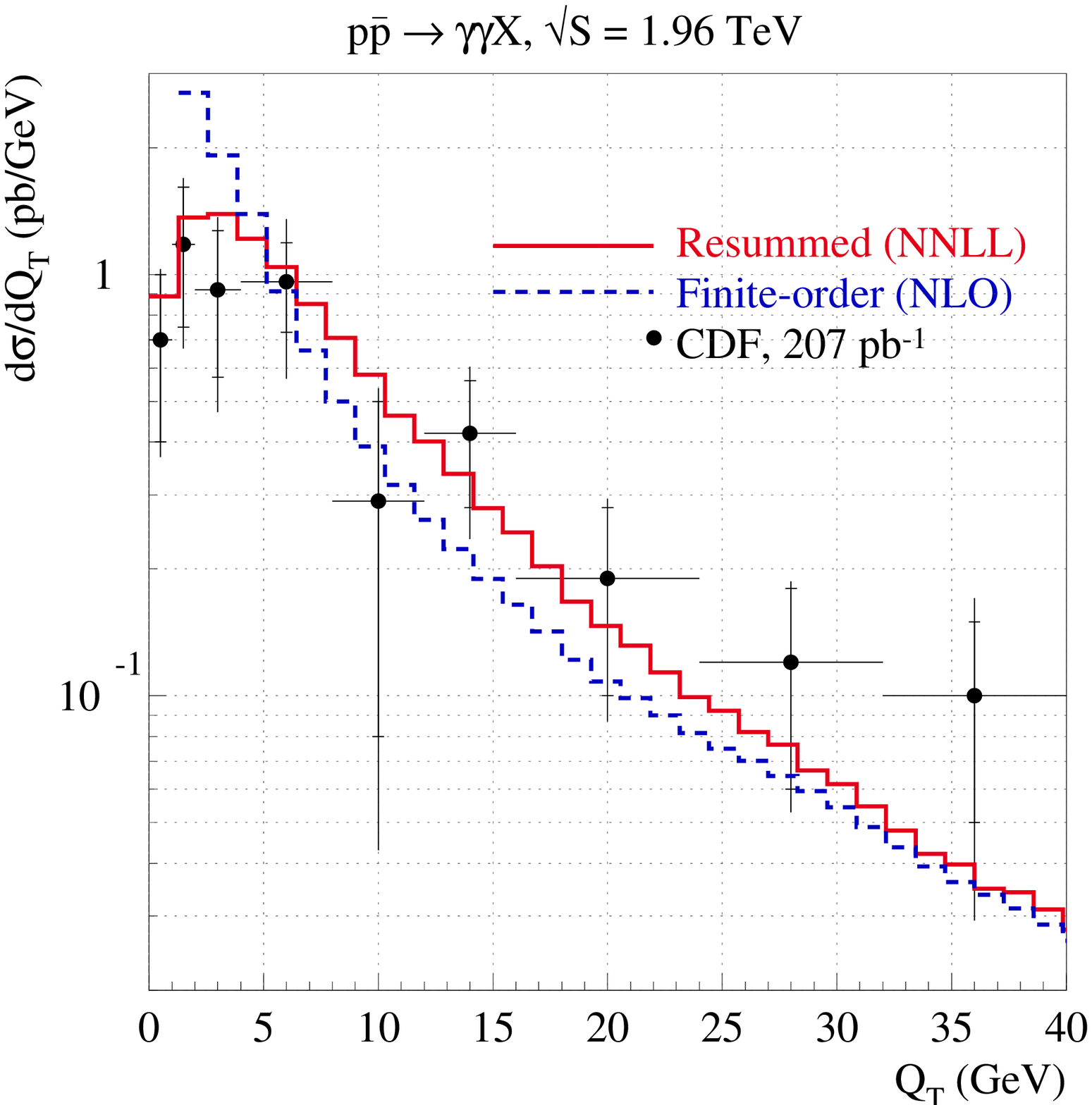}\\
 (a) \hspace{0.45\columnwidth} (b) \par\end{centering}

\caption{Transverse momentum distributions in $p\bar{p}\rightarrow\gamma\gamma X$
at $\sqrt{S}=1.96$~TeV along with the CDF data:
(a) the fixed-order prediction $P$ (dashes)
and its asymptotic approximation $A$ (dots);
(b) the full resummed cross section (solid), obtained by matching
the resummed $W+Y$ to the fixed-order prediction $P$ (dashed, same
as in (a)) at large $Q_{T}$. \label{Fig:QTCDF}}
\end{figure*}

The finite-order expectation for the transverse momentum distribution
$d\sigma/dQ_{T}$ (i.e., the integral of  
$P(Q,Q_{T},y,\Omega_{*})$ over $Q$, $y$, and $\Omega_{*}$, or $P$ for brevity)
is shown as a dashed curve in Fig.~\ref{Fig:QTCDF}(a). It exhibits an
integrable singularity in the small-$Q_{T}$ limit. 
Terms with inverse power and logarithmic 
dependence on $Q_{T}$, associated with initial-state radiation
as $Q_{T}\rightarrow0$, are extracted from $P$ and form the asymptotic
contribution, denoted as $A$ (dotted curve). 
In the figure, both $P$ and $A$
are truncated at a small value of $Q_{T}$, that is, not drawn all
the way to $Q_{T}=0$. The curves for $P$ and $A$ are close at small
values of $Q_{T}$, signaling that the initial-state logarithmic singularities
dominate the NLO distribution. The difference $Y$ between the $P$
and $A$ distributions includes the finite regular terms not included
in $A$ and logarithmic terms from the final-state fragmentation singularities,
with the latter subtracted when $Q_{T}<E_{T}^{iso}$, as described
in Sec.~\ref{subsection:Fragmentation-model}. The data clearly disfavor
the fixed-order prediction in the region of low $Q_{T}$.

Figure~\ref{Fig:QTCDF}(b) features the resummed $W+Y$ contribution
(solid curve). Resummation of the initial-state logarithmic terms
renders $W$ finite in the region of small $Q_{T}$. The sum of $W$
and $Y$ includes the resummed initial-state singular contributions
plus the remaining relevant terms in $P$. Since $P$ provides a reliable
fixed-order estimate at large $Q_{T}$, we present our final resummed
prediction by switching from $W+Y$ to $P$ at the point at which the
two differential cross sections (as functions of $Q$, $Q_{T}$ and
$y$) cross each other. In contrast to the fixed-order (dashed) curve
$P$ in Fig.~\ref{Fig:QTCDF}(b), the agreement with data is improved
at the lowest values of $Q_{T}$, where resummation brings the rate
down, and for $Q_{T}=12-32$ GeV, where the resummed logarithmic terms
increase the rate.

The resummed predictions for the Tevatron experiments are practically
insensitive to the choice of the resummation scheme and the nonperturbative
model~\cite{Nadolsky:2007ba}. About 75\% (25\%) of the total rate
at the Tevatron with CDF cuts imposed comes from the $q{\bar{q}}+qg+{\bar{q}}g$
($gg+gq_{S}$) initial state. The fractions for the cuts used by D\O~ are 
84\% and 16\%. The $gg+gq_{S}$ contribution falls steeply after
$Q_{T}>22$ GeV, because the gluon PDF decreases rapidly with parton
fractional momentum $x$ \cite{Nadolsky:2007ba}.

\begin{figure*}
\includegraphics[height=9.5cm]{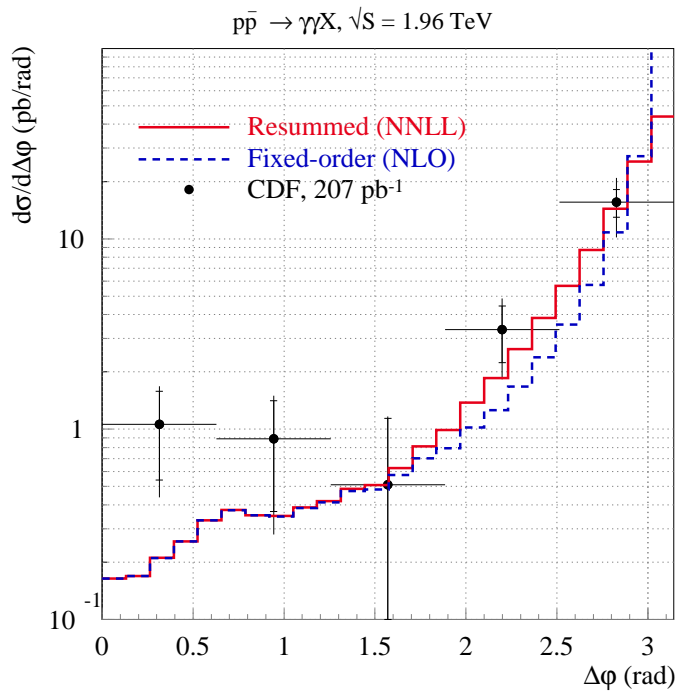}

\caption{The difference $\Delta\varphi$ in the azimuthal angles of the two
photons in the laboratory frame predicted by the resummed (solid)
and fixed-order (dashed) calculations, compared to the CDF data.}

\label{Fig:DeltaPhiCDF} 
\end{figure*}

The distribution in the difference $\Delta\varphi$ of the azimuthal
angles of the photons is shown in Fig.~\ref{Fig:DeltaPhiCDF}. As
is true for the transverse momentum distribution in the limit $Q_{T}\rightarrow0$,
the distribution computed at fixed order is ill-defined at $\Delta\varphi=\pi$.
The resummed distribution shows a larger cross section near $\Delta\varphi=2.5$~rad,
in better agreement with the data. In the region of small $\Delta\varphi\lesssim\pi/2$,
the fixed-order and the resummed predictions are the same, a result
of our matching of the resummed and fixed-order distributions at mid
to high values of $Q_{T}$. Although the cross section is not large
in the region $\Delta\varphi<\pi/2$, there is an indication of a
difference between our predictions and data in this region, a topic
we address below.

\subsubsection{The region $Q_{T}>Q$ \label{subsection:QT_gt_Q}}

It is evident from Fig.~\ref{Fig:KinCuts} that the $\Delta\varphi<\pi/2$
region is populated mostly by events with $Q_{T}>Q$. New types of
radiative contributions may be present in this region, including
various fragmentation contributions described 
in Sec.~\ref{subsection:Fragmentation-model} 
and enhancements at large $|\cos\theta_{*}|$ in the direct production rate. 

While experimental isolation generally suppresses long-distance fragmentation,
a greater fraction of fragmentation photons are expected to survive
isolation when $\Delta\varphi<\pi/2$. Besides single-photon `one-fragmentation'
and `two-fragmentation' contributions (with one photon per fragmenting
parton), one encounters additional logarithmic singularities of the
form $\log(Q/Q_{T})$. We noted in Sec.~\ref{subsection:Fragmentation-model}
that these logarithms are associated with the fragmentation of a parton
carrying large transverse momentum $Q_{T}$ into a system of small
invariant mass $Q$ \cite{Berger:1998ev,Berger:2001wr}, a light $\gamma\gamma$
pair in our case. Small-$Q$ $\gamma\gamma$ fragmentation of this
kind is not implemented yet in theoretical models. Therefore, we are
prepared for the possibility that both the fixed-order calculation
and our resummed calculation may be deficient in the region $Q_{T}\gg Q$.
A detailed experimental study of the region $Q_{T}>Q$ may offer the
opportunity to measure the parton to two-photon fragmentation function
$D_{\gamma\gamma}(z_{1},z_{2})$, provided that the single-photon
`one-fragmentation' function $D_{\gamma}(z)$ is determined by single-photon
data, and the low-$Q$ logarithmic terms are properly resummed theoretically.

In addition to the low-$Q$ fragmentation, the small-$\Delta\varphi$ region
may be sensitive to large higher-order contributions associated with
$\widehat{t}$- or $\widehat{u}$-channel exchanges in the $q\bar{q}\rightarrow\gamma\gamma$
and $gg\rightarrow\gamma\gamma$ subprocesses. In the Born processes in
Figs.~\ref{Fig:FeynDiag}(a) and (h),
the $\widehat{t}$- and $\widehat{u}$-channel singularities arise
at $\cos\theta_{*}\approx\pm1$ and $\Delta\varphi\approx\pi$.
These singularities are excluded by the $p_{T}^{\gamma}$
cuts in Eq.~(\ref{pTcutTev}), but related residual enhancements
in the NLO contributions may still persist at $|\cos\theta_{*}|\approx1$ and
$\Delta\varphi\rightarrow0$, not excluded by the cuts (cf.  Fig.~\ref{Fig:KinCuts}). Because $\left| \cos \theta_*\right|$ is large in such events,
they tend to have substantial $|\Delta y|$, so they are retained
by the $\Delta R_{\gamma\gamma}>0.3$ cut. In contrast, the low-$Q$
fragmentation contributions tend to be abundant at small $|\Delta y|$.
It may be therefore possible to distinguish between the large-$|\cos\theta_{*}|$
and fragmentation events at small $\Delta\varphi$ based on the distribution
in $|\Delta y|$.

We expect much better agreement of our predictions with data
if the selection $Q_{T}<Q$ is made. This selection preserves the
bulk of the cross section and assures that a fair comparison is made
in the region of phase space where the predictions are most valid.

\subsubsection{Fragmentation and comparison with the \textsc{DIPHOX} code}

One way to obtain an estimate of theoretical uncertainty is to compare
theoretical approaches in various parts of phase space, including
small $\Delta\varphi$. We handle the collinear final-state photon
singularities in the manner described in Sec.~\ref{Sec:Theory},
without including photon fragmentation functions explicitly. An alternative
calculation implemented in the \textsc{DIPHOX} code \cite{Binoth:1999qq}
includes NLO cross sections for single-photon fragmentation
processes. Neither code includes a term in which both photons are
fragmentation products of the same final-state parton, i.e., the diphoton
fragmentation function $D_{\gamma\gamma}(z_{1},z_{2})$.

\begin{figure*}
\begin{centering}\includegraphics[width=0.43\columnwidth]{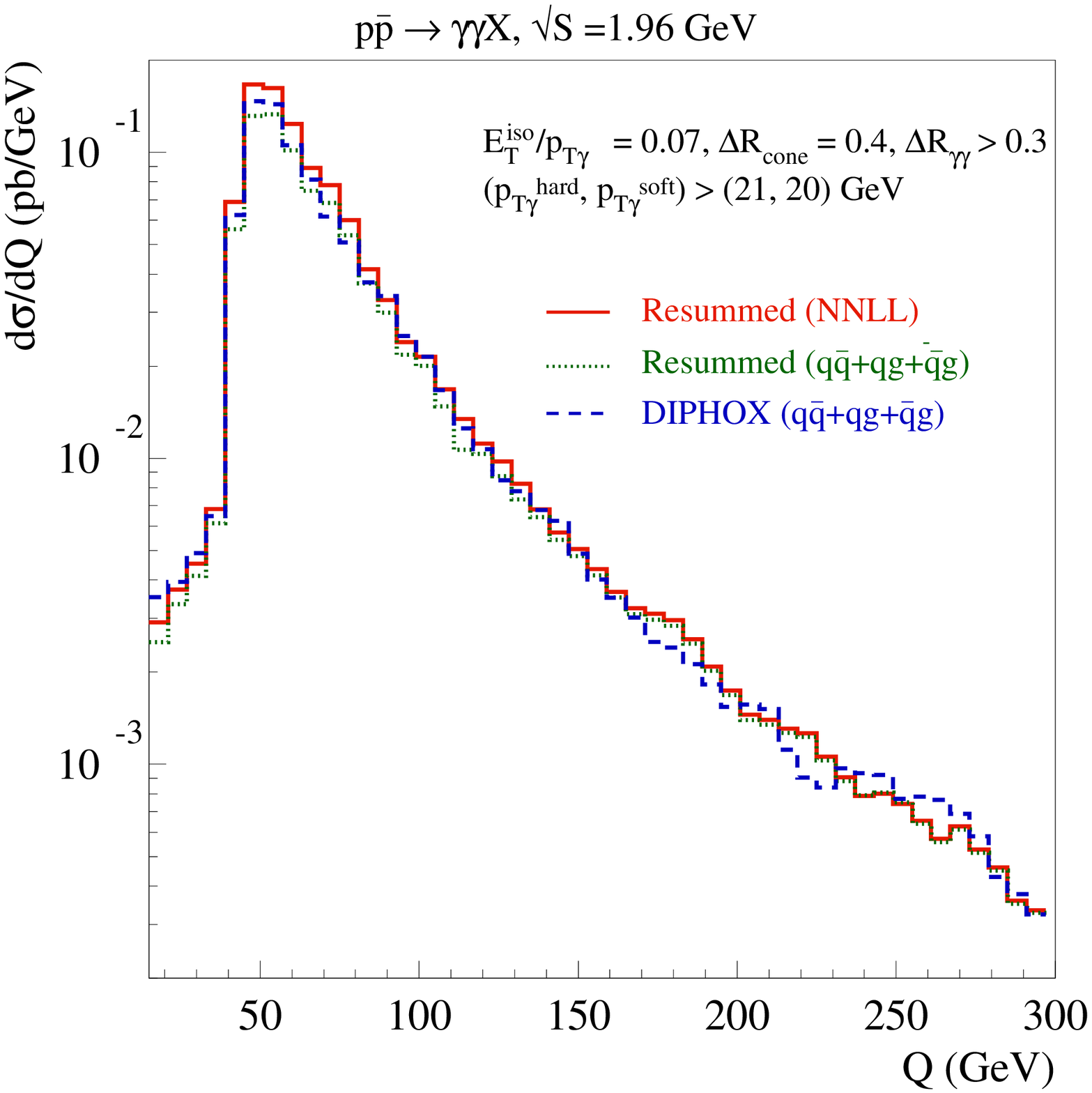}
\includegraphics[width=0.43\columnwidth]{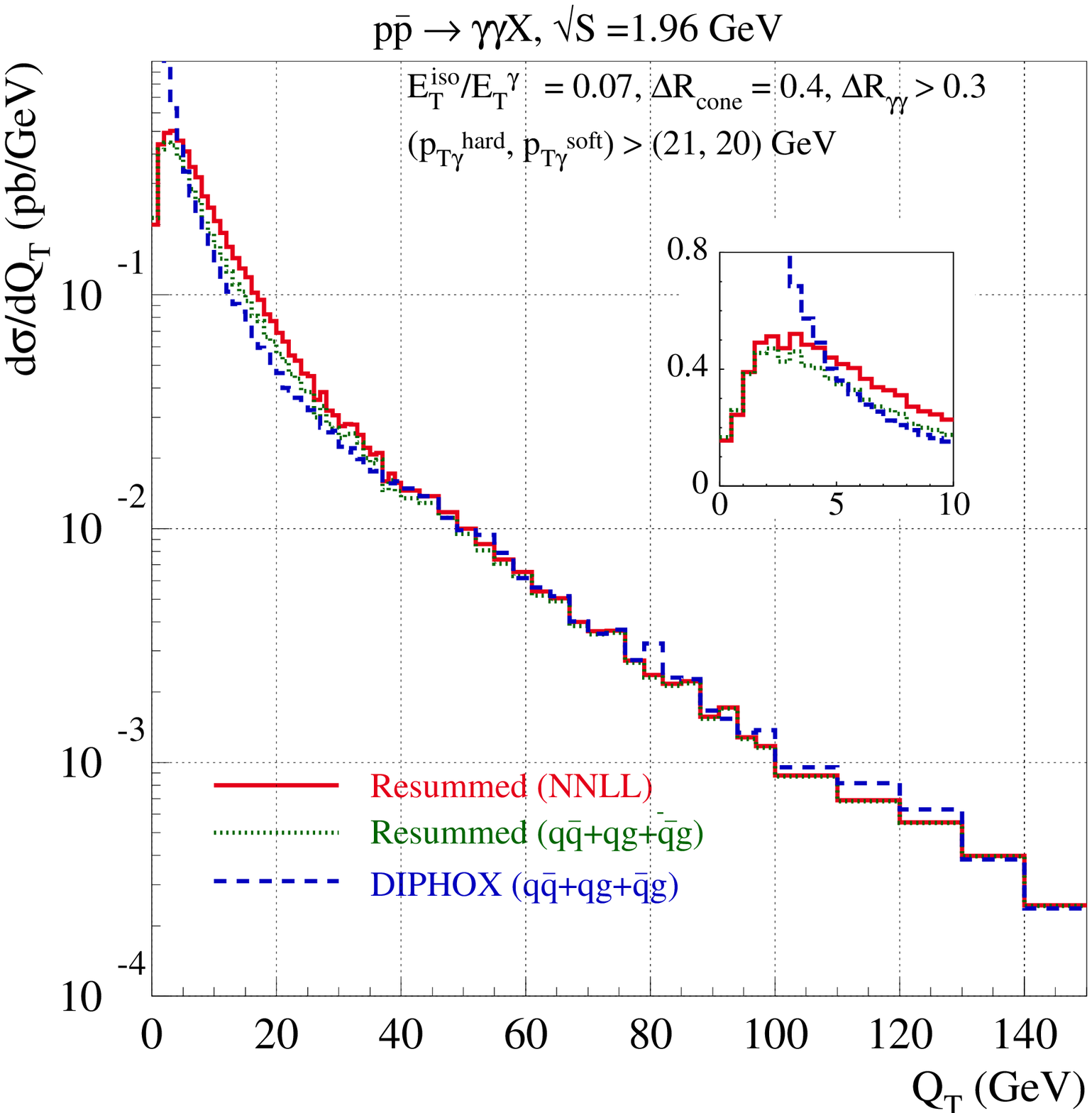}\\
 (a) \hspace{0.45\columnwidth} (b) \par\end{centering}

\caption{Comparison of our resummed and \textsc{DIPHOX} predictions for (a)
the invariant mass and (b) transverse momentum distributions of $\gamma\gamma$
pairs for D\O~kinematic cuts. The solid curves show our resummed
distributions with all channels included. The dashed and dotted curves
illustrate the resummed and DIPHOX distributions in the $q\bar{q}+qg$
channel. \label{Fig:QTD0}}
\end{figure*}

In Ref.~\cite{Balazs:2006cc} we show comparisons of our predictions
with those of \textsc{DIPHOX} along with the CDF data. Here in Fig.~\ref{Fig:QTD0},
we show analogous plots of the invariant mass and transverse momentum
distributions for D\O~cuts. We note that our fixed-order $q\bar{q}+qg$
contribution agrees well with the direct contribution in \textsc{DIPHOX}.
This agreement is particularly impressive in the region of large $Q_{T}$,
where both codes use the same fixed-order formalism to handle direct
contributions. A contribution from the $gg$ channel is also present
in both codes, computed at LO in \textsc{DIPHOX} but at NLO+NNLL in
our case. Since the $gg+gq_{S}$ contribution is not dominant (especially
in the high $Q_{T}$ region), this difference does not have a significant
impact on the comparison. 

The explicit single-photon fragmentation contributions in \textsc{DIPHOX}
(mostly `one-fragmentation' contribution) are quite small for the
nominal hadronic energy $E_{T}^{iso} \sim 1$~GeV in the cone around each photon. Exceptions
occur in the region $Q_{T}\leq E_{T}^{iso}$, where the fragmentation
contributions to $d\sigma/dQ_{T}$ have logarithmic singularities,
and in the $\Delta\varphi\rightarrow0$ region, where fragmentation
is comparable to the direct contributions. Our isolation prescription
reproduces the integrated \textsc{DIPHOX} rate well for $0\leq Q_{T}\leq E_{T}^{iso}$,
leading to close agreement between the resummed and \textsc{DIPHOX}
inclusive rates for most $Q$ values.

Returning to the CDF measurement, we remark that the resummed and
\textsc{DIPHOX} cross sections for the same $E_{T}^{iso}=1$ GeV underestimate
the data within two standard deviations for $Q\lesssim27$ GeV, $Q_{T}>25$
GeV, and $\Delta\varphi<1$ rad (cf.  the relevant figures in Ref.~\cite{Balazs:2006cc}).
The \textsc{DIPHOX} cross section can be raised to agree with data
in this {}``shoulder'' region, if a much larger isolation energy
($E_{T}^{iso}=4$ GeV) is chosen, and smaller factorization and renormalization
scales are used ($\mu_{F}=\mu_{R}=Q/2$). These are the choices made
in the CDF study \cite{Acosta:2004sn}. Since $E_{T}^{iso}$ is an 
approximate characteristic of the experimental isolation, one might
argue that both $E_{T}^{iso}=1$ and 4 GeV can be appropriate in a calculation
to match the conditions of the CDF measurement. The
direct contribution is weakly sensitive to $E_{T}^{iso}$, while the
one-fragmentation part of $d\sigma/dQ_{T}$ is roughly proportional to $E_{T}^{iso}$
(cf.  Section~\ref{subsection:Fragmentation-model}). The one-fragmentation
contribution is enhanced on average by 400\% if $E_{T}^{iso}$ is
increased in the calculation from 1 to 4 GeV. The rate in the shoulder
region is enhanced further if the factorization scale $\mu_{F}$ is reduced.

\begin{figure*}
\includegraphics[width=0.49\columnwidth]{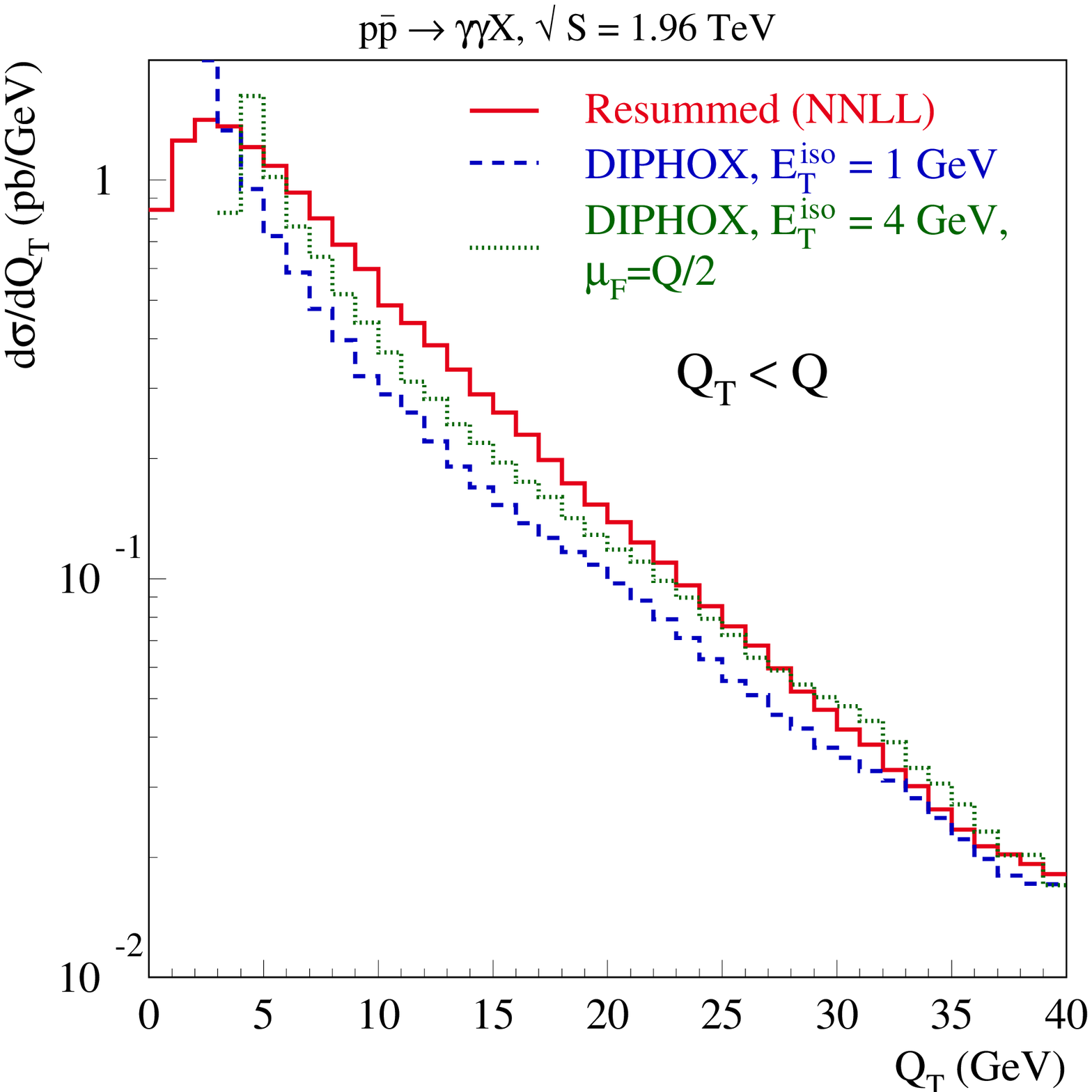}
\includegraphics[width=0.49\columnwidth]{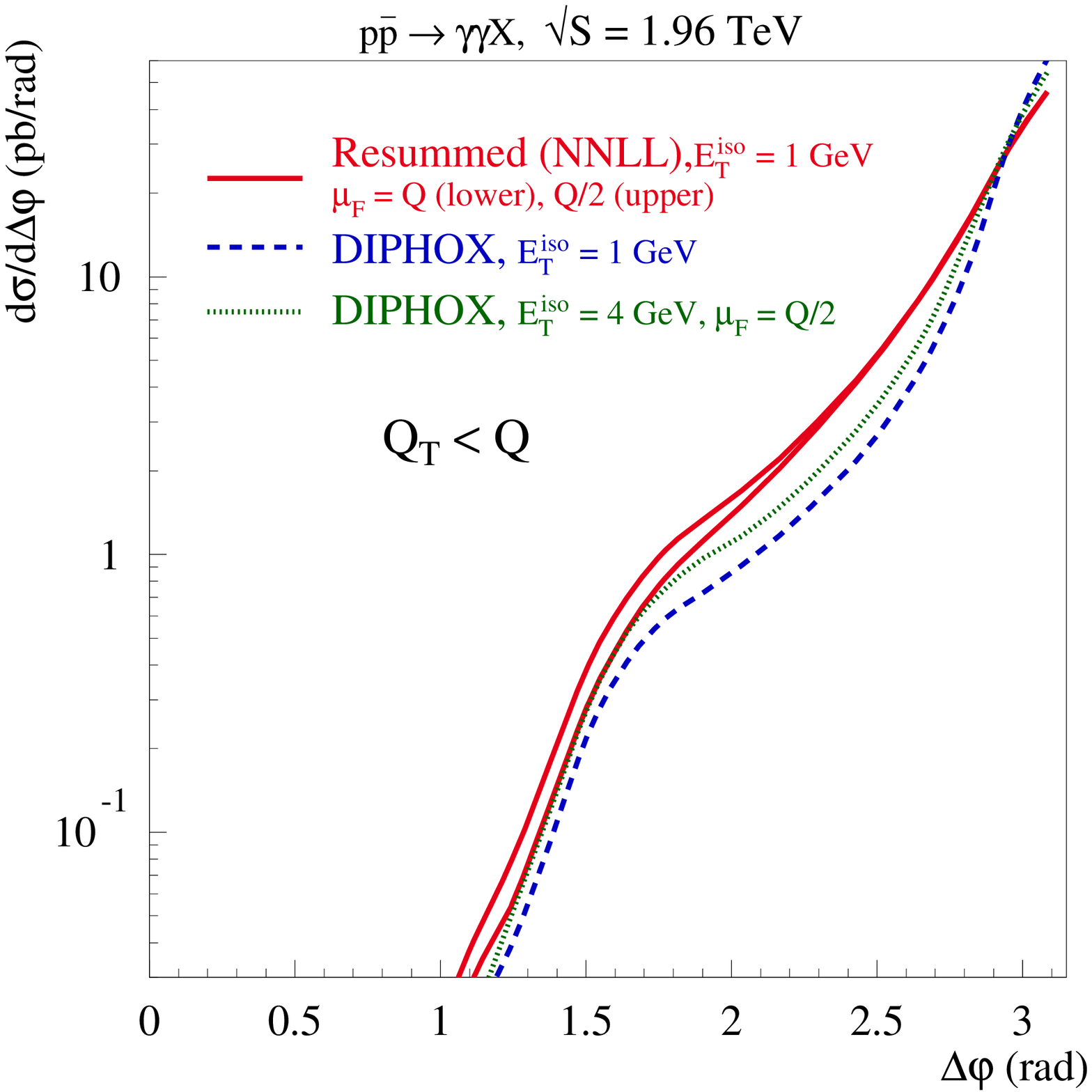}
(a) \hspace{0.45\columnwidth} (b)

\caption{Predicted cross sections for diphoton production in $p\bar{p}\rightarrow\gamma\gamma X$
at $\sqrt{S}=1.96$~TeV as a function of (a) the $\gamma\gamma$
pair transverse momentum $Q_{T}$ and (b) the difference $\Delta\varphi$
in the azimuthal angles of the two photons. Our resummed predictions
(solid) are shown together with \textsc{DIPHOX} predictions for the
default isolation energy $E_{T}^{iso}=1$ GeV and factorization scale
$\mu_{F}=Q$ (dashed), and for $E_{T}^{iso}=4$ GeV, $\mu_{F}=Q/2$
(dotted). We impose the condition $Q_{T}<Q$ to reduce theoretical
uncertainties associated with fragmentation.}

\label{Fig:DeltaPhiCDFResFox2} 
\end{figure*}

Since the theoretical specifications for isolation and for the fragmentation
contribution are admittedly approximate, we question whether great
importance should be placed on the agreement of theory and experiment
in the region of small $\Delta\varphi$ or in the shoulder
region in the $Q_{T}$ distribution. A straightforward way to reduce
sensitivity to fragmentation is to require $Q>27$ GeV or $Q_{T}<Q$, 
as discussed above.  The two cuts have similar effects on the event
distributions. Figure~\ref{Fig:DeltaPhiCDFResFox2} shows the effects
of the $Q_{T}<Q$ cut on the $Q_{T}$ and $\Delta\varphi$ distributions.
The cut $Q_{T}<Q$ is particularly efficient at suppressing the fragmentation
$Q_{T}$ shoulder (and the region of small $\Delta\varphi$
altogether), while only a small portion of the event sample is lost.
This cut is especially favorable, since it constrains the comparison
with data to a region where the theory is well understood and has
a small uncertainty. Furthermore, with the requirement of $Q_{T}<Q$,
the dependence of differential cross sections on the choices of isolation
energy $E_{T}^{iso}$ and factorization scale $\mu_{F}$ is greatly
reduced to the typical size of higher-order corrections. We predict
that if a $Q_{T}<Q$ cut, or a $Q>27$ GeV cut, is applied to the
Tevatron data, the enhancement at low $\Delta\varphi$ and intermediate
$Q_{T}$ associated with the fragmentation contribution will disappear.
This is an important conclusion of our study, and we urge the CDF
and D\O\, collaborations to apply these cuts in their future analyses
of the diphoton data.

\subsubsection{Average transverse momentum \label{subsubsection:AverageQT}}

\begin{figure*}
\includegraphics[width=0.49\textwidth,keepaspectratio]{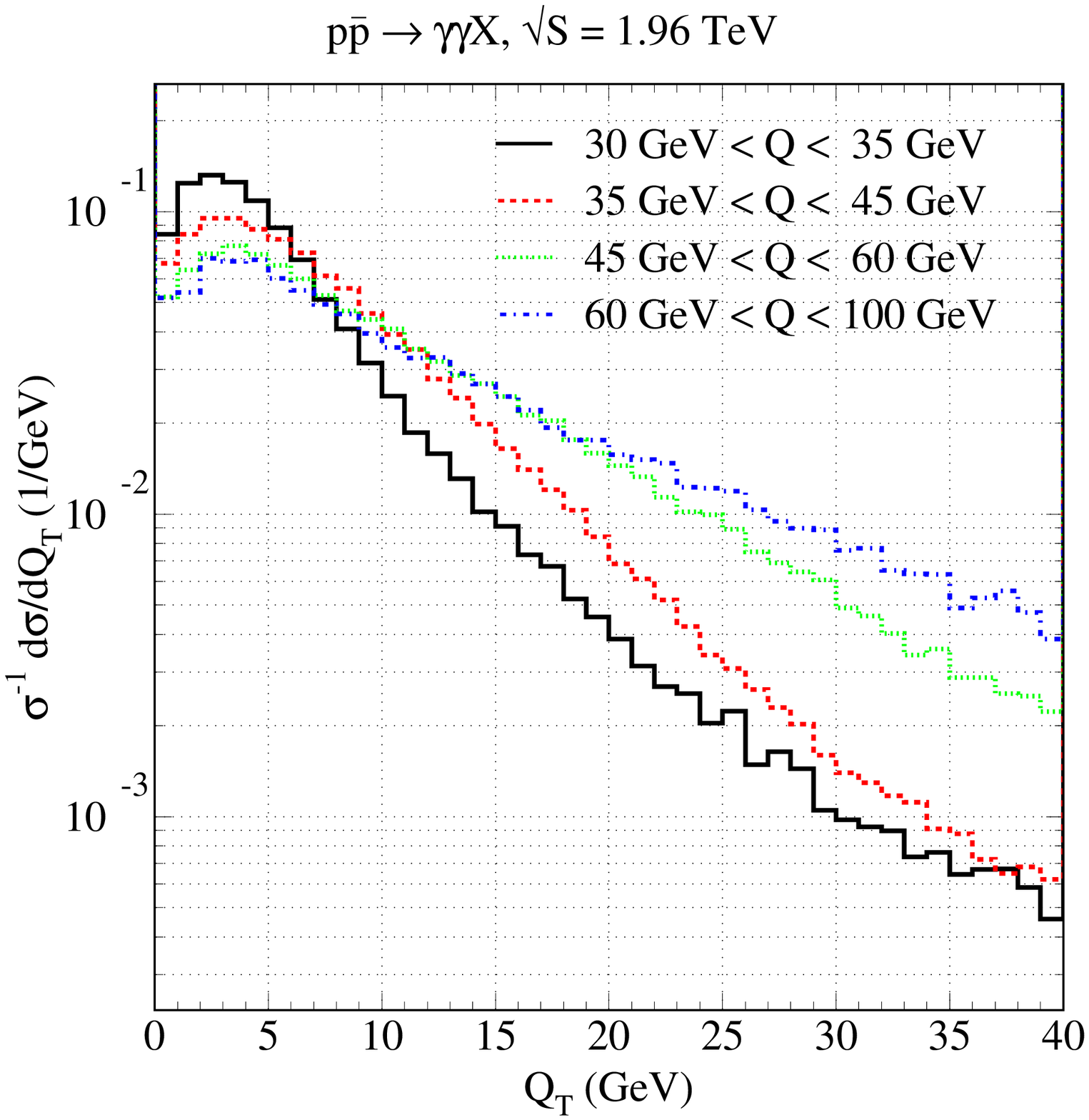}\includegraphics[width=0.49\textwidth]{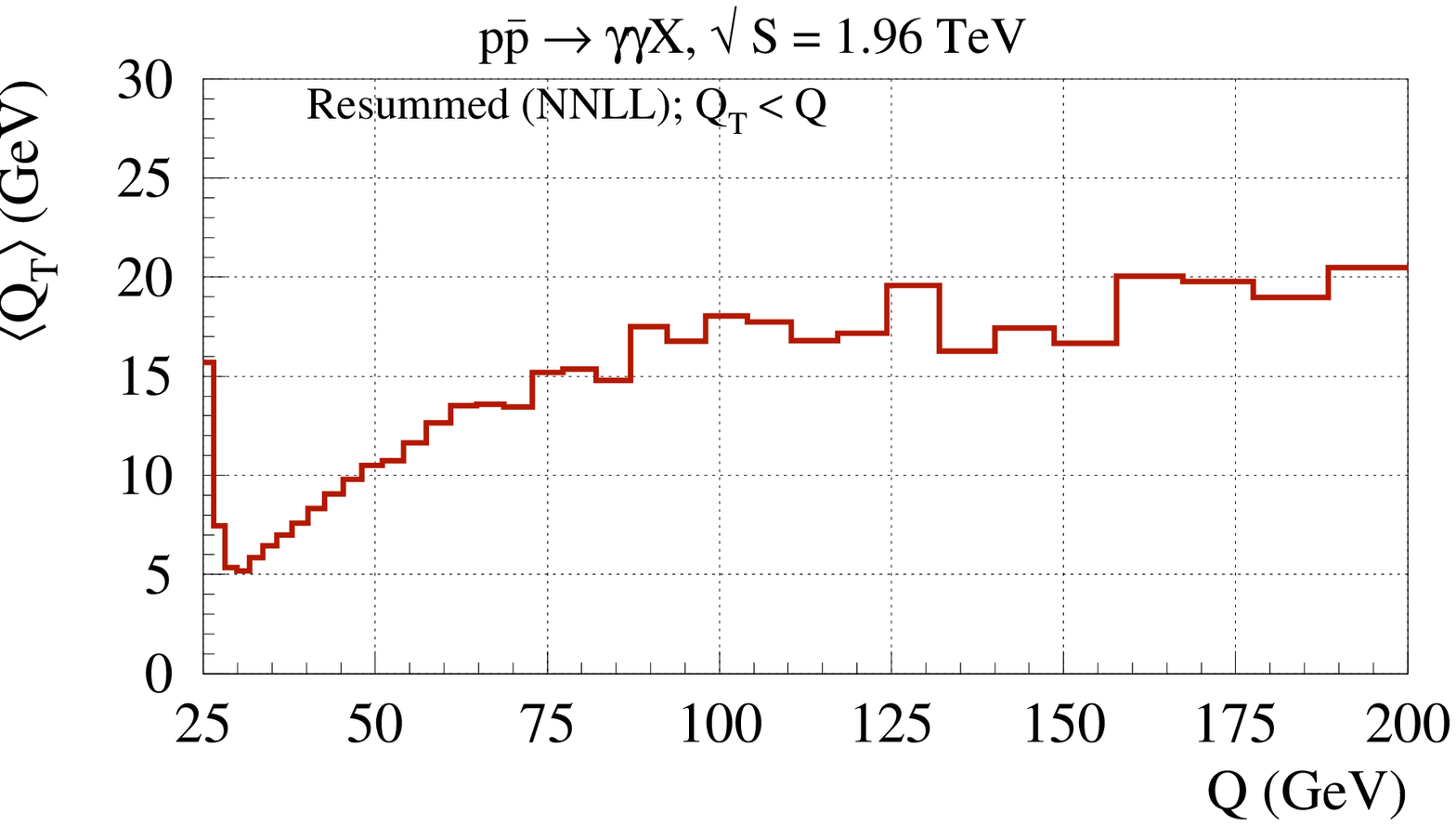}

\begin{centering}(a) \hspace{0.45\columnwidth} (b) \par\end{centering}

\caption{(a) Resummed transverse momentum distributions of photon pairs in
various invariant mass bins used in the CDF measurement, normalized
to the total cross section in each $Q$ bin. (b) The average $Q_{T}$
as a function of the $\gamma\gamma$ invariant mass, computed for
$Q_{T}<Q$. }

\label{Fig:QTMassBinsCDF} 
\end{figure*}

An important prediction of the resummation formalism is the change
of the transverse momentum distribution with the diphoton invariant
mass. This dependence comes, in part, from the $\ln Q^{2}$ dependence
in the Sudakov exponent, Eq.~(\ref{Sudakov}), and it is desirable
to identify this feature amid other influences. In Fig.~\ref{Fig:QTMassBinsCDF}(a),
we show normalized resummed transverse momentum distributions for
various selections of the invariant mass of the photon pairs. Without
kinematical constraints on the decay photons, the
$Q_{T}$ distribution is expected to broaden with increasing $Q$, and
the position of the peak in $d\sigma/dQ_T$ to shift to larger $Q_T$
values. The shift of the peak may or may not be observed in the data
depending on the chosen lower cuts on $p_T$ of the photons, which
suppress the event rate at low $Q$ and $Q_T$. The interplay of the
Sudakov 
broadening of the $Q_T$ distribution and kinematical suppression by
the photon $p_T$ cuts is reflected in the shape of $d\sigma/dQ_{T}$
in different $Q$ bins.

According to dimensional analysis, the average $\langle Q_{T}\rangle$
in the interval $Q_{T}\leq Q$ may be expected to behave as \begin{equation}
\langle Q_{T}\rangle_{Q_{T}\leq Q}=Qf(Q/\sqrt{S}),\label{mean}\end{equation}
 where the scaling function $f(Q/\sqrt{S}$) reflects phase space
constraints, dependence on 
the Sudakov logarithm, and the $x$ dependence
of the PDFs. Figure~\ref{Fig:QTMassBinsCDF}(b)
shows our calculated diphoton mass dependence of $\langle Q_{T}\rangle$.
The linear increase shown in Eq. ({\ref{mean}) is observed over the
range $30<Q<80$~GeV. For values of $Q$ below the kinematic cutoff
at about 30 GeV, the cuts shown in Fig. 3 suppress diphoton production
at small $Q_{T}$, and $\langle Q_{T}/Q\rangle$ grows toward 1 as
$Q$ decreases (corresponding to production only at $Q_{T}$ close
to $Q$). For $Q\sim80$~GeV and above, we see a saturation of the
growth of $\langle Q_{T}\rangle$, a reflection of the influences
of the $x$ dependence of the PDFs and other factors. Similar saturation
behavior is observed in calculations of $\langle Q_{T}\rangle$ in
other processes \textbf{}\cite{Berger:2002ut}\textbf{.} It would
be interesting to see a comparison of our prediction with data from
the CDF and D\O~collaborations.

\subsection{Results for the LHC}

\subsubsection{Event selection}

To obtain predictions for $pp$ collisions at the LHC at $\sqrt{S}=14$~TeV,
we employ the cuts on the individual photons used by the ATLAS collaboration
in their simulations of Higgs boson decay, $h\rightarrow\gamma\gamma$~\cite{ATLAS:1999fr}.
We require \begin{eqnarray}
 &  & {\textrm{transverse momentum}}~p_{T}^{\gamma}>40~(25)~{\textrm{GeV for the harder (softer) photon, }}\label{pTcutLHC}\\
 &  & {\textrm{and rapidity}}~|y^{\gamma}|<2.5~{\textrm{for each photon}}.\label{ycutLHC}\end{eqnarray}
 In accord with ATLAS specifications, we impose a looser isolation
restriction than for our Tevatron study, requiring less than $E_{T}^{iso}=15$
GeV of hadronic and extra electromagnetic transverse energy inside
a $\Delta R=0.4$ cone around each photon. We also require the separation
$\Delta R_{\gamma\gamma}$ between the two isolated photons to be
above 0.4.

The cuts listed above, optimized for the Higgs boson search, may require 
adjustments
in order to test perturbative QCD predictions in the full $\gamma\gamma$
invariant mass range accessible at the LHC. The values of the $p_{T}^{\gamma}$
cuts on the photons in Eq.~(\ref{pTcutLHC}) preserve a large fraction
of Higgs boson events with $Q>115$ GeV. These cuts may be too restrictive
in studies of $\gamma\gamma$ production at smaller $Q$, considering
that the two final-state photons most likely originate from a $\gamma\gamma$
pair with small $Q_{T}$ and have similar values of $p_{T}^{\gamma}$
of about $Q/2$. The $p_T$ cuts interfere with the expected Sudakov
broadening of $Q_T$ distributions with increasing diphoton invariant mass, as
discussed in Section~\ref{subsubsection:AverageQT}. We further note that the asymmetry between the $p_{T}^{\gamma}$
cuts on the harder and softer photons is necessary in a fixed-order
perturbative QCD calculation, but it is not required in the resummed
calculation. At a fixed order of $\alpha_{s}$, asymmetry in the $p_{T}^{\gamma}$
cuts prevents instabilities in $d\sigma/dQ$ caused by logarithmic
divergences in $d\sigma/dQ_{T}$ at small $Q_{T}$. Such instabilities
are eliminated altogether once the small-$Q_{T}$ logarithmic terms
are resummed to all orders of $\alpha_{s}$. Here we do not consider
alternative $p_{T}^{\gamma}$ cuts, although experimental collaborations
are encouraged to employ relaxed and/or symmetric cuts to increase the $\gamma\gamma$
event sample in their data analysis.

\subsubsection{Resummed $Q_{T}$ distributions and average transverse momentum}

\begin{figure*}
\includegraphics[width=0.7\textwidth]{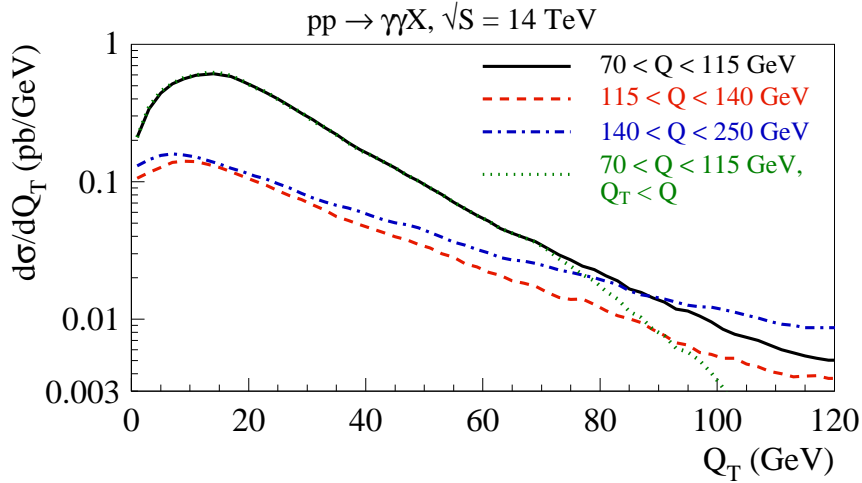}

\caption{Resummed transverse momentum distributions of photon pairs in various
invariant mass bins at the LHC. The cuts listed in Eqs.~(\ref{pTcutLHC})~and~(\ref{ycutLHC})
are imposed.The $Q_T$ distribution for $70 < Q < 115$ GeV with an additional 
constraint $Q_T < Q$ is shown as a dotted line. \label{Fig:QTinQBinsLHC}}
\end{figure*}

Figure~\ref{Fig:QTinQBinsLHC} shows transverse momentum distributions
of the photon pairs for various invariant masses. The average $\gamma\gamma$
transverse momentum grows with $Q$, as demonstrated by
Fig. \ref{Fig:QTaveLHC}. However, the rate of the growth decreases
monotonically with $Q,$ for similar reasons as at the Tevatron. 
\begin{figure}
\includegraphics[width=0.7\textwidth]{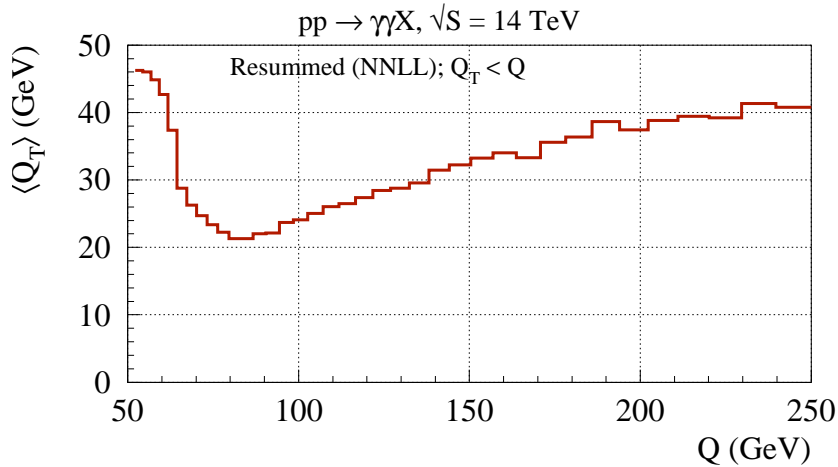}

\caption{The average $Q_{T}$ at the LHC as a function of the $\gamma\gamma$
invariant mass $Q.$ \label{Fig:QTaveLHC}}
\end{figure}

The $\gamma\gamma$ distributions in $Q$ and $\Delta\varphi$ for
different combinations of scattering subchannels and choices of theoretical
parameters are discussed in Refs.~\cite{Balazs:2006cc,Nadolsky:2007ba}.
In all ranges of $Q$, the $\gamma\gamma$ production rate is dominated
by a large $qg$ contribution, accounting for about 50\% of the
fixed-order (NLO)
rate. Although this number depends on the choice 
of the factorization scheme and scale, and, on the other hand, separate
treatment of  the $q\bar q$ and $qg$ cross sections is not meaningful 
in the resummation calculation
\cite{Nadolsky:2007ba}, it nonetheless reflects, in a crude way,
the increased relative importance of the $qg$ cross section. 
The $gg+gq_{S}$ channel contributes about 25\% at $Q\sim80$
GeV (the location of the cutoff in $d\sigma/dQ$ due to the cuts on 
$p_{T}^{\gamma}$) and less at larger $Q.$ As at the Tevatron, the dependence
of the cross sections on the resummation scheme is small~\cite{Nadolsky:2007ba}.
The dependence on the nonperturbative model can also be neglected,
as long as the nonperturbative function does not vary strongly with
$x$~\cite{Nadolsky:2007ba}.

\subsubsection{Final-state fragmentation and comparison with DIPHOX}

\begin{figure}
\includegraphics[width=0.49\textwidth,keepaspectratio]{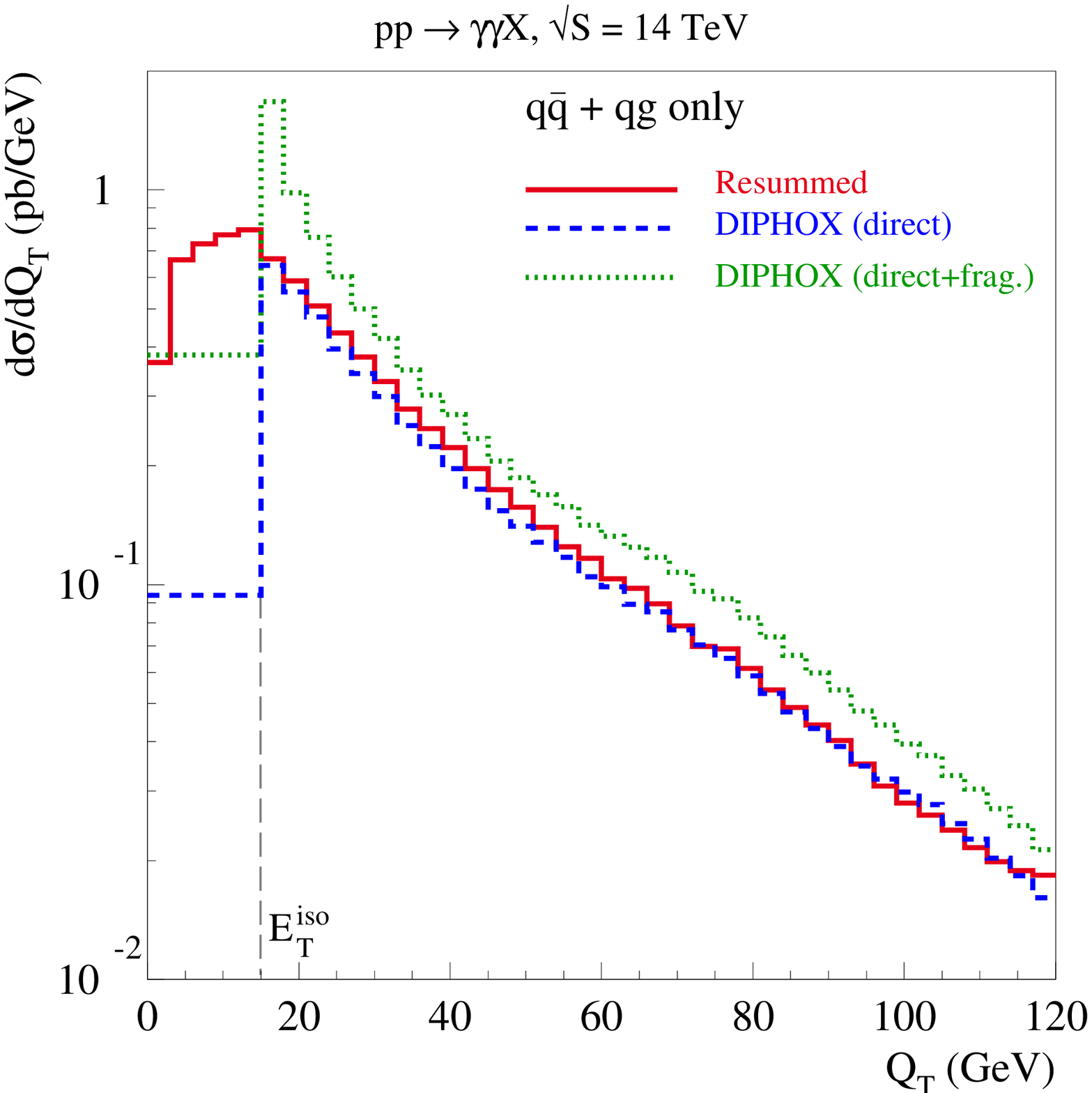}
\includegraphics[width=0.49\textwidth,keepaspectratio]{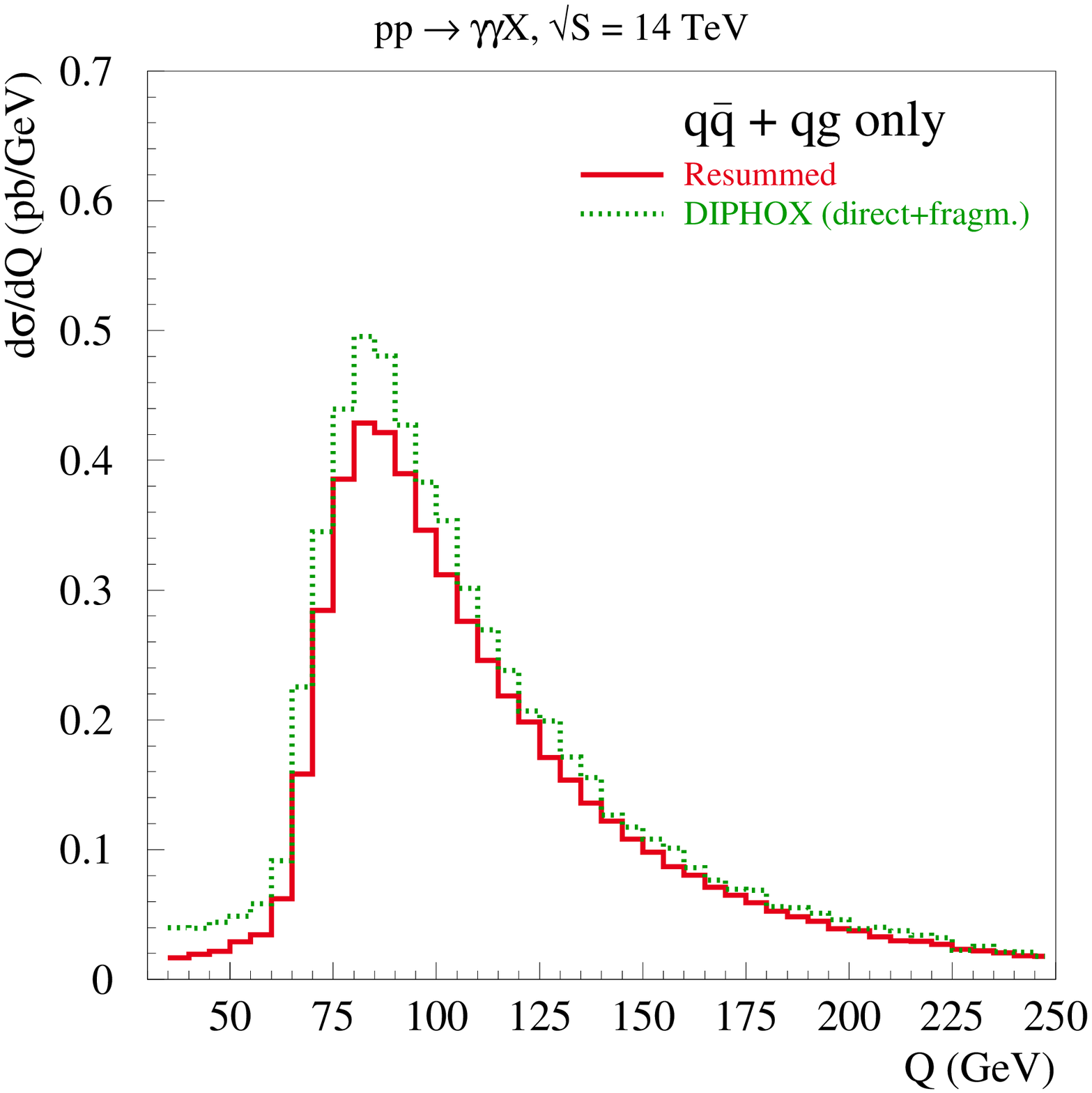}\\
 (a)\hspace{3in}(b)

\caption{Transverse momentum and invariant mass distributions $d\sigma/dQ$
in the $q\bar{q}+qg$ channel obtained in the resummation (solid)
and DIPHOX (dotted) calculations.}

\label{Fig:ResFoxQQLHC} 
\end{figure}

The impact of the final-state fragmentation at the LHC can be evaluated
if we compare our results with DIPHOX predictions. The transverse
momentum and invariant mass distributions in the $q\bar{q}+qg$ channel
from the two approaches are shown in Fig.~\ref{Fig:ResFoxQQLHC}.
In both calculations, quasi-experimental isolation removes direct
NLO events with collinear final-state photons and partons when $Q_{T}>E_{T}^{iso}=15$
GeV, but not when $Q_{T}$ is below $E_{T}^{iso}$.

Concentrating first on $\gamma\gamma$ events with $Q_{T}>E_{T}^{iso}$,
we observe that, at $Q_{T}>80$ GeV\textbf{,} the resummed $q\bar{q}+qg$
cross section reduces to the direct fixed-order cross section, evaluated
in the same way as in the DIPHOX code. Our resummed and the direct
DIPHOX cross sections, shown as solid and dashed curves, respectively,
in Fig.~\ref{Fig:ResFoxQQLHC}(a) consequently agree well at large
$Q_{T}$. At smaller $Q_{T},$ the resummed cross section is enhanced
by towers of higher-order logarithmic contributions. On the other
hand, the full $q\bar{q}+qg$ DIPHOX rate (shown as a dotted line) also
includes single-photon fragmentation contributions, which add to
the direct production cross section.  For the nominal isolation parameters,
the explicit fragmentation contribution constitutes about 25\% of
the full DIPHOX rate for $60<Q_{T}<120$ GeV. Its magnitude increases
approximately linearly with the assumed $E_{T}^{iso}$ value.

For $Q_{T}<E_{T}^{iso}$, the final-state collinear region of the
direct contribution is regulated by the collinear subtraction prescription
adopted in the resummation calculation, whereas the fragmentation
singularity is subtracted from the direct contribution and replaced
by photon fragmentation functions in the DIPHOX calculation. Subtraction
of singularities in DIPHOX introduces integrable singularities in
$d\sigma/dQ_{T}$ at different values of $Q_{T}$ below $E_{T}^{iso}$.
The origin of the final-state logarithmic singularities at values
of $Q_{T}$ below $E_{T}^{iso}$ is discussed in Refs.~\cite{Berger:1996vy,Catani:1998yh,Catani:2002ny}.
For $Q_{T}<E_{T}^{iso}$, the DIPHOX curves represent the average
over singular contributions in this $Q_{T}$ interval.

After the fragmentation singularity is subtracted, the DIPHOX direct
contribution (dashed line) is on average below our resummed $q\bar{q}+qg$
rate (solid line) over most of the range of $Q_{T}$ shown 
in Fig.~\ref{Fig:ResFoxQQLHC}(a).  After integration over all
$Q_{T}$, our resummed and DIPHOX $q\bar{q}+qg$ cross sections agree
within 10-20\% at most values of $Q$ (cf.~Fig.~\ref{Fig:ResFoxQQLHC}(b)),
with our resummed rate being below the DIPHOX rate at all $Q$. The
largest difference occurs at the lowest values of $Q$ (below the
cutoff), where the rates can differ by a factor of 2. In this region,
corresponding to diphoton events with small $\Delta\varphi$ and $Q_{T}$
larger than $Q$, the photon fragmentation contributions included
in the DIPHOX calculation are large in comparison to the direct rate.
Finally, we note that the integrated rate in DIPHOX is more stable
with respect to variations in $E_{T}^{iso}$ than the differential
distributions in DIPHOX, because $E_{T}^{iso}$ dependence for $Q_{T}>E_{T}^{iso}$
is canceled to a good degree by $E_{T}^{iso}$ dependence for $Q_{T}<E_{T}^{iso}.$

\begin{figure*}
\includegraphics[width=0.49\textwidth,keepaspectratio]{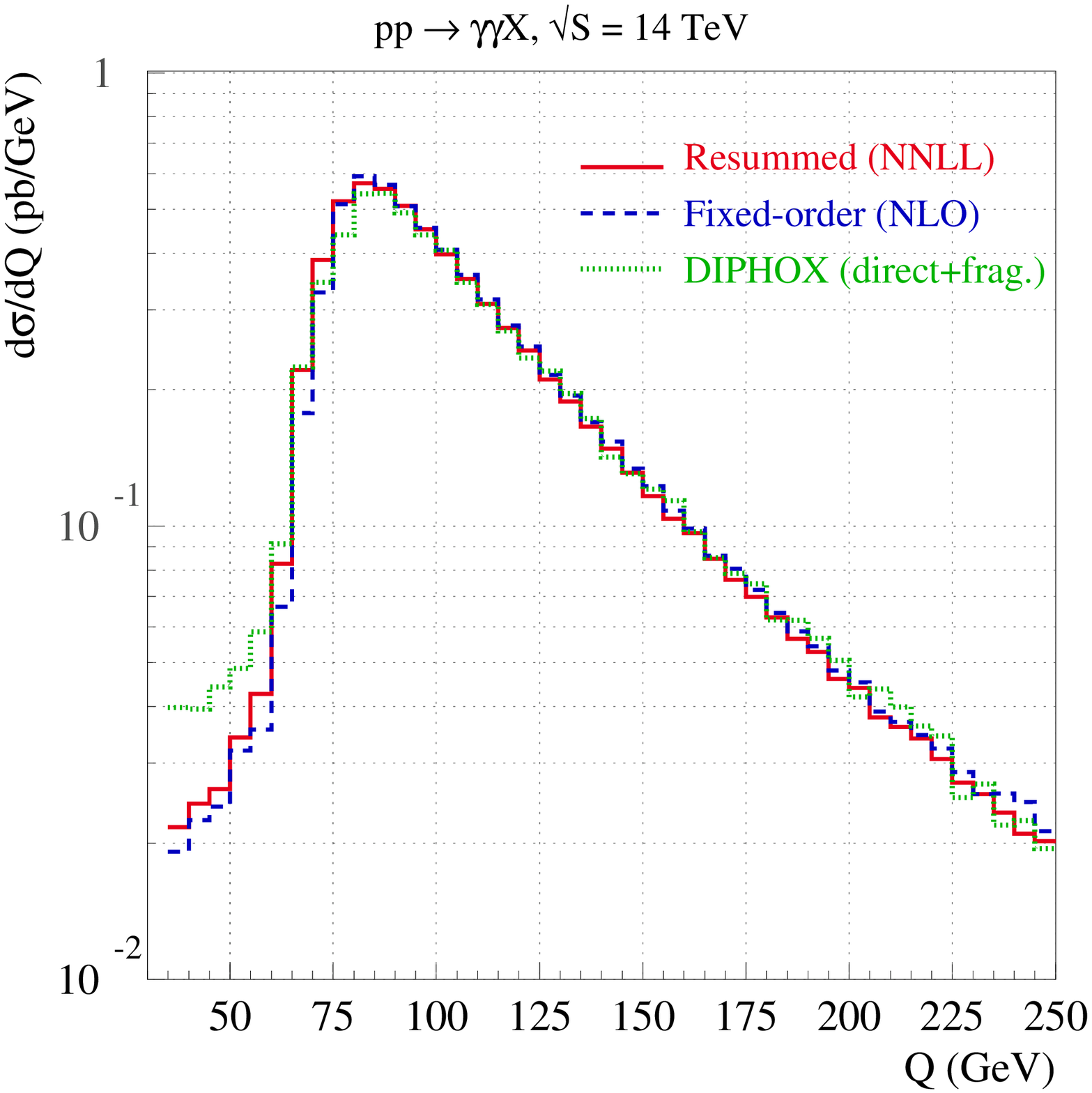}
\includegraphics[width=0.49\textwidth,keepaspectratio]{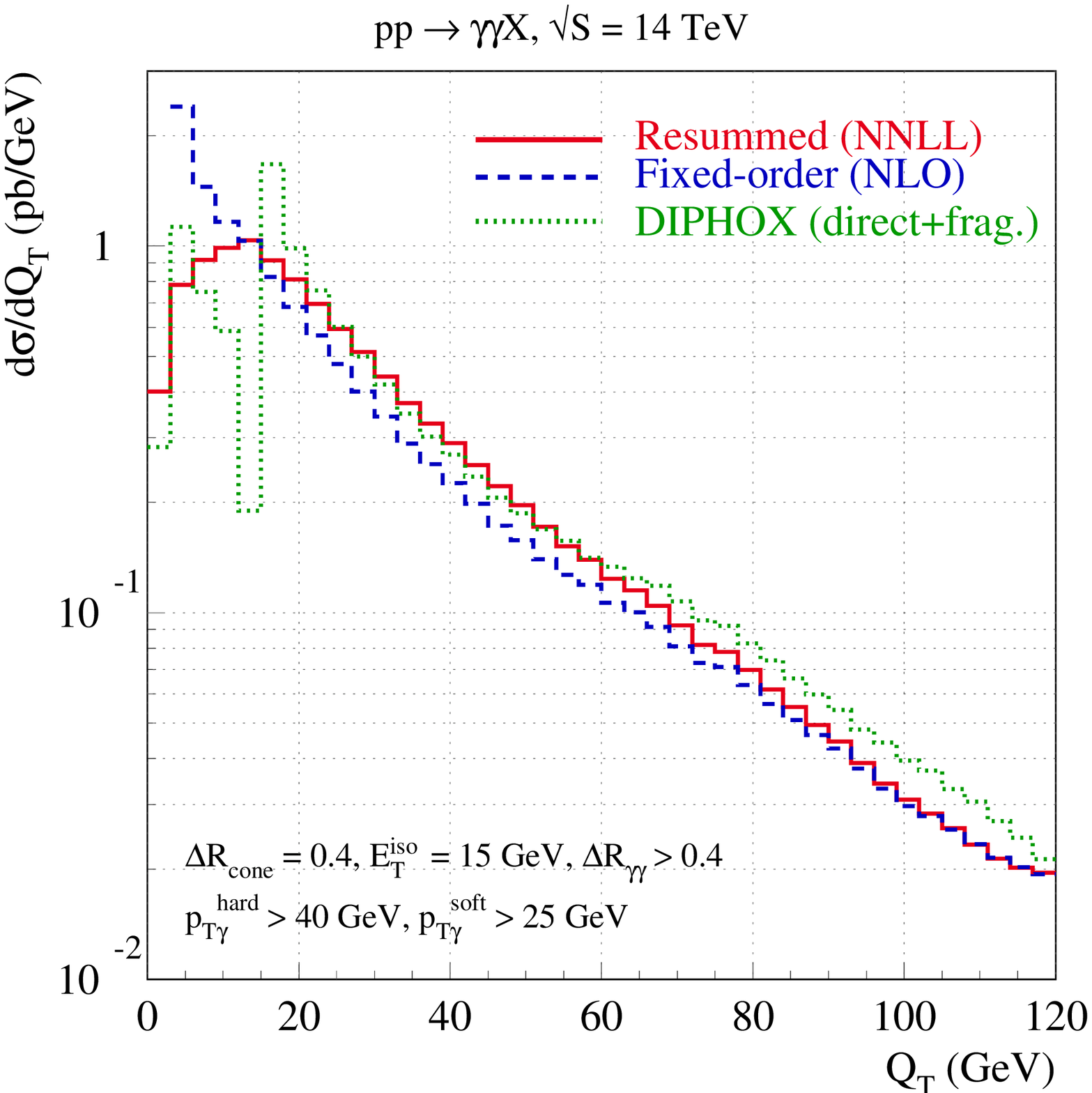}\\
(a)\hspace{3in}(b)
\includegraphics[scale=0.49]{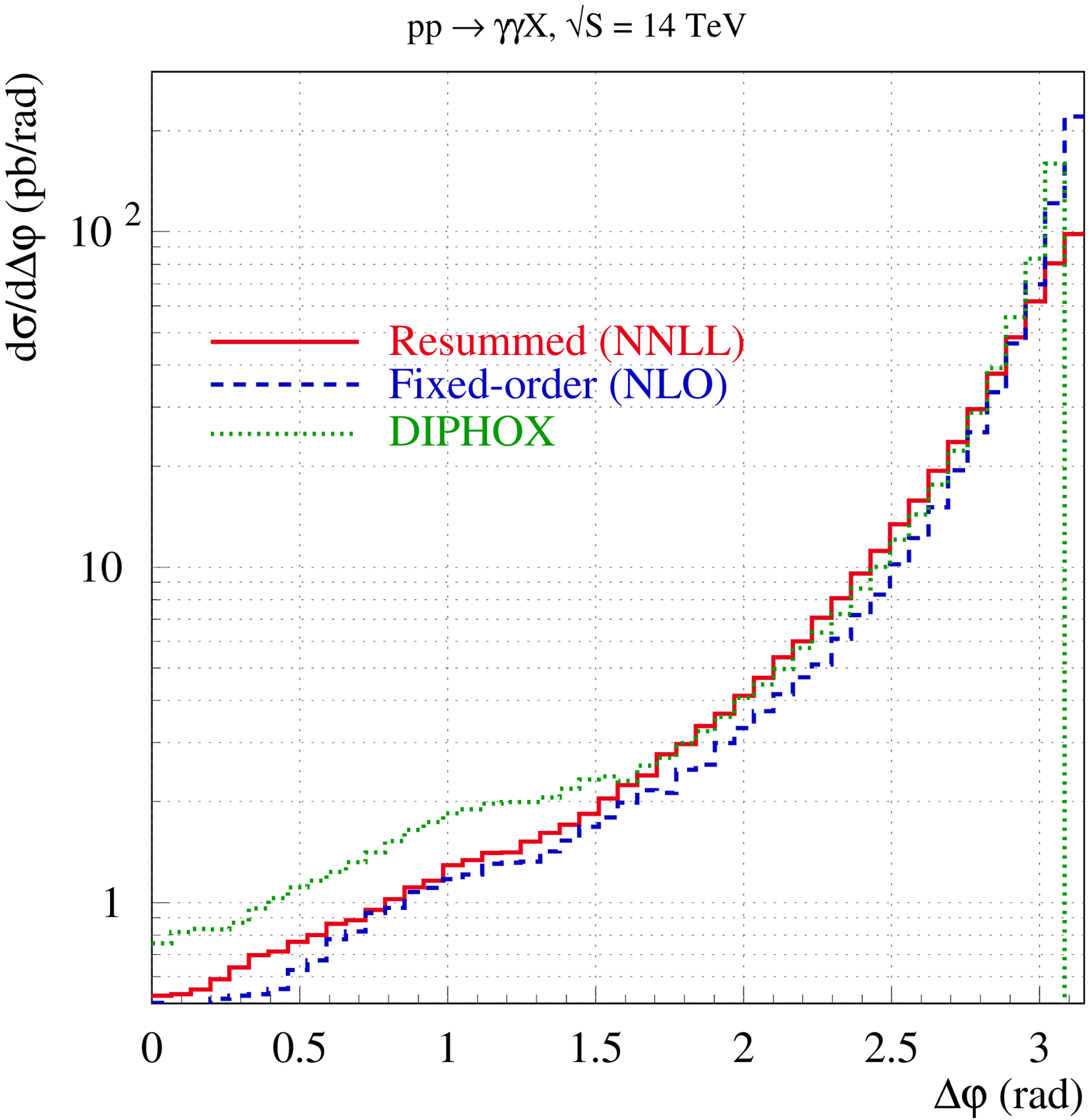}\\
(c)

\caption{Invariant mass, transverse momentum, and $\Delta\varphi$ distributions
from our resummed calculation and from \textsc{DIPHOX} at the LHC.
We show our fixed-order (dashed) and resummed (solid) distributions.
All initial states are included in both calculations, and the
single-$\gamma$ fragmentation contributions are included in \textsc{DIPHOX}.}

\label{Fig:ResFoxLHC} 
\end{figure*}

To obtain the final $\gamma\gamma$ production cross sections, after
inclusion of all channels, we combine the respective $q\bar{q}+qg$
results with the resummed NLO $gg+gq_{S}$ cross section in our case
and with the LO $gg$ cross section in the DIPHOX case. The distributions
in the $\gamma\gamma$ invariant mass $Q,$ the transverse momentum
$Q_{T}$, and the azimuthal angle separation $\Delta\varphi$ in the
lab frame are shown in Fig.~\ref{Fig:ResFoxLHC}. For the cuts chosen,
the LO $gg$ and the resummed $gg+gq_{S}$ total rates constitute
about 9\% and 20\% of the total rate. The resummed and DIPHOX invariant
mass distributions (Fig.~\ref{Fig:ResFoxLHC}(a)) are brought closer
to one another as a result of the inclusion of the $gg+gq_{S}$ contribution
in the resummed calculation. For $Q_{T}\neq0$, the full DIPHOX $Q_{T}$
distribution in Fig.~\ref{Fig:ResFoxLHC}(b) is determined entirely
by direct plus fragmentation contributions (the same as in Fig.~\ref{Fig:ResFoxQQLHC}(a)),
because the LO $gg$ cross section contributes at $Q_{T}=0$ only.
In contrast, our resummed $gg+gq_{S}$ contribution modifies the event
rate at all $Q_{T}$.

The resummed and DIPHOX rates are in a reasonable agreement for $1.5\lesssim\Delta\varphi\lesssim2.5$,
as shown in Fig.~\ref{Fig:ResFoxLHC}(c). In the $\Delta\varphi\rightarrow\pi$
limit, the fixed-order rates in DIPHOX diverge because of the singularities
at small $Q_{T}$, while our resummed rate yields a finite value.
For $\Delta\varphi<1.5$, the DIPHOX cross section is enhanced by
photon fragmentation contributions. As at the energy of the Tevatron,
theoretical uncertainties are greater at small $\Delta\varphi$. 

Predictions are most reliable when $Q_{T}<Q$ (and the angles 
$\theta_*$ and $\varphi_*$ are away from 0 or $\pi$). 
With the $Q_{T}<Q$ cut imposed, the uncertain  large-$Q_T$ 
photon fragmentation contributions 
are suppressed, and the resummed and
DIPHOX cross sections  agree well at large $Q_T$
(cf. Fig.~\ref{Fig:ResFoxLHCQTltQ}(b)). The $Q_T$
distribution  in the interval $70 < Q <
115 $ GeV with the  $Q_{T}<Q$ constraint 
is shown in Fig.~\ref{Fig:QTinQBinsLHC} by a dotted
curve. Distributions in the other two mass bins in  
Fig.~\ref{Fig:QTinQBinsLHC} are essentially not affected by this cut
in the $Q_T$ range presented.

\begin{figure*}
\includegraphics[width=0.49\textwidth,keepaspectratio]{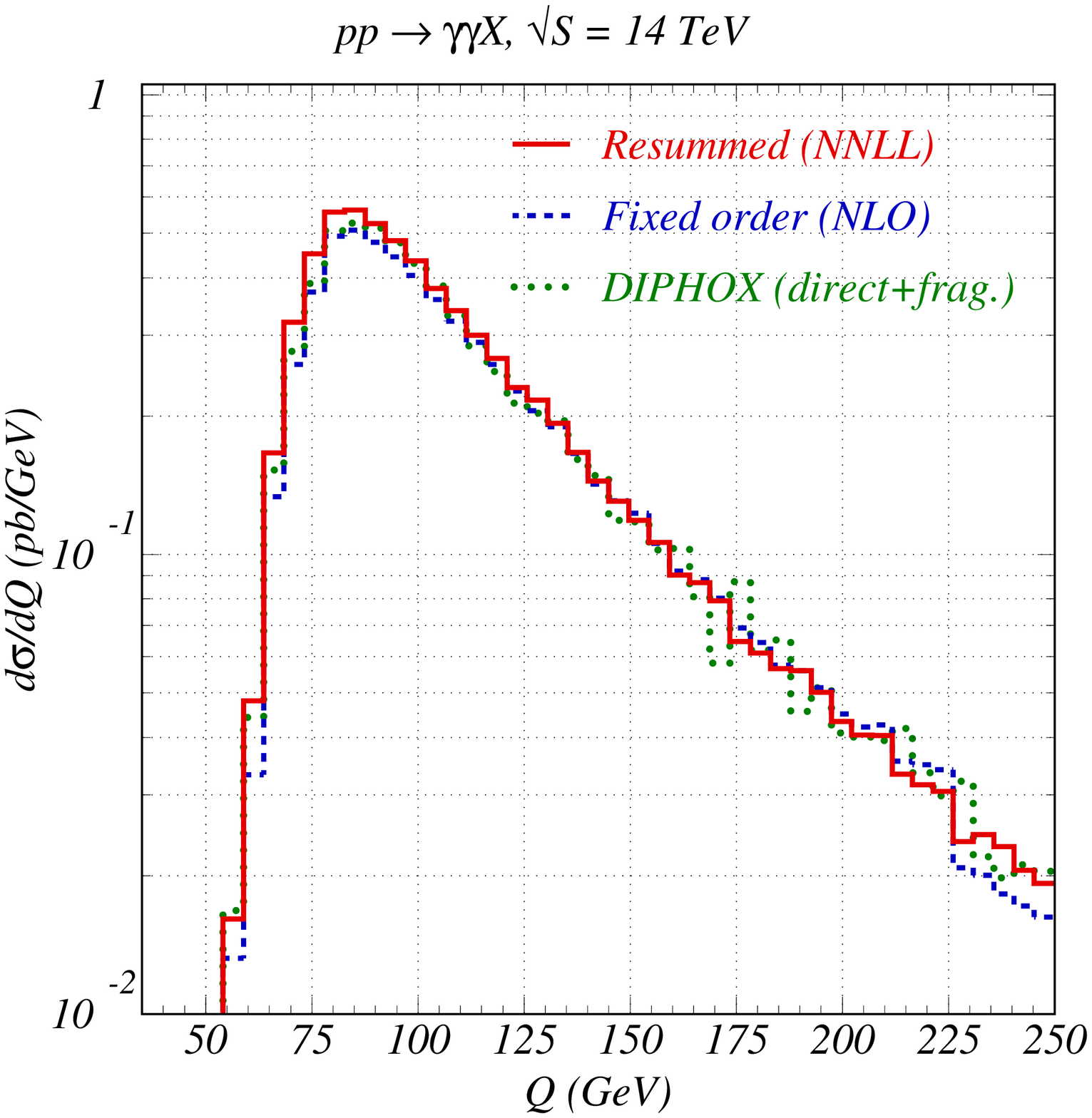}
\includegraphics[width=0.49\textwidth,keepaspectratio]{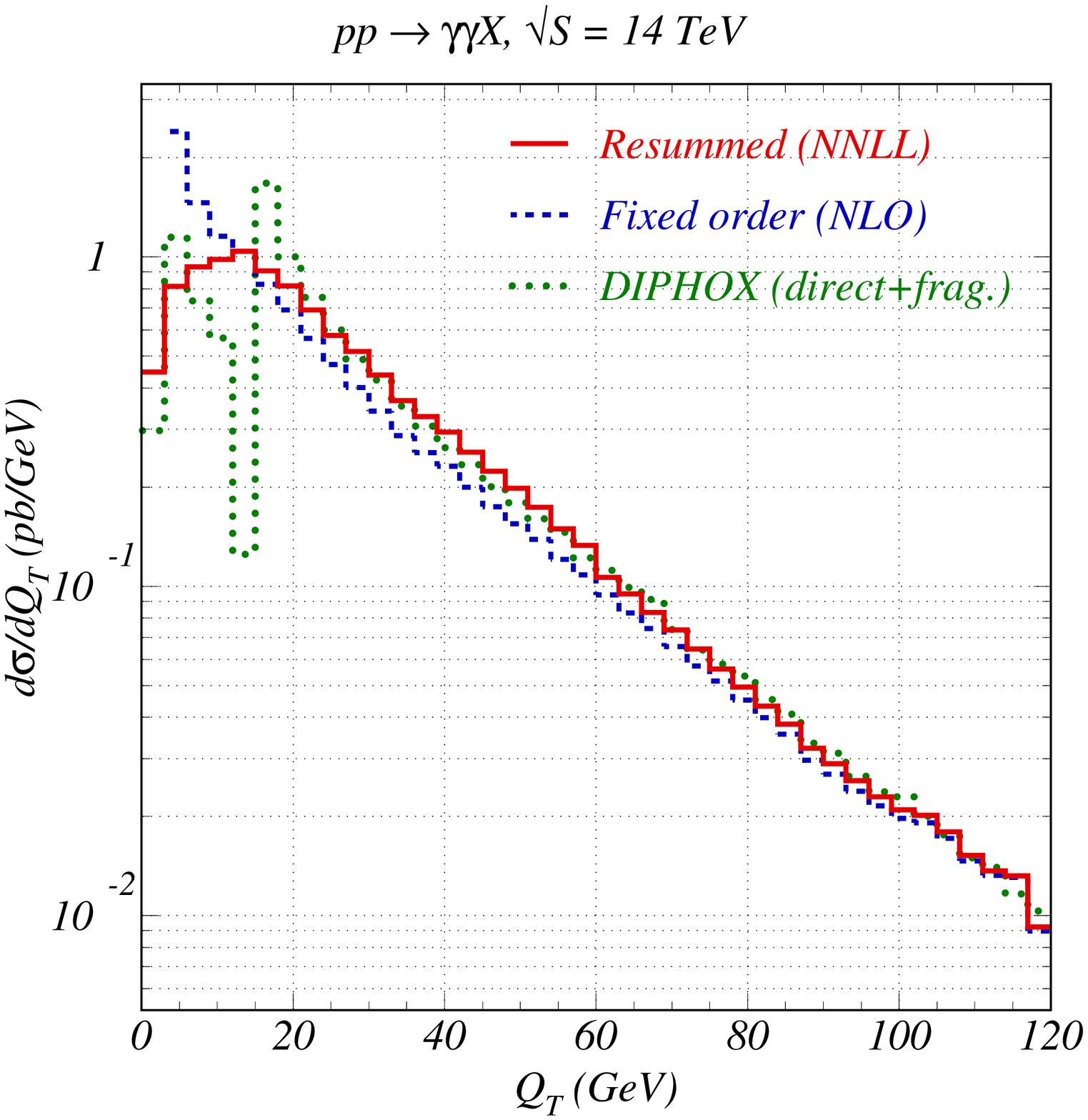}\\
(a)\hspace{3in}(b)

\caption{Invariant mass and transverse momentum distributions
from our resummed, NLO, and DIPHOX calculations at the LHC,
with the $Q_T < Q$ constraint imposed.}

\label{Fig:ResFoxLHCQTltQ} 
\end{figure*}

Our calculation captures the dominant contributions to $\gamma\gamma$
production at the LHC. However, as we noted, direct $qg$ scattering,
evaluated at order ${\mathcal{O}}(\alpha_{s})$ in our calculation,
is the leading scattering channel in the region relevant 
for the Higgs boson search at the LHC.
It is important to emphasize that the final-state collinear radiation
is not the main reason behind the enhancement of the $qg$ rate, which
is increased predominantly by contributions from non-singular phase
space regions. Consequently, the $q\bar{q}+qg$ direct rate is only
weakly sensitive to adjustments in the isolation parameters $E_{T}^{iso}$
and $\Delta R$ \cite{Bern:2002jx}. The unknown ${\mathcal{O}}(\alpha_{s}^{2})$
contributions to $qg$ scattering may be non-negligible, and it would
be valuable to compute them in the future when LHC data are available.

\begin{figure*}
\includegraphics[width=0.49\columnwidth,keepaspectratio]{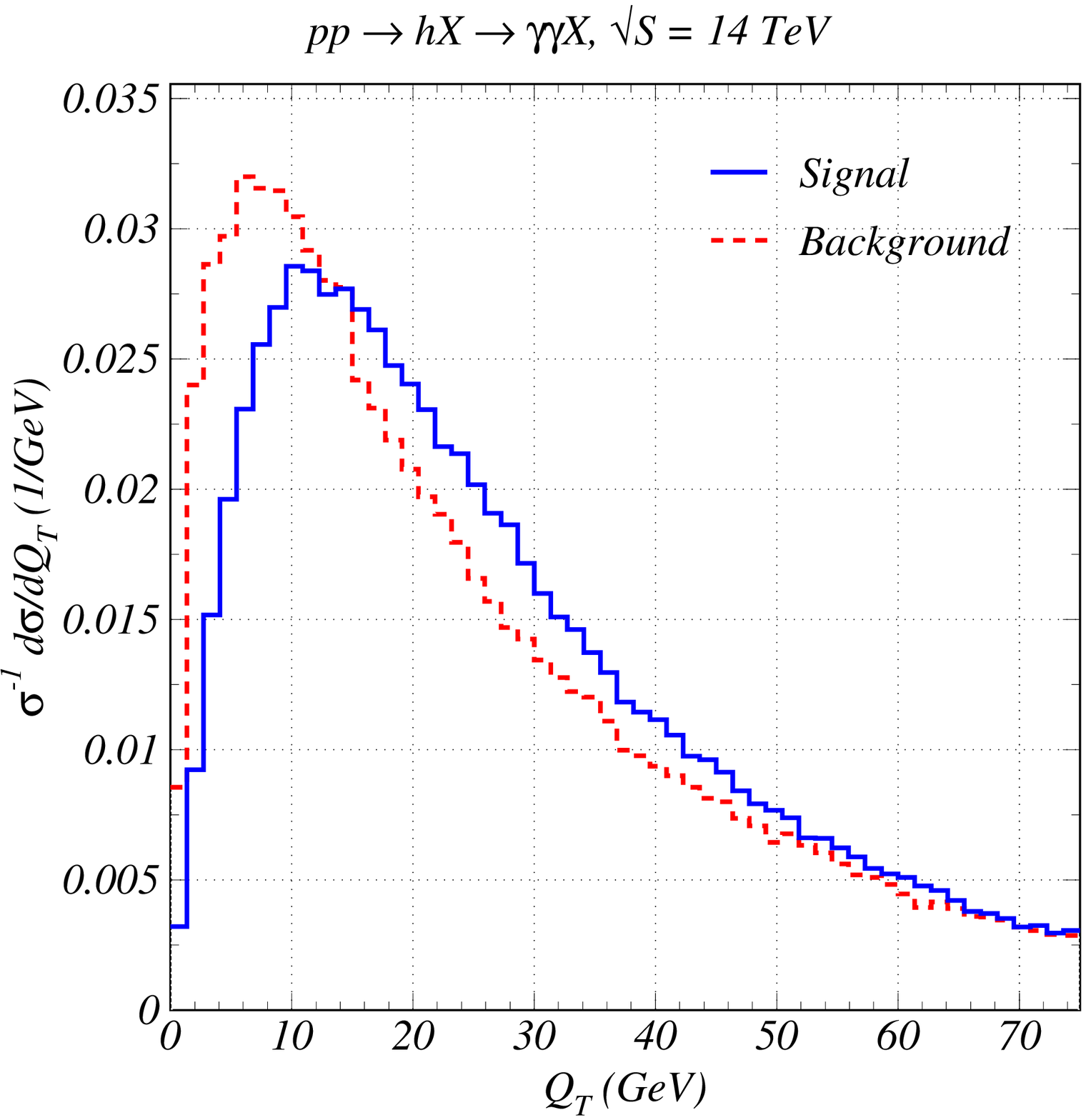}
\includegraphics[width=0.49\columnwidth,keepaspectratio]{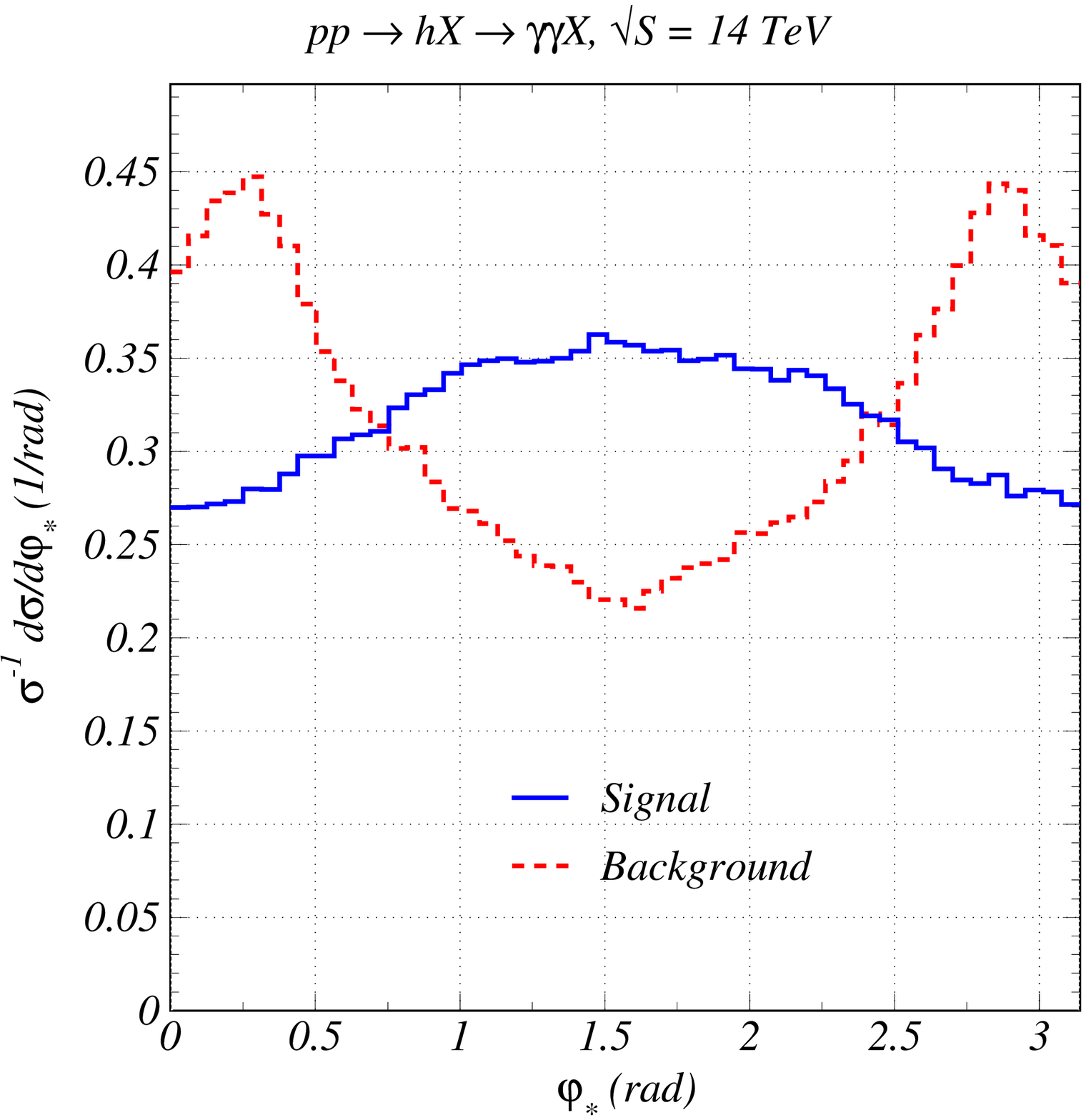}
\includegraphics[width=0.49\columnwidth,keepaspectratio]{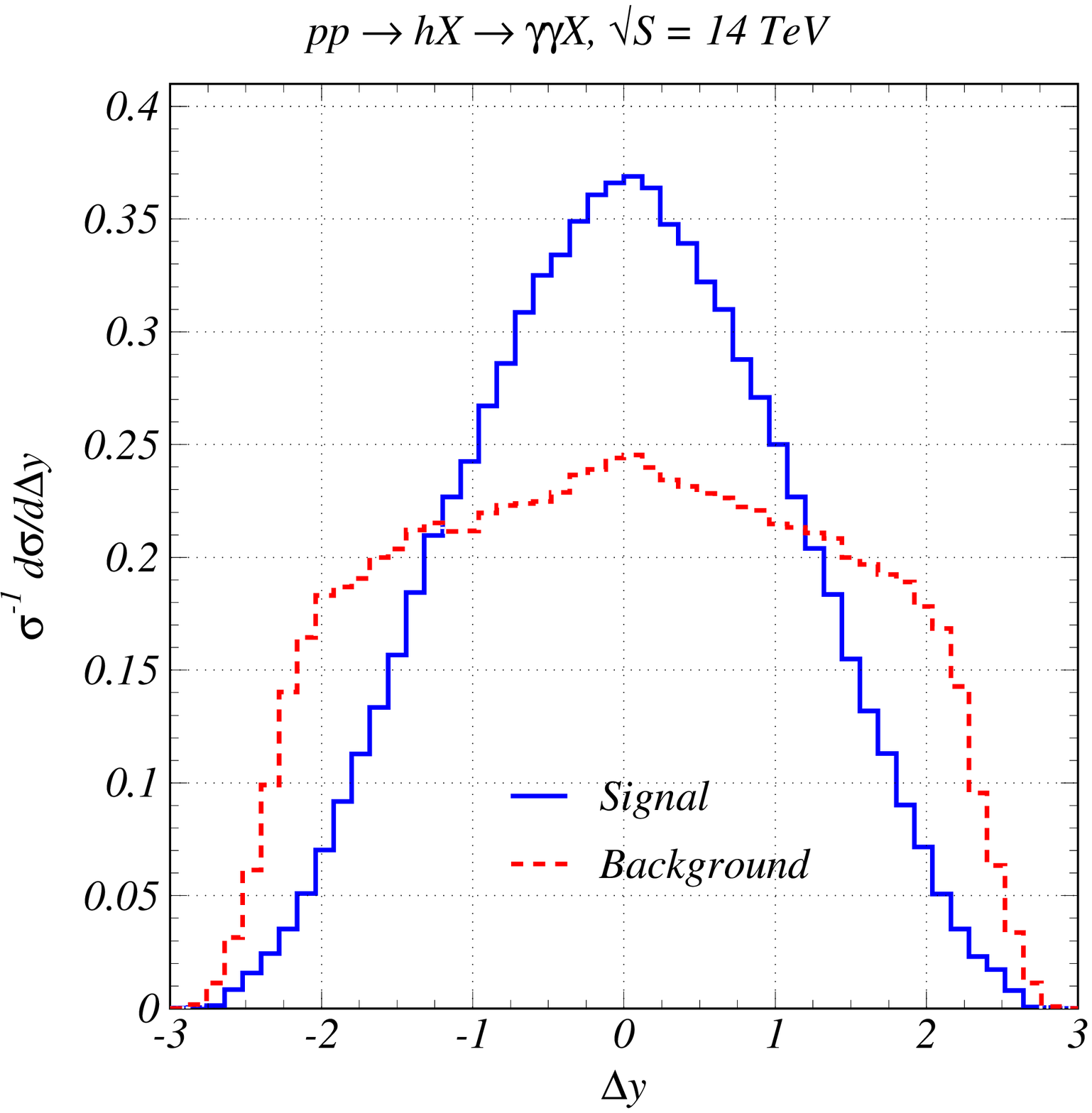}
\includegraphics[width=0.49\columnwidth]{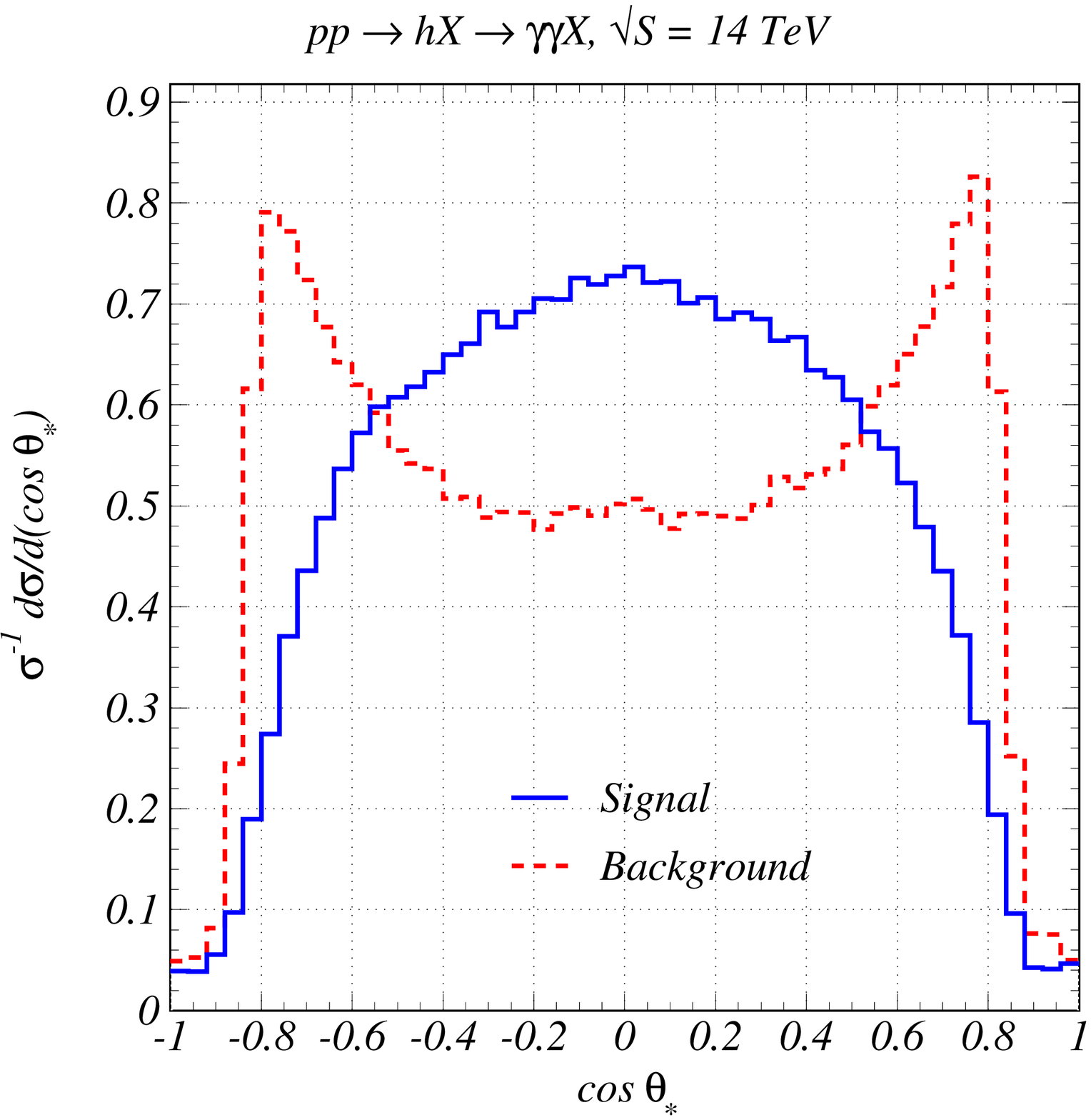}

\caption{Comparison of the normalized Higgs boson signal and diphoton background
distributions at the LHC, both computed at NNLL accuracy. The Higgs
boson mass is taken to be $m_{H}=130$ GeV, and the background is
calculated for $128<Q<132$ GeV.}

\label{Fig:SignalBg} 
\end{figure*}

\subsection{Comparison with Higgs boson signal distributions}

We highlight some similarities and differences between the production
spectra for the Higgs boson signal and the QCD background discussed
in this paper. We focus on the diphoton decay mode of a SM Higgs boson
produced from the dominant gluon-fusion mechanism, $gg\rightarrow h^{0}\rightarrow\gamma\gamma$,
where the Higgs boson production cross section is calculated at the
same order of precision as the QCD continuum background. We include
initial-state QCD contributions at $O(\alpha_{s}^{3})$ (NLO) and
resummed contributions at NNLL accuracy. These contributions are also
coded in \textsc{ResBos}~\cite{Balazs:2000wv}, and we can apply
the same cuts on the momenta of the photons to the signal and background.
Our findings should remain broadly applicable after the NNLO corrections to
Higgs boson production \cite{Catani:2003zt,Bozzi:2005wk} are included.
We compute the background in the range $128<Q<132$ GeV, and the signal
at a fixed Higgs boson mass $m_{H}=130$ GeV. We impose the kinematic
selection $Q_{T}<Q$, but its influence is not important at the large
values of diphoton mass of interest here. %

The cross section times branching ratio for the Higgs boson signal
is substantially smaller than the QCD continuum. To better illustrate
their differences, Fig.~\ref{Fig:SignalBg} presents distributions
normalized to the respective total rates. The top-left panel shows
normalized transverse momentum distributions of photon pairs. The
signal and background peak at about 12 and 5 GeV, respectively. The
average values of $Q_{T}$ are 26 and 23 GeV, computed over the range
0 to 75 GeV.

Differences in the shapes of these $Q_{T}$ spectra can be attributed
to the different structure of the leading terms in the initial-state
Sudakov exponents and to the effects of final-state photon fragmentation.
The Higgs boson signal is controlled by the characteristics of the
$gg+gq_{S}$ initial state, whereas the continuum is controlled primarily
by the $q\bar{q}+qg$ initial state. Because the dominant Sudakov
coefficient $\mathcal{A}_{q}^{(k)} \propto C_{F}$ in the $q\bar{q}$ case
is smaller than $\mathcal{A}_{g}^{(k)}= (C_{A}/C_{F})\mathcal{A}_{q}^{(k)}$ 
in the $gg$ case,
the resummed $q\bar{q}+qg$ initial-state radiation produces narrower
$Q_{T}$ distributions than $gg+gq_{S}$ initial-state radiation.
About 80\% of the diphoton rate is provided by the $q\bar{q}+qg$
channel, implying a narrower $Q_{T}$ distribution of the background,
if based on the value of $\mathcal{A}^{(k)}$ alone.

The continuum background contribution is also enhanced by final-state
radiation in $qg$ scattering. The $Q_{T}$ profile of the final-state
collinear terms depends more on the isolation model (including $E_{T}^{iso}$
and $\Delta R$) than on the initial-state Sudakov exponent. For the
nominal ATLAS cuts, the final-state collinear contribution in our
calculation hardens the background $Q_{T}$ distribution, diminishing
its difference from the Higgs boson signal distribution. More effective
isolation may reduce the impact of the final-state radiation on $Q_{T}$
distributions.

Another difference between the signal and continuum is observed in
the distribution in the azimuthal angle of the photons, such as the
angle $\varphi_{*}$ in the Collins-Soper frame shown in the top-right 
panel of Fig.~\ref{Fig:SignalBg}.
This distribution is qualitatively the same if integrated over all
$Q_{T}$, as in Fig.~\ref{Fig:SignalBg}, or integrated above some
minimal $Q_{T}$ value, as in an experimental measurement.
Without isolation imposed, the spin-$0$ Higgs boson signal must be
flat in $\varphi_{*},$ but the QCD background peaks toward $\varphi_{*}=0$
and $\pi$ (i.e., $\sin\varphi_{*}=0$) as a result of the final-state
$qg$ singularity.%
\footnote{By definition, the recoil parton 5 always lies in the $Oxz$ plane (has zero azimuthal
angle) in the Collins-Soper frame. For the final-state singularity
to occur at NLO, the photons should be in the same
plane with 5, i.e., have $\sin\varphi_{*}=0$.}${}^,$ 
\footnote{One of the resummed structure functions for the $gg$ background is
modulated by $\cos2\varphi_{*}$ (see Sec.~\ref{subsection:ISRResummation}),
but we neglect this modulation in our present calculation.%
} Isolation cuts suppress both the signal and the background for $\sin\varphi_{*}<\sin\Delta R$.
The result is a signal distribution with a broad peak near $\varphi_{*}=\pi/2$,
while the background favors values of $\varphi_{*}$ near $0$ and
$\pi$. A selection of events with $\varphi_{*}$ in the vicinity
of $\pi/2$, and $Q_{T}$ large enough, helps to reduce the impact
of the $qg$ background.In the lab frame, a related distribution 
is in the
variable $\left|\varphi_{3T}-\varphi_{4T}\right|,$where $\varphi_{iT}$
is the azimuthal angle between $\vec{p}_{T}^{\gamma_{i}}$ and $\vec{Q}_{T}$.
The signal (background) processes tend to have more events with large
(small) magnitude of $\left|\varphi_{3T}-\varphi_{4T}\right|$.

A third potential discriminator between the signal and background
is the difference in the rapidities $\Delta y=y_{hard}-y_{soft}$
of the photons with harder and softer values of $p_{T}^{\gamma}$
in the lab frame, calculated on an event by event basis. This distribution
is displayed in the lower-left frame of Fig.~\ref{Fig:SignalBg}.
The background distribution peaks at the origin, while the signal
is almost flat over a wide range of $\Delta y$. Different spin correlations
in the decay of a spin-0 Higgs boson from those characteristic of
QCD background processes are the source of this distinction. Discrimination
based on this difference can improve the statistical significance
of the signal~\cite{Bern:2002jx}. We note that our resummed calculation
does not exhibit the kinematic singularity at $\Delta y\approx2$
present in the finite-order cross section and obvious in Fig. 10 
of Ref.~\cite{Bern:2002jx}, where the distribution with respect to
$y^*\equiv\Delta y/2$ is shown. The
discontinuity in $d\sigma/dQ_{T}$ caused by the finite-order approximation
is resummed in our calculation, yielding a smooth result.

The rapidity difference is related to the scattering angle in the
Collins-Soper frame: $\tanh(\Delta y/2)=\cos\theta_{*}$ when $Q_{T}$
is zero. The $\cos\theta_{*}$ distribution is shown in the lower-right
frame of Fig.~\ref{Fig:SignalBg}. The difference between the signal
and background rates is even more pronounced in this variable, clearly
expressing the difference in the spin correlations of the systems producing the
photons.

A comparison of $Q_{T}$ distributions in the top-left panel of Fig.~\ref{Fig:SignalBg}
suggests that the signal versus background ratio would be enhanced
if a cut is made to restrict $Q_{T}>10$~GeV. After applying this
cut, we may again examine the distributions in the rapidity difference
of the two photons, the scattering angle in the Collins-Soper frame,
and the azimuthal angle distribution of the photons in the Collins-Soper
frame. The results are qualitatively similar to those in Fig.~\ref{Fig:SignalBg}
and are not shown here. A more efficient procedure to increase the
Higgs boson discovery significance is to apply a simultaneous likelihood
analysis to several kinematic distributions. Based on the present
discussion, we would argue that the resummed $Q_{T}$, $\varphi_{*}$,
and $\cos\theta_{*}$ distributions are good discriminators between
the Higgs boson signal and background in such an analysis.

\section{Conclusions \label{Sec:conclusions}}

The theoretical study of continuum diphoton production in hadron collisions is 
interesting and valuable for several reasons{\bf:}   
there are data from the CDF and D\O~ collaborations at Fermilab 
with the promise of larger event samples; there are new 
theoretical challenges associated with all-orders soft-gluon 
resummation of two-loop amplitudes; and continuum diphotons are 
a large standard-model background above which one may observe 
the products of Higgs boson decay into a pair of photons at the 
LHC.  

In this paper and Refs.~\cite{Balazs:2006cc,Nadolsky:2007ba}, we present our calculation of the fully differential cross 
section $d\sigma/(dQdQ_{T}dyd\Omega_{*})$ as a function of the 
mass $Q$, transverse momentum $Q_T$, and rapidity $y$ of the diphoton 
system, and of the polar and azimuthal angles of the individual photons 
in the diphoton rest frame.  Our basic QCD hard-scattering subprocesses 
are all computed at next-to-leading order (NLO) in the strong coupling strength 
$\alpha_s$, and we include the state-of-art resummation of initial-state gluon radiation 
to all orders in $\alpha_s$, valid to next-to-next-to-leading logarithmic 
accuracy (NNLL). Resummation is essential for a realistic and reliable 
calculation of the $Q_T$ dependence in the region of small and intermediate 
values of $Q_T$, where the cross section is greatest.  It is also needed 
for stable estimates of the effects of experimental acceptance on distributions 
in the diphoton invariant mass and other variables.

Our analytical results are included in  a fully updated \textsc{ResBos} 
code~\cite{Balazs:1997xd,Balazs:1999gh}. 
This numerical program allows us to impose selections on the transverse 
momenta and angles of the final photons, in order to match those employed 
by the CDF and D\O~ collaborations, as well as those anticipated in experiments at the LHC.  Our predictions are especially 
pertinent in the region $Q_{T}\lesssim Q$.  We show that our results at 
the Tevatron and at the LHC are insensitive to the choice of the resummation 
scheme and of the nonperturbative functions required by the integration 
into the region of large impact parameter.  

The published collider data are presented in the form of singly
differential distributions.  We follow suit in order to make comparisons, and we 
find excellent agreement with data, as shown in Sec.~\ref{Sec:Phenomenology}.  
We recommend that more differential studies be
made, and, to motivate these, we present predictions for the changes expected 
in the $Q_T$ distribution as a function of mass $Q$, and for the dependence of 
the mean transverse momentum on $Q$.  

We make predictions for continuum diphoton mass, transverse momentum, and angular 
distributions at the energy of the LHC. Moreover, we contrast in Fig.~\ref{Fig:SignalBg} 
the shapes of some of these distributions with those expected from the decay of a 
Higgs boson. The distinct features of the signal and background suggest that   
that the Higgs boson discovery significance can be increased via a simultaneous 
likelihood analysis of several kinematic distributions, particularly the 
resummed $Q_{T}$, $\varphi_{*}$, and $\cos\theta_{*}$ distributions.

Another approach to the computation of continuum diphoton production is 
presented by the \textsc{DIPHOX} collaboration~\cite{Binoth:1999qq}.  This  
calculation includes both the direct production of photons from hard-scattering 
processes and the photons produced from fragmentation of (anti-)quarks or gluons. 
It is valid at NLO, except for the $gg$ subprocess, which is included at leading 
order only. The \textsc{DIPHOX} code is useful for rates integrated 
over transverse momentum, but it is not designed to predict the transverse 
momentum distribution or other distributions 
sensitive to the region in which the transverse momentum of the diphoton pair 
is small.  Compared to a fixed-order calculation, such as direct photon pair 
calculation in \textsc{DIPHOX}, our calculation improves
the theoretical prediction for event distributions which are sensitive
to the region of low $Q_{T}$.   Furthermore, our calculation includes the NLO 
contribution from the combined $gg+gq_{S}$ channel, leading to more accurate 
predictions at the LHC, where the $gg+gq_{S}$ contribution is generally not small.

Only {\em isolated}, not inclusive, photons are identified experimentally. 
While it is straightforward to define an isolated photon in a given experiment, 
it is challenging to devise a theoretical prescription that can match the  
experimental definition, short of first understanding the long-distance dynamics 
of QCD.  The isolated diphoton production rate is modeled in the 
\textsc{DIPHOX} code by including explicit photon fragmentation function 
contributions at 
NLO accuracy.  A shortcoming of this approach (as well as of our method for 
treating isolation) is that one cannot accurately represent 
photon fragmentation without including final-state parton showering in 
the presence of isolation constraints.  There is   
inevitable ambiguity and uncertainty in the choice of the  
``isolation energy'' used to define an isolated photon 
theoretically for comparison with the isolated photon measured experimentally.  
As shown in Sec.~\ref{Sec:Phenomenology}, the \textsc{DIPHOX} cross section 
can vary by a large factor in some regions of phase space at the Tevatron when $E_{T}^{iso}$ 
is changed from 1\,GeV to 4\,GeV.

Our approach is to concentrate on physical observables which are less
sensitive to the fragmentation contributions. We apply the {}``collinear
subtraction'' prescription or the {}``smooth-cone isolation'' prescription
to define an isolated photon in our calculation.  We find good agreement 
with the data, except in the region with small $Q$ and $\Delta\varphi<\pi/2$, 
consistent with our theoretical expectation that higher-order direct photon 
production
and photon fragmentation contributions can strongly modify the rate of 
diphoton pairs in this region.  We 
suggest that much better agreement with current and future data will be 
obtained if an addition requirement of $Q_{T}<Q$ is applied. With this cut, 
the fragmentation contributions are largely suppressed.  With the cut 
$Q_{T}<Q$ cut applied to the Tevatron data, the enhancement at low 
$\Delta\varphi$ and intermediate $Q_{T}$ (the shoulder region) should 
disappear. We urge the CDF and D\O\, collaborations to apply these cuts in
future analyses of the diphoton data.  

In our calculation, we identify an interesting spin-flip
contribution (with $\cos2\varphi_{*}$ dependence) in the $gg$
channel, cf. Ref.~\cite{Nadolsky:2007ba}, and we suggest that measurements be 
made of the 
distribution of $\varphi_{*}$ as a function of $Q_{T}$. All-orders resummation 
of the gluon spin-flip contribution may be needed when a 
larger statistical sample of diphoton data is available.  

The contributions from $qg+{\bar{q}}g$ processes become more important at 
the LHC than at the Tevatron, and calculations at a higher order of precision 
may be warranted eventually. To improve the theoretical
prediction in the region of phase space with $Q_{T}<E_{T}^{iso}$
and $\varphi_{*}\sim0$ or $\pi$, a joint resummation calculation
is needed in which the effects of both the initial- and final-state multiple 
parton emissions are treated simultaneously.  

Although we emphasize that better agreement of our predictions with
data should be apparent if the selection $Q_{T}<Q$ is made, we also point 
out that the region $Q_{T} > Q$ should manifest very interesting physics of 
a different sort. Additional logarithmic singularities of the form 
$\log(Q/Q_{T})$ are encountered in the region $Q_{T} \gg Q$.  These
logarithms are associated with the fragmentation of a parton carrying
large transverse momentum $Q_{T}$ into a system of small invariant
mass $Q$ \cite{Berger:1998ev,Berger:2001wr}, a light $\gamma\gamma$
pair in our case.  Small-$Q$ $\gamma\gamma$ fragmentation of this
kind is not implemented yet in theoretical models. 
Experimental study of the region $Q_{T} \gg Q$ may offer the
opportunity to measure the parton to two-photon fragmentation function
$D_{\gamma\gamma}(z_{1},z_{2})$. 

\section*{Acknowledgments}

Research in the High Energy Physics Division at Argonne is supported
in part by US Department of Energy, Division of High Energy Physics,
Contract DE-AC02-06CH11357. The work of C.-P. Y. is supported by the
U. S. National Science Foundation under grant PHY-0555545. C. B. thanks
the Fermilab Theoretical Physics Department, where a part of this
work was done, for its hospitality and financial support. The diagrams 
in Figs.~\ref{Fig:FeynDiag} and~\ref{fig:LowQfrag} were drawn with 
aid of the program \textsc{JaxoDraw}
\cite{Binosi:2003yf}.

\appendix

\section{Summary of perturbative coefficients \label{Appendix:Summary}}

In this appendix we present an overview of the perturbative QCD expressions
for the resummed and asymptotic cross sections used in our computation.

The functions $\mathcal{A}_{a}(C_{1},\bar{\mu}),$ $\mathcal{B}_{a}(C_{1},C_{2},\bar{\mu}),$
$\mathcal{C}_{a/a_{1}}(x,b;C_{1}/C_{2},\mu)$, and $h_{a}(Q,\theta_{*})$
are introduced in Sec.~\ref{Sec:Theory}. These functions are derived
as perturbative expansions in the small parameter $\alpha_{s}/\pi$:\begin{eqnarray*}
 &  & \mathcal{A}_{a}(C_{1},\bar{\mu})=\sum_{n=1}^{\infty}\mathcal{A}_{a}^{(n)}(C_{1})\left(\frac{\alpha_{s}(\bar{\mu})}{\pi}\right)^{n};\,\mathcal{B}_{a}(C_{1},C_{2},\bar{\mu})=\sum_{n=1}^{\infty}\mathcal{B}_{a}^{(n)}(C_{1},C_{2})\left(\frac{\alpha_{s}(\bar{\mu})}{\pi}\right)^{n};\\
 &  & \mathcal{C}_{a/a_{1}}\left(x,b;\frac{C_{1}}{C_{2}},\mu\right)=\sum_{n=0}^{\infty}\mathcal{C}_{a/a_{1}}^{(n)}(x,b\mu,\frac{C_{1}}{C_{2}})\left(\frac{\alpha_{s}(\mu)}{\pi}\right)^{n};\, h_{a}(Q,\theta_{*})=\sum_{n=0}^{\infty}h_{a}^{(n)}(\theta_{*})\left(\frac{\alpha_{s}(Q)}{\pi}\right)^{n}.\end{eqnarray*}
 The value of a perturbative coefficient $F^{(n)}$ for a set of scales
$C_{1}/b$ and $C_{2}Q$ can be expressed in terms of its value $F^{(n,c)}$
obtained for the {}``canonical'' combination $C_{1}=c_{0}$ and
$C_{2}=1.$ Here $c_{0}\equiv2e^{-\gamma_{E}}\approx1.123$, where $\gamma_{E}=0.5772\dots$ 
is the Euler constant.  The
relationships between $F^{(n)}$ and $F^{(n,c)}$ take the form \begin{eqnarray}
\mathcal{A}_{a}^{(1)}(C_{1}) & = & \mathcal{{\mathcal{A}}}_{a}^{(1,c)};\\
\mathcal{A}_{a}^{(2)}(C_{1}) & = & \mathcal{{\mathcal{A}}}_{a}^{(2,c)}-\mathcal{{\mathcal{A}}}_{a}^{(1,c)}\beta_{0}\ln\frac{c_{0}}{C_{1}};\\
\mathcal{A}_{a}^{(3)}(C_{1}) & = & \mathcal{A}_{a}^{(3,c)}-2\mathcal{A}_{a}^{(2,c)}\beta_{0}\ln\frac{c_{0}}{C_{1}}-\frac{\mathcal{A}_{a}^{(1,c)}}{2}\beta_{1}\ln\frac{c_{0}}{C_{1}}+\mathcal{A}_{a}^{(1,c)}\beta_{0}^{2}\left(\ln\frac{c_{0}}{C_{1}}\right)^{2};\\
\mathcal{B}_{a}^{(1)}(C_{1},C_{2}) & = & \mathcal{B}_{a}^{(1,c)}-\mathcal{A}_{a}^{(1,c)}\ln\frac{c_{0}^{2}C_{2}^{2}}{C_{1}^{2}};\\
\mathcal{B}_{a}^{(2)}(C_{1},C_{2}) & = & \mathcal{B}_{a}^{(2,c)}-\mathcal{{\mathcal{A}}}_{a}^{(2,c)}\ln\frac{c_{0}^{2}C_{2}^{2}}{C_{1}^{2}}\nonumber \\
 & + & \beta_{0}\left[\mathcal{{\mathcal{A}}}_{a}^{(1,c)}\ln^{2}\frac{c_{0}}{C_{1}}+\mathcal{B}_{a}^{(1,c)}\ln C_{2}-\mathcal{{\mathcal{A}}}_{a}^{(1,c)}\ln^{2}C_{2}\right];\\
\mathcal{C}_{a/a_{1}}^{(1)}(x,b\mu,\frac{C_{1}}{C_{2}}) & = & \mathcal{C}_{a/a_{1}}^{(1,c)}(x)+\delta_{aa_{1}}\delta(1-x)\left(\frac{\mathcal{B}_{a}^{(1,c)}}{2}\ln\frac{c_{0}^{2}C_{2}^{2}}{C_{1}^{2}}-\frac{\mathcal{A}_{a}^{(1,c)}}{4}\left(\ln\frac{c_{0}^{2}C_{2}^{2}}{C_{1}^{2}}\right)^{2}\right)\nonumber \\
 & - & P_{a/a_{1}}(x)\ln\frac{\mu b}{c_{0}}.\end{eqnarray}
 They depend on the QCD beta-function coefficients $\beta_{0}=(11N_{c}-2N_{f})/6$,
$\beta_{1}=(17N_{c}^{2}-5N_{c}N_{f}-3C_{F}N_{f})/6$ for $N_{c}$
colors and $N_{f}$ active quark flavors, with $C_{F}=(N_{c}^{2}-1)/(2N_{c})=4/3$
for $N_{c}=3$. The relevant ${\mathcal{O}}(\alpha_{s})$ splitting
functions $P_{a/a_{1}}(x)$ are\begin{eqnarray}
 &  & P_{q/q}=C_{F}\left(\frac{1+z^{2}}{1-x}\right)_{+};\, P_{q/g}=\frac{1}{2}(1+2x+2x^{2});\, P_{g/q_{S}}=C_{F}\frac{(1-x)^{2}+1}{x};\\
 &  & P_{g/g}=2C_{A}\left[\frac{x}{(1-x)_{+}}+\frac{1-x}{x}+x(1-x)\right]+\beta_{0}\delta(1-x).\end{eqnarray}

The coefficients $h^{(1)}(\theta_{*})$, ${\cal B}^{(2)}$, and
$\mathcal{C}^{(1)}$ depend on the resummation scheme. The hard-scattering
function is

\begin{equation}
h_{a}(Q,\theta_{*})=1+\delta_{s}\frac{\alpha_{s}(Q)}{\pi}\frac{\mathcal{V}_{a}(\theta_{*})}{4}+...,\end{equation}
 where $\delta_{s}=0$ in the CSS scheme and $\delta_{s}=1$ in the
CFG scheme. The functions $\mathcal{V}_{q}(\theta_{*})$ for $q\bar{q}\rightarrow\gamma\gamma$
scattering and $\mathcal{V}_{g}(\theta_{*})$ for $gg\rightarrow\gamma\gamma$
scattering are derived in Refs.~\cite{Balazs:1997hv} and \cite{Nadolsky:2002gj},
respectively.

For the $q\bar{q}+qg$ initial state, we obtain the following expressions
for the coefficients $\mathcal{A}$, $\mathcal{B}$, and ${\mathcal{C}}$:
\begin{eqnarray}
\mathcal{A}_{q}^{(1,c)} & = & C_{F};\nonumber \\
\mathcal{A}_{q}^{(2,c)} & = & C_{F}\left[\left(\frac{67}{36}-\frac{\pi^{2}}{12}\right)C_{A}-\frac{5}{9}T_{R}N_{f}\right];\\
\mathcal{A}_{q}^{(3,c)} & = & \frac{C_{F}^{2}N_{f}}{2}\left(\zeta(3)-\frac{55}{48}\right)-\frac{C_{F}N_{f}^{2}}{108}+C_{A}^{2}C_{F}\left(\frac{11\zeta(3)}{24}+\frac{11\pi^{4}}{720}-\frac{67\pi^{2}}{216}+\frac{245}{96}\right)\nonumber \\
 & + & C_{A}C_{F}N_{f}\left(-\frac{7\zeta(3)}{12}+\frac{5\pi^{2}}{108}-\frac{209}{432}\right);\nonumber \\
\mathcal{B}_{q}^{(1,c)} & = & -\frac{3}{2}C_{F};\nonumber \\
\mathcal{B}_{q}^{(2,c)} & = & -\frac{1}{2}\left[{C_{F}}^{2}\,\left(\frac{3}{8}-\frac{\pi^{2}}{2}+6\zeta(3)\right)\right.+C_{F}C_{A}\left(\frac{17}{24}+\frac{11\pi^{2}}{18}-3\zeta(3)\right)\nonumber \\
 & - & \left.C_{F}N_{f}T_{R}\left(\frac{1}{6}+\frac{2\pi^{2}}{9}\right)\right]+\beta_{0}\left[\frac{C_{F}\pi^{2}}{12}+(1-\delta_{s})\frac{\mathcal{V}_{q}(\theta_{*})}{4}\right];\nonumber \\
\mathcal{C}_{j/k}^{(0)}(x) & = & \delta_{jk}\delta(1-x);\,\,\mathcal{C}_{j/g}^{(0)}(x)=0;\nonumber \\
\mathcal{C}_{j/k}^{(1,c)}(x) & = & \delta_{jk}\left\{ \frac{C_{F}}{2}(1-x)+\delta(1-x)(1-\delta_{s})\frac{\mathcal{V}_{q}(\theta_{*})}{4}\right\} ;\nonumber \\
\mathcal{C}_{j/g}^{(1,c)}(x) & = & {\frac{1}{2}}x(1-x).\end{eqnarray}
 Here $C_{A}=N_{c},$ $T_{R}=1/2$, and the Riemann constant $\zeta(3)=1.202\dots$
. The ${\mathcal{C}}$ functions are given for $j,k=u,\bar{u},d,\bar{d},\dots$.
These coefficients are taken from \cite{Balazs:1997hv,deFlorian:2000pr,Moch:2004pa}.

Similarly, the ${\mathcal{A}}$, ${\mathcal{B}}$, and
${\mathcal{C}}$ coefficients in the $gg+gq_{S}$ channel are \begin{eqnarray}
\mathcal{A}_{g}^{(k,c)} & = & (C_{A}/C_{F})\mathcal{A}_{q}^{(k,c)},\mbox{ for }k=1,2,3;\nonumber \\
\mathcal{B}_{g}^{(1,c)} & = & -\beta_{0};\nonumber \\
\mathcal{B}_{g}^{(2,c)} & = & -\frac{1}{2}\Biggl[C_{A}^{2}\left(\frac{8}{3}+3\zeta(3)\right)-C_{F}T_{R}N_{f}-\frac{4}{3}C_{A}T_{R}N_{f}\Biggr]+\beta_{0}\left[\frac{C_{A}\pi^{2}}{12}+(1-\delta_{s})\frac{\mathcal{V}_{g}(\theta_{*})}{4}\right];\nonumber \\
\mathcal{C}_{g/a}^{(0)}\left(x\right) & = & \delta_{ga}\delta(1-x);\,\,\mathcal{C}_{g/g}^{(1,c)}\left(x\right)=\delta(1-x)(1-\delta_{s})\frac{\mathcal{V}_{g}(\theta_{*})}{4};\,\,\mathcal{C}_{g/q_{S}}^{(1,c)}\left(x\right)=\frac{C_{F}}{2}x.\end{eqnarray}
 These coefficients are taken from Refs.~\cite{Balazs:1997hv,Nadolsky:2002gj,Yuan:1991we,Vogt:2004mw}.

\section{Components of the asymptotic cross sections \label{Appendix:Asymptotic}}

In Sec.~\ref{subsection:ISRResummation} we introduce asymptotic
small-$Q_{T}$ approximations for the $q\bar{q}+qg$ and $gg+gq_{S}$
NLO cross sections,\begin{equation}
A_{q\bar{q}}(Q,Q_{T},y,\Omega_{*})=\sum_{i=u,\bar{u},d,\bar{d},...}\frac{\Sigma_{i}(\theta_{*})}{S}\left\{ \delta(\vec{Q}_{T})F_{i,\delta}(Q,y,\theta_{*})+F_{i,+}(Q,y,Q_{T})\right\} ,\label{ASYqqbar2}\end{equation}
 and\begin{eqnarray}
A_{gg}(Q,Q_{T},y,\Omega_{*}) & = & \frac{1}{S}\Biggl\{\Sigma_{g}(\theta_{*})\left[\delta(\vec{Q}_{T})F_{g,\delta}(Q,y,\theta_{*})+F_{g,+}(Q,y,Q_{T})\right]\nonumber \\
 &  & \hspace{12pt}+\Sigma_{g}^{\prime}(\theta_{*},\varphi_{*})F_{g}^{\prime}(Q,y,Q_{T})\Biggr\}.\label{ASYgg2}\end{eqnarray}
 The functions $F$ in these equations are defined as\begin{eqnarray}
 &  & F_{i,\delta}(Q,y,\theta_{*})\equiv f_{q_{i}/h_{1}}(x_{1},\mu_{F})f_{\bar{q}_{i}/h_{2}}(x_{2},\mu_{F})\left(1+2\frac{\alpha_{s}}{\pi}h_{q}^{(1)}(\theta_{*})\right)\nonumber \\
 & + & \frac{\alpha_{s}}{\pi}\Biggl\{\left(\left[\mathcal{C}_{q_{i}/a}^{(1,c)}\otimes f_{a/h_{1}}\right](x_{1},\mu_{F})-\left[P_{q_{i}/a}\otimes f_{a/h_{1}}\right](x_{1},\mu_{F})\,\ln\frac{\mu_{F}}{Q}\right)\, f_{\bar{q}_{i}/h_{2}}(x_{2},\mu_{F})\nonumber \\
 &  & \hspace{26pt}+f_{q_{i}/h_{1}}(x_{1},\mu_{F})\left(\left[\mathcal{C}_{\bar{q}_{i}/a}^{(1,c)}\otimes f_{a/h_{2}}\right](x_{2},\mu_{F})-\left[P_{\bar{q}_{i}/a}\otimes f_{a/h_{2}}\right](x_{2},\mu_{F})\,\ln\frac{\mu_{F}}{Q}\right)\Biggr\};\label{FqDelta}\end{eqnarray}
 \begin{eqnarray}
F_{q,+} & = & \frac{1}{2\pi}\frac{\alpha_{s}}{\pi}\Biggl\{ f_{q_{i}/h_{1}}(x_{1},\mu_{F})f_{\bar{q}_{i}/h_{2}}(x_{2},\mu_{F})\left(\mathcal{A}_{q}^{(1,c)}\left[\frac{1}{Q_{T}^{2}}\ln\frac{Q^{2}}{Q_{T}^{2}}\right]_{+}+\mathcal{B}_{q}^{(1,c)}\left[\frac{1}{Q_{T}^{2}}\right]_{+}\right)\nonumber \\
 & + & \left[\frac{1}{Q_{T}^{2}}\right]_{+}\Bigl(\left[P_{q_{i}/a}\otimes f_{a/h_{1}}\right](x_{1},\mu_{F})\, f_{\bar{q}_{i}/h_{2}}(x_{2},\mu_{F})\nonumber \\
 &  & \hspace{47pt}+f_{q_{i}/h_{1}}(x_{1},\mu_{F})\left[P_{\bar{q}_{i}/a}\otimes f_{a/h_{2}}\right](x_{2},\mu_{F})\Bigr)\Biggr\};\label{FqPlus}\end{eqnarray}
 \begin{eqnarray}
 &  & F_{g,\delta}\equiv f_{g/h_{1}}(x_{1},\mu_{F})f_{g/h_{2}}(x_{2},\mu_{F})\left(1+2\frac{\alpha_{s}}{\pi}h_{g}^{(1)}(\theta_{*})\right)\nonumber \\
 & + & \frac{\alpha_{s}}{\pi}\Biggl\{\left(\left[\mathcal{C}_{g/a}^{(1,c)}\otimes f_{a/h_{1}}\right](x_{1},\mu_{F})-\left[P_{g/a}\otimes f_{a/h_{1}}\right](x_{1},\mu_{F})\,\ln\frac{\mu_{F}}{Q}\right)\, f_{g/h_{2}}(x_{2},\mu_{F})\nonumber \\
 &  & \hspace{27pt}+f_{g/h_{1}}(x_{1},\mu_{F})\left(\left[\mathcal{C}_{g/a}^{(1,c)}\otimes f_{a/h_{2}}\right](x_{2},\mu_{F})-\left[P_{g/a}\otimes f_{a/h_{2}}\right](x_{2},\mu_{F})\,\ln\frac{\mu_{F}}{Q}\right)\Biggr\};\label{FgDelta}\end{eqnarray}
 \begin{eqnarray}
F_{g,+} & = & \frac{1}{2\pi}\frac{\alpha_{s}}{\pi}\Biggl\{ f_{g/h_{1}}(x_{1},\mu_{F})f_{g/h_{2}}(x_{2},\mu_{F})\left(\mathcal{A}_{g}^{(1,c)}\left[\frac{1}{Q_{T}^{2}}\ln\frac{Q^{2}}{Q_{T}^{2}}\right]_{+}+\mathcal{{\mathcal{B}}}_{g}^{(1,c)}\left[\frac{1}{Q_{T}^{2}}\right]_{+}\right)\nonumber \\
 & + & \left[\frac{1}{Q_{T}^{2}}\right]_{+}\Bigl(\left[P_{g/a}\otimes f_{a/h_{1}}\right](x_{1},\mu_{F})\, f_{g/h_{2}}(x_{2},\mu_{F})\nonumber \\
 &  & \hspace{47pt}+f_{g/h_{1}}(x_{1},\mu_{F})\left[P_{g/a}\otimes f_{a/h_{2}}\right](x_{2},\mu_{F})\Bigr)\Biggr\};\label{FgPlus}\end{eqnarray}
 and\begin{eqnarray}
F_{g,+}^{\prime} & = & \frac{1}{2\pi}\frac{\alpha_{s}}{\pi}\left[\frac{1}{Q_{T}^{2}}\right]_{+}\Bigl(\left[P_{g/g}^{\prime}\otimes f_{g/h_{1}}\right](x_{1},\mu_{F})\, f_{g/h_{2}}(x_{2},\mu_{F})\nonumber \\
 &  & \hspace{77pt}+f_{g/h_{1}}(x_{1},\mu_{F})\left[P_{g/g}^{\prime}\otimes f_{g/h_{2}}\right](x_{2},\mu_{F})\Bigr)\Biggr\}.\label{FgPlusPrime}\end{eqnarray}
 Expressions for the coefficients $\mathcal{A}_{a}^{(1,c)},$ $\mathcal{B}_{a}^{(1,c)}$,
$h_{a}^{(1)}(\theta_{*}),$ $\mathcal{C}_{a/a^{\prime}}^{(1,c)}(x),$
and splitting functions $P_{a/c}(x)$, are listed in Appendix~\ref{Appendix:Summary}.
Summation over all relevant parton flavors $a^{\prime}=g,u,\bar{u,}d,\bar{d},...$
for $a=q$ and $a^{\prime}=g,q_{S}$ for $a=g$ is assumed. In addition,
the $\varphi_{*}$-dependent part $\Sigma_{g}^{\prime}(\theta_{*},\varphi_{*})F_{g}^{\prime}(Q,y,Q_{T})$
of the $gg+gq_{S}$ asymptotic cross section $A_{gg}$ contains a
splitting function\begin{equation}
P_{gg}^{\prime}(x)=2C_{A}(1-x)/x,\label{Pggprime}\end{equation}
 contributed by the interference of splitting amplitudes with opposite
gluon polarizations in the helicity amplitude formalism \cite{Bern:1993mq,Bern:1993qk,Bern:1994zx,Bern:1994fz}.
The origin and behavior of this spin-flip function are discussed in
Ref.~\cite{Nadolsky:2007ba}.
}


\end{document}